\documentclass[a4paper,11pt]{article}
\pdfoutput=1
\usepackage{jheppub}
\usepackage[T1]{fontenc}
\usepackage{amsmath,amssymb,graphicx,multirow,float,nicefrac,booktabs,array}
\usepackage{orcidlink}

\newcolumntype{P}[1]{>{\centering\arraybackslash}p{#1}}
\newcolumntype{M}[1]{>{\centering\arraybackslash}m{#1}}

\title{A Unified Neural-Network Framework for Nucleon Imaging from Numerical Simulations of QCD}

\author[a,b]{Min-Huan Chu \orcidlink{0000-0003-4445-5448}}
\author[a]{Krzysztof Cichy \orcidlink{0000-0002-5705-3256}}
\author[c]{Martha Constantinou \orcidlink{0000-0002-6988-1745}}
\author[b]{Pawe\l{} Sznajder \orcidlink{0000-0002-2684-803X}}
\author[b]{Jakub Wagner \orcidlink{0000-0001-8335-7096}}

\affiliation[a]{Faculty of Physics and Astronomy, Adam Mickiewicz University, ul. Uniwersytetu Pozna\'nskiego 2, 61-614 Pozna\'n, Poland}
\affiliation[b]{National Centre for Nuclear Research, NCBJ, 02-093 Warsaw, Poland}
\affiliation[c]{Department of Physics, Temple University, Philadelphia, PA 19122 - 1801, USA}

\emailAdd{minhuan.chu@amu.edu.pl}

\abstract{
Parton distributions encode the momentum-space structure and, in their generalizations, the spatial tomography of quarks and gluons inside hadrons, the building blocks of visible matter. We present a unified neural-network approach that learns these distributions directly from matrix elements calculated via numerical simulations of quantum chromodynamics (QCD) on the lattice by fitting two complementary inputs simultaneously: data matched to physical quantities via known momentum-space and coordinate-space formalisms. Utilizing data from both methods stabilizes the extraction and mitigates biases that can arise when either is used alone. We validate the method on controlled mock data and apply it to lattice-QCD matrix elements to extract parton distribution functions (PDFs). We show benefits of such an approach for determining the physical quantities. We further extend the framework to zero-skewness generalized parton distributions and demonstrate nucleon tomography within the same neural-network parameterization. Our results provide an adaptable and systematically improvable approach for extracting partonic distributions from Euclidean correlators. It can incorporate polarization, additional channels, and future experimental constraints from current and future facilities, such as the Electron-Ion Collider.
}

\begin{document}
\maketitle
\flushbottom

\section{Introduction}
\label{sec:intro}

Nucleons -- the proton and the neutron -- are the basic building blocks 
of matter, providing almost the entire mass of the visible universe. 
They are the simplest hadrons composed of three quarks, yet our understanding of their internal structure remains very limited. 
Achieving a fundamental description of this structure in terms of 
the nucleon’s elementary constituents -- quarks and gluons, 
collectively known as partons -- is one of the central goals of 
the particle physics community.

In principle, all nucleon properties can be calculated from quantum chromodynamics (QCD), one of the pillars of the Standard Model of elementary particles and their interactions. However, in practice, the nucleon's structure is a difficult problem that requires nonperturbative tools and as such, it is far from a full solution. Thus, the combined theoretical efforts need to be aided by experimental progress.
For many years, the nucleon's structure was primarily investigated through inclusive processes, i.e., scattering events where only some of the final-state particles are measured, while others are unobserved.
This leads to a description in terms of parton distribution functions (PDFs). These functions, which depend on the hard scale of the process and the fraction of the hadron’s longitudinal momentum $x$ carried by the partons, encode a wealth of information and are essential for making any meaningful interpretation of measurements at hadron colliders (for a review, see, e.g.,~\cite{Gao:2017yyd}). 

In the late 1990s, a new approach was proposed \cite{Mueller:1998fv, Ji:1996ek,Ji:1996nm,Radyushkin:1996nd} to access a more comprehensive picture of hadron structure -- one that goes beyond the essentially one-dimensional information encoded in PDFs. This framework, used to describe exclusive processes, where all scattering products are measured, and based on generalized parton distributions (GPDs), allows for a unified description of both the longitudinal momentum and the transverse spatial distribution of partons, thereby enabling hadron tomography \cite{Burkardt:2002hr}. Importantly, GPDs also provide unique insights into the ``mechanical'' properties and spin structure of hadrons. Namely, they allow the determination of the pressure and shear stress induced by quarks and gluons inside the nucleon~\cite{Polyakov:2018zvc}, and, through Ji’s sum rule~\cite{Ji:1996ek}, provide access to the total angular momentum carried by partons, including their orbital contributions. This additional information carried by GPDs comes with a more involved formalism. At leading twist (LT) and leading order (LO) in the description of deeply virtual Compton scattering (DVCS), four distinct GPDs are required. Each of these depends on four variables: the hard scale of the process, the average longitudinal momentum fraction of the parton $x$, the longitudinal momentum transfer (skewness) $\xi$, and the squared four-momentum transfer~$t$.

While the scale dependence of both PDFs and GPDs is accessible via perturbative techniques in QCD, their dependence on the remaining variables must be extracted by other means. In the case of PDFs, more than four decades of dedicated global analyses, spanning data from fixed-target experiments, HERA (at DESY), the Tevatron (at Fermilab), and the Large Hadron Collider (at CERN), have led to the development of diverse theoretical frameworks and fitting methodologies \cite{Gao:2017yyd}. For GPDs, the situation is more challenging due to their more complex structure and the experimental difficulty of accessing exclusive processes. Despite these challenges, measurements such as DVCS constitute the cornerstone of the experimental programs at facilities like Jefferson Lab and the forthcoming Electron-Ion Collider (EIC). 
In particular the latter, a multi-billion-dollar flagship project of the nuclear physics community, aims to answer fundamental questions about the structure of the nucleon and other hadrons \cite{NAP25171}.
To describe exclusive scattering, several fitting strategies have been employed, ranging from model-dependent parametrizations to approaches based on conformal moments \cite{Kumericki:2009uq, Kroll:2012sm, Moutarde:2018kwr, Kumericki:2016ehc,Guo:2025muf}. More recently, artificial neural networks (ANNs) have been explored as a flexible and largely model-independent tool, enabling improved uncertainty control \cite{Dutrieux:2021wll, Moutarde:2019tqa,Cuic:2020iwt}.

In parallel, lattice QCD (LQCD) provides a complementary, first-principles approach to exploring hadronic structure. 
The framework of LQCD discretizes the QCD path integral and places the relevant degrees of freedom on a spacetime grid, with quark fields on lattice sites and gluon fields on links between them.
In this way, the formally infinite-dimensional path integral becomes finite and amenable to numerical evaluation using Monte Carlo methods.
However, PDFs and GPDs cannot be accessed directly in LQCD either, because these distributions are defined in Minkowski spacetime, via light-front correlations.
Meanwhile, the spacetime metric that needs to be employed on the lattice is Euclidean, for reasons of numerical stability. 
In Minkowski spacetime, the path-integral weight is complex and highly oscillatory.
Hence, it is necessary to Wick-rotate the temporal direction and replace real time with imaginary time, thus prohibiting access to light-front correlations.
The Euclidean QCD path integral then involves a purely real exponential factor, i.e., a Boltzmann-like weight, making its numerical evaluation well-behaved.
Despite this mismatch, access to PDFs/GPDs is still possible by using matrix elements that can be considered as Euclidean counterparts of Minkowski-spacetime definitions.
The essence of this powerful and breakthrough idea, outlined in two papers by 
Ji \cite{Ji:2013dva,Ji:2014gla}, is to replace light-front correlations with purely spatial ones, in a momentum-boosted hadron.
This approach is known as quasi-distributions or large momentum effective theory (LaMET).

Crucially, such Euclidean objects differ from their light-front counterparts only in the ultraviolet regime.
As such, the difference between these objects and physical distributions can be calculated perturbatively.
Formally, the Euclidean object can be factorized into the light-front one, i.e., expressed as a convolution of the latter with a known perturbative kernel.
This factorization, however, can be performed in two ways, which gives rise to two related approaches: the aforementioned quasi-distributions (LaMET) of Ji and the pseudo-distributions introduced by Radyushkin \cite{Radyushkin:2017cyf,Radyushkin:2019mye}, also known as short-distance expansion (SDE).
Both rely on the same lattice-evaluated matrix elements, but are analyzed differently and are therefore subject to different systematic effects.
A natural idea is, thus, to treat both approaches as complementary and use both of them in a single analysis.
This idea was discussed by Ji in a recent paper \cite{Ji:2022ezo}.
In that paper, a simple illustrative example shows how a quasi-reconstructed distribution can benefit from including information about the second moment of the distribution obtained from pseudo-PDFs.

In our work, we apply the idea of combining the LaMET and SDE approaches -systematically, using not only information from a single moment, but the full set of pseudo-PDF/GPD data.
This is one of the two essential aspects of our approach, which significantly improves on the standard practice of the community to apply exclusively one of the approaches or, at best, both of them, but in independent analyses.
The effective combination of quasi- and pseudo-distribution methods, leveraging their complementary strengths, is enabled by the second essential aspect of our work, the flexibility of artificial neural networks.
Apart from enabling a meaningful and consistent combination of Ji's and Radyushkin's approaches, it yields an efficient and robust determination of PDFs and GPDs over a broad kinematic range. As an illustration, we focus in this article on the case of unpolarized PDFs and on unpolarized zero-skewness ($\xi=0$) GPDs, which provide access to hadron tomography. This work is quite timely, as future experimental data will sharply constrain partonic imaging, increasing the need for lattice-based, systematically controlled reconstructions that connect Euclidean correlators to physical quantities.
The importance of this thread of LQCD research is illustrated by the vast amount of calculations in recent years, see, e.g.,  ~\cite{Lin:2014zya,Alexandrou:2015rja,Chen:2016utp,Alexandrou:2016jqi,Alexandrou:2017huk,Chen:2017mzz,Orginos:2017kos,Lin:2017ani,Zhang:2017zfe,Zhang:2017bzy,Bali:2018spj,Xu:2018mpf,Alexandrou:2018pbm,Lin:2018pvv,Alexandrou:2018eet,Liu:2018uuj,Zhang:2018nsy,Sufian:2019bol,Alexandrou:2019lfo,Izubuchi:2019lyk,Joo:2019jct,Joo:2019bzr,Cichy:2019ebf,Bringewatt:2020ixn,DelDebbio:2020rgv,Chai:2020nxw,Joo:2020spy,Bhat:2020ktg,Alexandrou:2020uyt,Alexandrou:2020qtt,Lin:2020ssv,Fan:2020nzz,Gao:2020ito,Lin:2020fsj,Bhattacharya:2020cen,Bhattacharya:2020xlt,Bhattacharya:2020jfj,Zhang:2020gaj,Hua:2020gnw,LatticeParton:2020uhz,Shanahan:2020zxr,Zhang:2020dbb,Bhattacharya:2021boh,Bhattacharya:2021moj,Karpie:2021pap,Alexandrou:2021oih,Egerer:2021ymv,HadStruc:2021qdf,Li:2021wvl,Shanahan:2021tst,Schlemmer:2021aij,Detmold:2021qln,Gao:2021dbh,LatticePartonLPC:2022eev,LatticePartonCollaborationLPC:2022myp,Zhang:2022xuw,Gao:2022iex,LatticeParton:2022xsd,HadStruc:2022yaw,HadStruc:2022nay,Gao:2022uhg,LatticeParton:2022zqc,Gao:2022vyh,Holligan:2023rex,Shindler:2023xpd,Gao:2023lny,Delmar:2023agv,Alexandrou:2023ucc,LatticeParton:2023xdl,LatticePartonLPC:2023pdv,LatticeParton:2024mxp,Holligan:2024umc,Holligan:2024wpv,LatticeParton:2024vck,LatticeParton:2024zko}. Notably, apart from the quasi- and pseudo-distribution approaches mentioned above, other methods to address hadron structure on the lattice exist~\cite{Liu:1993cv,Braun:1994jq,Aglietti:1998ur,Detmold:2005gg,Braun:2007wv,Chambers:2017dov,Ma:2014jla,Ma:2017pxb}, although are comparatively less popular.
A more detailed account of the progress of LQCD is reported, e.g., in the reviews~\cite{Cichy:2018mum,Ji:2020ect,Constantinou:2020pek,Cichy:2021lih,Cichy:2021ewm}.

 The rest of the paper is organized as follows. Sec.~\ref{sec:theory} establishes the theoretical framework and specifies the lattice setup and inputs for the observables under study. Sec.~\ref{sec:ann} introduces the neural-network framework with matching to the light cone, Fourier transforms, and training. Sec.~\ref{sec:pdf} presents mock-data validation of our neural-network reconstruction and demonstrates it in the determination of PDFs. Sec.~\ref{sec:gpd} generalizes the framework to zero-skewness GPDs and reports nucleon imaging and related observables. Finally, Sec.~\ref{sec:summary} summarizes our main findings and outlines prospects for future work. The appendices present full details of the framework and an extended discussion of the results.

\section{Theoretical and lattice framework}
\label{sec:theory}

\subsection{Generalities}
\label{sec:generalities}

\begin{figure}[!ht]
\centering
\includegraphics[scale=0.5]{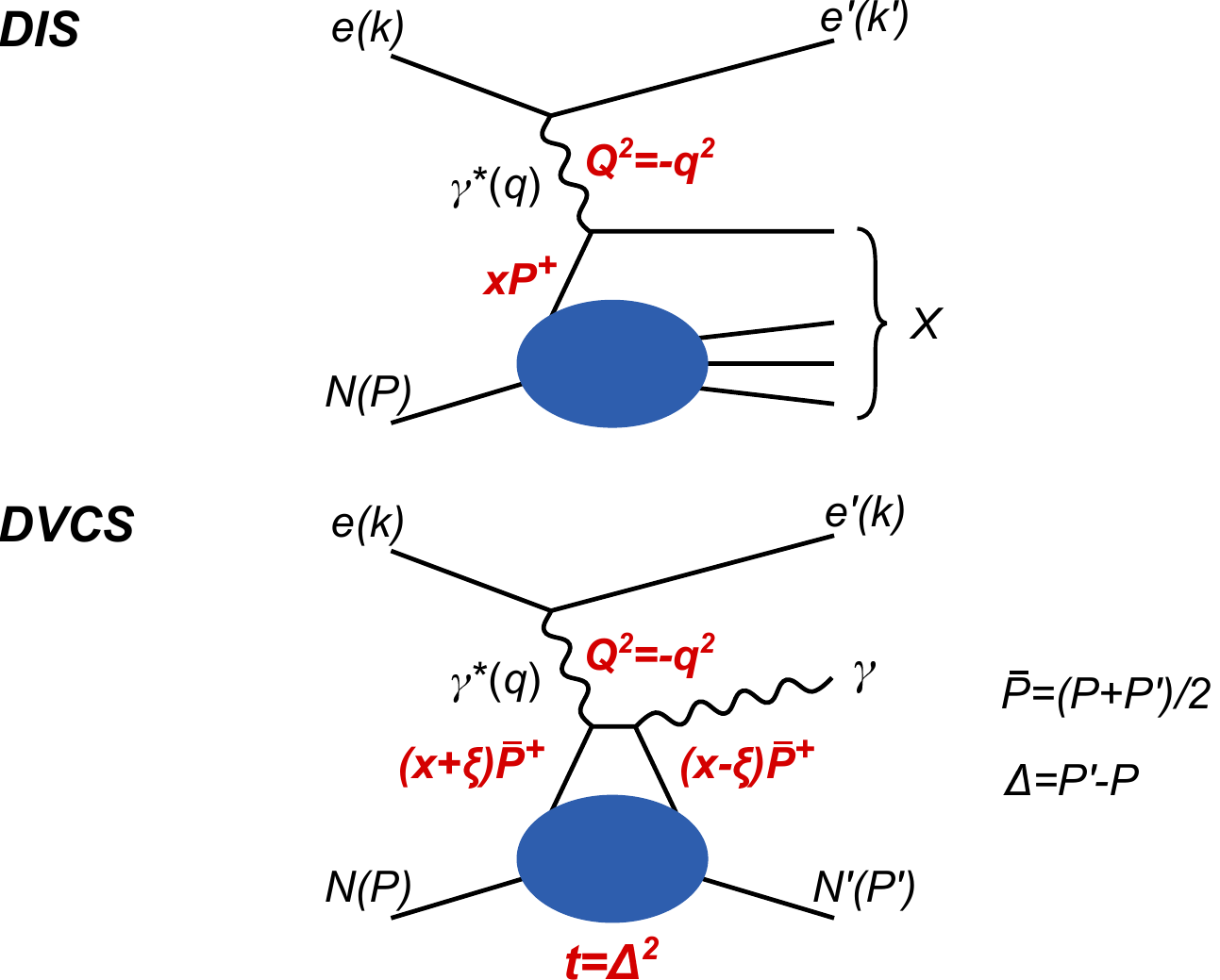}
\caption{QCD interpretation of DIS ($e+N \to e'+X$) and DVCS ($e+N \to e'+N'+\gamma$) processes, which are sensitive to PDFs and GPDs, respectively. Here, $X$ denotes the set of all unmeasured final-state particles. The four-momenta of particles are given in parentheses, and relevant kinematic variables are marked in red, with the superscript ``$+$'' denoting the light-cone plus-component.}
\label{fig:dis_dvcs}
\end{figure}

A basic feature underlying the predictive power of perturbative QCD is the factorization property, which enables the separation of short-distance, perturbatively calculable dynamics from long-distance, nonperturbative physics. In inclusive processes, such as deep inelastic scattering (DIS) illustrated in Fig.~\ref{fig:dis_dvcs}, in the Bjorken limit, collinear factorization enables the structure functions to be expressed as convolutions of perturbative coefficient functions with PDFs, which are defined through light-cone correlations of quark and gluon fields. 
In exclusive processes, such as deeply virtual Compton scattering (DVCS), also shown in Fig.~\ref{fig:dis_dvcs}, a similar factorization theorem holds. In this case, the amplitude factorizes into a hard scattering kernel and GPDs, which are non-forward matrix elements.

For example, the unpolarized PDFs or GPDs of quark flavor $q$ (we consider $q=u,d$ in this work, for up/down quarks), denoted here generically as $q(x)$ with only the dependence on $x$ listed, are given by the matrix elements (MEs)
\begin{align}
q(x)\!=\!\!\int \!\frac{dz^-}{4\pi} \, e^{i x \bar{P}^+ z^-} \!\left\langle P' \left| \bar{\psi}^q(0) \gamma^+ \mathcal{W}(0, z^-) \psi^q(z^-) \right| P \right\rangle,
\label{eq:pdfdef}
\end{align}
where $P/P'$ are initial/final four-momenta, $2\bar{P} = P+P'$, $\psi/\bar{\psi}$ are quark/antiquark fields, $\mathcal{W}(0, z^-)$ denotes a Wilson line along the light cone ($v^\pm=(v^0\pm v^3)/\sqrt{2}$ in general) ensuring gauge invariance, and the integration is performed at a fixed light-front time $z^+ = 0$ and transverse position $\vec{z}_\perp = 0$.
PDFs are obtained when $P=P'$, while the nonforward case $P\neq P'$ corresponds to GPDs.
For the latter, the distribution depends on two additional variables, the invariant momentum transfer $t = \Delta^2$, with $\Delta = P' - P$ and the skewness parameter $\xi = -\Delta^+ / (P^++P'^+)$, related to the longitudinal momentum transfer.
Thus, it will be denoted as $q(x,\xi,t)$. For simplicity, the factorization scale, which for DIS and DVCS is usually identified with the virtuality of the photon emitted by the electron, $Q^2$, is suppressed. The dependence on this scale is fully calculable and is governed by the Dokshitzer-Gribov-Lipatov-Altarelli-Parisi (DGLAP) evolution equations~\cite{Dokshitzer:1977sg,Gribov:1972ri,Altarelli:1977zs} for PDFs, and by the DGLAP and Efremov-Radyushkin-Brodsky-Lepage (ERBL)~\cite{Efremov:1979qk,Lepage:1979zb} equations for GPDs.

The longitudinal momentum fraction variable $-1<x<1$, where positive/negative values correspond to quarks/antiquarks.
The nucleon consists of three valence quarks (e.g., two up and a down quark in the proton), but QCD implies that additional quark-antiquark pairs are dynamically produced. The latter are referred to as sea quarks/antiquarks.
One can, thus, consider the distributions of only valence or only sea partons, $q_{\rm val}(x)$ or $q_{\rm sea}(x)$, with $q(x)=q_{\rm val}(x)+q_{\rm sea}(x)$.
Due to the crossing relation between quarks and antiquarks,
the sea- and valence-quark distributions at $x>0$ are defined as
\begin{equation}
\label{eq:qsea}
q_{\rm sea}(x)=-q(-x),
\end{equation}
\begin{equation}
\label{eq:qval}
q_{\rm val}(x)=q(x)-q_{\rm sea}(x).
\end{equation}

The unpolarized GPDs can be further decomposed in terms of two invariant functions, $H^q(x, \xi, t)$ and $E^q(x, \xi, t)$,
\begin{flalign}
q(x, \xi, t) = \bar{u}(P') \left[ \gamma^+ H^q(x, \xi, t) + \frac{i \sigma^{+\mu} \Delta_\mu}{2M} E^q(x, \xi, t) \right] u(P),
\end{flalign}
where $u(P)$ and $\bar{u}(P')$ are Dirac spinors for the incoming and the outgoing nucleon, $M$ is the nucleon mass, and $\sigma^{+\mu} = \frac{i}{2} [\gamma^+, \gamma^\mu]$. 

The GPDs $H^q(x,\xi,t)$ reduce to the usual PDFs in the forward limit,
\begin{gather}
    H^q(x,0,0) \equiv q(x)\,.
    \label{eq:forward_limit_H}
\end{gather}
 The limits of the GPDs $E^q(x,\xi,t)$,
\begin{gather}
    E^q(x,0,0) \equiv e_q(x)\,,
    \label{eq:forward_limit_E}
\end{gather}
are, on the other hand, not accessible in DIS. This is because probing them would require the nucleon state to stay coherent after the interaction, whereas in DIS the nucleon generally breaks up.

Another important feature confirming generalizing capabilities of GPDs is their relation to Dirac, $F_1(t)$, and Pauli, $F_2(t)$, elastic form factors,
\begin{align}
    \int_{-1}^{1} dx\, H^q(x,\xi,t) \equiv F_1^q(t)\,,  \label{eq:theory:eff_H}\\
    \int_{-1}^{1} dx\, E^q(x,\xi,t) \equiv F_2^q(t) \,. \label{eq:theory:eff_E}
\end{align}
Here, on the right hand-side the superscript ``$q$'' denotes the contribution of a specific quark flavor to a given elastic form factor. For instance, for the proton one has $\nicefrac{2}{3}F_1^u(t) - \nicefrac{1}{3}F_1^d(t) = F_1(t)$ (and similarly for $F_2(t)$), with the fractions reflecting the quark electric charges. The relations \eqref{eq:theory:eff_H} and~\eqref{eq:theory:eff_E} are not sensitive to the values of $\xi$, which is a consequence of the Lorentz invariance of GPDs.

The GPD formalism provides several distinct insights into nucleon imaging. Among the most notable ones is nucleon tomography \cite{Burkardt:2002hr}, allowing for a determination of the position of the active parton in the plane perpendicular to the nucleon's motion,
\begin{equation}
q(x,{\bf b_\perp}) =
\int \frac{d^2{\bf \Delta}_\perp}{(2\pi)^2}  
e^{-i{\bf b_\perp}\cdot {\bf \Delta}_\perp}
H^q(x,0,-{\bf \Delta}_\perp^2) \,.
\label{eq:nt:H}
\end{equation}
Here, ${\bf b_\perp} = (b_x, b_y)$ is the impact parameter, defined in the coordinate system whose origin is set by the center of momentum of all partons (for details regarding the interpretation of nucleon tomography, see Appendix~\ref{app:tomography}).

When the proton is transversely polarized, for instance along the $X$-axis, rather than longitudinally polarized along its direction of motion (the $Z$-axis), the parton distribution becomes distorted (along the $Y$-axis) and takes the form:
\begin{equation}
q_X(x,{\bf b_\perp}) = 
q(x,{\bf b_\perp})
-\frac{1}{2M} \frac{\partial}{\partial b_y}
e_q(x,{\bf b_\perp})\,,
\label{eq:nt:E}
\end{equation}
where
\begin{equation}
e_q(x,{\bf b_\perp})  =
\int \frac{d^2{\bf \Delta_\perp}}{(2\pi)^2}
e^{-i{\bf b_\perp}\cdot {\bf \Delta}_\perp}
E_q(x,0,-{\bf \Delta}_\perp^2) \,.
\end{equation}

Another notable aspect of GPDs is their relation to the elements of the energy–momentum tensor. This relation has attracted significant attention in recent years because it provides access to ``mechanical'' properties of the nucleon~\cite{Goeke:2007fp, Polyakov:2018zvc}. For the purposes of this work, we note that the relation between GPDs and the energy–momentum tensor gives rise to the so-called Ji’s sum rule \cite{Ji:1996ek},
\begin{equation}
\int_{-1}^{1}dx\, x \Big(H^q(x, \xi, 0) + E^q(x, \xi, 0)\Big) = 2J^{q}\,,
\label{eq:basics_ji}
\end{equation}
which allows the determination of the total angular momentum carried by specific partons, $J^{q}$. This relation plays a central role in addressing the long-standing problem of nucleon spin decomposition, which originated from the famous measurement by the EMC collaboration at CERN~\cite{EuropeanMuon:1987isl}.

\subsection{PDFs and GPDs from lattice QCD}
\label{sec:pdfs_and_gpds_from_lattice}

As argued above, PDFs and GPDs cannot be accessed directly on the lattice. 
The reason is that the vector connecting the quark and antiquark fields is chosen along the light-cone minus direction, i.e., 
MEs of PDFs/GPDs probe light-front correlations.
Euclidean LQCD, in turn, can only calculate spatial correlations. Thus, it can access MEs of the form
\begin{align}
\label{eq:latticeME}
F^q(z, P, P') = \langle P'|\,\bar{\psi}^q(0)\,\gamma_t\, \mathcal{W}(0,z)\,\psi^q(z)\,|P \rangle,
\end{align}
where $P=(P_t,0,0,P_z)$ and $z$ is along the $Z$-direction;\footnote{We choose to adopt the convention that Euclidean vectors are written with lower indices.}
$P=P'$ for PDFs and $P\neq P'$ for GPDs. The $\gamma^+$ matrix used in Minkowski spacetime is replaced with $\gamma_t$.\footnote{Inclusion of $\gamma_+$ is also possible, but its $\gamma_z$ component leads to undesired mixing on the lattice \cite{Constantinou:2017sej}.}
Obviously, the choice of the $Z$-direction for probing spatial correlations is conventional, but crucially, the nucleon is boosted along the same direction, i.e., the four-momentum vector has only the $P_z$ spatial component non-zero.

Below, we turn to the quasi-distribution (LaMET) \cite{Ji:2013dva,Ji:2014gla} and pseudo-distribution (SDE) \cite{Radyushkin:2017cyf,Radyushkin:2019mye} frameworks, concentrating on the PDF case, where $P=P'$ in Eq.~\eqref{eq:latticeME} and this ME can be written as $F^q(z,P_z)$.
The generalization to the GPD case is straightforward.
Bare MEs, as defined above, contain standard logarithmic and power divergences, the latter resulting from the presence of the Wilson line \cite{Ji:2015jwa,Ishikawa:2017faj, Constantinou:2017sej}. In this work, we adopt two standard ways of non-perturbative renormalization of these divergences, namely regularization-independent momentum subtraction (RI/MOM)~\cite{Martinelli:1994ty} in a variant suitable for non-local operators~\cite{Alexandrou:2017huk} for quasi-distributions, and the double ratio scheme~\cite{Orginos:2017kos} for pseudo-distributions.    
More details on both methods are given in Appendix~\ref{app:lattice}.
We note that our current work concentrates on introducing the ANN framework for reconstructing $x$-dependent PDFs/GPDs and on the synergy one can obtain by combining the quasi- and pseudo-distribution methods.
Hence, at this stage, we leave incorporation of latest developments in the renormalization of quasi-distributions, namely the inclusion of the hybrid scheme~\cite{Ji:2020brr} for further work.

Flavor-$q$ MEs obtained at some boost $P_z$ are related to momentum-space quasi-distri\-bu\-tions via a Fourier transform,
\begin{align}
\label{eq:Fourier_quasi}
\widetilde{F}^q(z,P_z) =\int_{-1}^{1}dx\,e^{-ixzP_z}\,\widetilde{q}(x,P_z)\,,
\end{align}
where 
$\widetilde{F}^q(z,P_z)$ is the coordinate-space renormalized ME, and $\widetilde{q}(x,P_z)$ is its momentum-space counterpart. The latter is the quasi-PDF -- an Euclidean version of the physical Minkowski-space PDF. The tilde in $\widetilde{F}$ denotes that the ME was renormalized in the RI$'$ scheme. For clarity, in Eq.~\eqref{eq:Fourier_quasi} we list only the relevant variables of the underlying objects. We also note that, given the definitions of valence/sea distributions in Eqs.~\eqref{eq:qsea} and \eqref{eq:qval}, the real part of MEs is related solely to the valence component via the cosine part of the Fourier transform, while the imaginary part probes $q_{\rm val}(x)+2q_{\rm sea}(x)$ via its sine part.

The next step is the factorization, again written schematically with only the relevant variables,
\begin{align}
\label{eq:factorization_quasi}
\widetilde{q}(x,P_z)=\int_{-1}^{1}\frac{dy}{|y|}C\left(\frac{x}{y}\right)q(y)+\mathcal{O}\left(\frac{\Lambda^2_{\rm{QCD}}}{x^2P_z^2},\frac{\Lambda^2_{\rm{QCD}}}{(1-x)^2P_z^2}\right),
\end{align}
where $C(x/y)$ is the momentum-space matching kernel calculated in perturbation theory.
According to this equation, the quasi-PDF is matched into its light-front counterpart $q(x)$.
Note that the light-front distribution no longer depends on $P_z$, i.e., quasi-distributions obtained at different $P_z$ should be matched into the same light-front function.\footnote{Technical details of the matching are given in Appendix~\ref{app:matching}.}
This statement is true up to power-suppressed corrections in the inverse of the boost momentum, with $\Lambda_{\rm QCD}$ being the non-perturbative scale generated in QCD.
The denominator of power corrections includes the scales $xP_z$ and $(1-x)P_z$, which implies that only the intermediate-$x$ region can be accessed reliably,
\begin{align}
\label{eq:xquasi}
x\in[x_{\rm{min}}\sim\frac{\Lambda_{\rm{QCD}}}{P^z},\; x_{\rm{max}}\sim 1-x_{\rm{min}}]\,.
\end{align}

In turn, the pseudo-distribution (SDE) approach, while utilizing the same MEs, reverses the order of $x$ dependence reconstruction and factorization.
Namely, the latter comes as the first step, i.e., the renormalized ME (dubbed Ioffe time distribution (ITD) in this context and written as a function of the Ioffe time \cite{Ioffe:1969kf}, $\nu=zP_z$ in our notation) is factorized already in coordinate space, by means of a short-distance operator product expansion (commonly abbreviated to SDE),
\begin{align}
\bar{F}^q(z,\nu) =\int_{-1}^{1}dy\, C_\nu\left(y\right) \, \mathcal{F}^q(y\nu)+\mathcal{O}\left(z^2\Lambda^2_{\rm{QCD}}\right),
\end{align}
where the bar in $\bar{F}$ denotes the double ratio renormalization.
Here, $C_\nu(y)$ is the coordinate-space matching kernel and $\mathcal{F}^q(\nu)$ is the light-cone counterpart of the Euclidean ITD.
Note that the matched ITD is independent of the initial scale $z$ and boost $P_z$ separately, with only their product being of relevance.
Analogously to momentum-space factorization, this statement is true up to power corrections, determined by the relation between $z$ and $\Lambda_{\rm QCD}$.
This means that the pseudo-distribution approach accesses reliable information only from a segment of the light-cone distance $\nu$,
\begin{align}
\label{eq:zpseudo}
[0,\;\nu_{\rm{max}}\sim z_{\rm{max}}P_z]\,,
\end{align}
where $z_{\rm{max}}$ should be in the perturbative regime, i.e., much smaller than $\Lambda^{-1}_{\rm QCD}$.
After the coordinate-space matching, the final step of the pseudo-distribution method to arrive at a physical distribution is to reconstruct the $x$ dependence, formally given by the Fourier transform
\begin{align}
\label{eq:Fourier_pseudo}
\mathcal{F}^q(\nu) =\int_{-1}^{1}dx\,e^{-ix\nu}\,q(x)\,,
\end{align}
with the real or imaginary parts of matched ITDs related to $q_{\rm val}(x)$ or $q_{\rm val}(x)+2q_{\rm sea}(x)$, respectively.

In principle, the final extracted object, the light-cone PDF $q(x)$ should be the same from both the quasi- and the pseudo-distribution approach.
However, we argued above that both approaches have different ranges of applicability, expressed by Eqs.~\eqref{eq:xquasi} and \eqref{eq:zpseudo}.
In practice, the interchange of the reconstruction/factorization steps and in particular, performing the latter either in momentum or coordinate space, leads to a situation where $q(x)$ can considerably differ after the whole procedure, even though the input lattice data are the same.
The difference of the two evaluations of $q(x)$ reflects the different systematic effects of both procedures, predominantly power corrections that influence the momentum-space or coordinate-space factorizations.
As hinted already in the introduction, this feature actually enables us to treat them as complementary.
The ensuing synergy between them is one of essential aspects of our framework, which we discuss in the next section.

For convenience of the reader, we provide a summary of our notation for MEs and $x$-dependent light-cone distributions, given in Tab.~\ref{tab:notation}.
The latter are always expressed in the $\overline{\text{MS}}$ scheme at the scale of 2~GeV.
\begin{table}[h!]
\caption{Notation for matrix elements in the quasi- and pseudo-distribution approaches and for the corresponding $x$-dependent light-cone distributions. The symbol $q$, either as a superscript or the function label, can take the values $u$ or $d$ for up or down quarks, respectively. The light-cone distributions can further have a subscript val/sea referring to their valence/sea components.}
\vspace*{-2mm}
\begin{center}
\begin{tabular}{
P{0.25\columnwidth}
P{0.2\columnwidth}
P{0.2\columnwidth}
P{0.23\columnwidth}
}
\toprule
\multirow{2}{*}{type} & \multirow{1}{0.2\columnwidth}{\centering quasi-ME \\ (LaMET)} & \multirow{1}{0.2\columnwidth}{\centering pseudo-ITD \\ (SDE)} & \multirow{1}{0.233\columnwidth}{\centering light-cone \\ $x$-distribution}\\
&&&\\
\midrule
PDF & $\widetilde F^q(z,P_z)$ & $\bar F^q(z,\nu)$ & $q(x)$ \\\addlinespace
GPD $H$ at $\xi=0$ & $\widetilde F_H^q(z,P_z,t)$ & $\bar F_H^q(z,\nu,t)$ & $H^q(x,t)$\\\addlinespace
GPD $E$ at $\xi=0$ & $\widetilde F_E^q(z,P_z,t)$ & $\bar F_E^q(z,\nu,t)$ & $E^q(x,t)$\\
\bottomrule
\end{tabular}
\label{tab:notation}
\end{center}
\end{table}

We finalize this section by a short outline of the actual lattice data used in our work.
The starting point is a set of gauge field configurations, obtained in a separate simulation \cite{Alexandrou:2018egz}, with some choice of parameters that determine the lattice spacing and quark masses.
Namely, we use $N_f=2+1+1$ twisted-mass fermions \cite{Frezzotti:2000nk,Frezzotti:2003ni} supplemented with a clover term \cite{Sheikholeslami:1985ij}, utilizing an Iwasaki-improved gluon action \cite{Iwasaki:1983iya}. The lattice has a spatial volume of $32^3$ and a temporal extent of $64$, with a lattice spacing of 0.093 fm and quark mass parameters corresponding to a pion mass of 260 MeV. 
The outcome of the lattice calculation are, finally, bare MEs corresponding to PDFs and GPDs.
The computational methodology employed to obtain the required MEs is well-established and has been used in several works aimed at extraction of PDFs and GPDs, see e.g.\ Refs.~\cite{Alexandrou:2020zbe,Bhattacharya:2020cen,Bhattacharya:2022aob}.
We also provide a short summary of this methodology in Appendix \ref{app:lattice}.
We use the up and down quark data separately.
In principle, this would require us to take into account disconnected diagrams in the evaluation of MEs and the mixing with gluons.
However, both effects have been shown to be negligible at the current level of precision \cite{Alexandrou:2020uyt,Alexandrou:2021oih,Delmar:2023agv} and hence, we neglect them.
\section{Neural networks as a tool to unify lattice calculations of PDFs and GPDs}
\label{sec:ann}

The essence of our approach is to achieve a robust reconstruction of the $x$ dependence of PDFs and $(x,t)$ dependence of GPDs (as noted in the introduction, the $\xi$ dependence of GPDs is not our focus here and will be addressed in the future.).
From the theoretical point of view, this is enabled by unification of two novel approaches of accessing parton physics on a Euclidean lattice -- quasi-distributions/LaMET and pseudo-distributions/SDE, discussed in the previous section.
In our framework, both are used in their respective domains of applicability, which differ due to factorization performed either in momentum or coordinate space.
However, another crucial aspect of our method, is to use artificial neural networks (ANNs) as practical means of performing the reconstruction.
This is essential for several reasons that we discuss below.

The importance of using an advanced reconstruction method is related to the step of the Fourier transform in both approaches, see Eqs.~\eqref{eq:Fourier_quasi} and \eqref{eq:Fourier_pseudo}.
The distributions to be determined appear on the right-hand sides of these equations, under an integral over all momentum fractions.
Rearranging these relations to express $x$-dependent distributions in terms of lattice-extracted coordinate-space objects (renormalized MEs or matched ITDs) leads to the inverse problem, meaning that the momentum-space distributions are not well-defined.
The reason for this is that MEs are evaluated at discrete values of $z$ and available only until some maximum value of $z$, i.e., truncated.
The latter aspect is particularly evident for pseudo-distributions, where the $\mathcal{O}(z^2\Lambda^2_{\rm QCD})$ power corrections imply that only short separations $z$ can be used in the factorization.
Formally, infinitely many momentum-space distributions can satisfy the equation for the inverted Fourier transform from discrete and truncated data, and one needs a criterion to choose one of the solutions.
The inverse problem in the context of lattice PDFs was discussed in detail in Ref.~\cite{Karpie:2019eiq}, wherein neural networks were one of the proposed solutions, however, demonstrated only with mock data.
Later, neural networks were indeed applied to actual lattice data (see, e.g., \cite{Cichy:2019ebf,DelDebbio:2020rgv}), but their usage in the field is still not common.

One of the most widely used approaches is the Backus-Gilbert (BG) reconstruction \cite{BackusGilbert}.
A necessary criterion for assessing the robustness of any reconstruction method is its ability to reproduce a known distribution from which mock data have been generated. Below, we examine whether the BG method satisfies this criterion by means of such a closure test.

We generated mock data for the valence $u$ quark from the Goloskokov-Kroll (GK) model \cite{Goloskokov:2006hr}, which utilizes a custom parameterization of the CTEQ6m PDF set~\cite{Pumplin:2002vw}.\footnote{The use of a slightly outdated PDF set is not important for this test but ensures consistency with the later analysis of GPDs.}
We took an inverse Fourier transform of the GK data to coordinate space and applied the inverse matching to get to the stage of Euclidean MEs. To mimic the situation in an actual lattice calculation, 10 data points were produced for the real part of the ME (i.e., this corresponds to data at $z/a=0,\,1,\,\ldots,z_{\rm cut}$, with $z_{\rm cut}=9$), with statistical variation constructed to parallel the variation of the actual lattice data (at $P_z=1.67$ GeV). This ensures identical correlations between MEs at different $z$ values, a key feature of the lattice data (see Ref.~\cite{Riberdy:2023awf} for a discussion). These 10 points were then subjected to the BG procedure to get from MEs to the quasi-PDF and the LaMET matching procedure to arrive at the light-cone PDF, which should be consistent with the original model.
{\color{red}Our BG implementation followed the one typically used in several earlier papers on PDFs and GPDs, see e.g.~\cite{Bhat:2020ktg,Alexandrou:2020zbe,Alexandrou:2020qtt,Alexandrou:2021bbo,Bhat:2022zrw,Bhattacharya:2022aob,Bhattacharya:2023jsc}.
We used Tikhonov regularization with $\rho=10^{-3}$ and no preconditioning.
We refer to the above papers for more details on this method.}

\begin{figure}[t!]
\centering
\includegraphics[width=0.49\columnwidth]{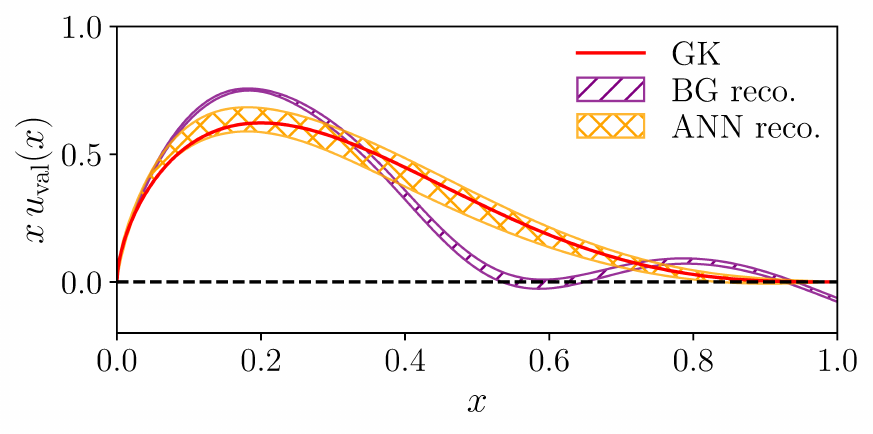}
\caption{Closure test of the Backus-Gilbert reconstruction method. Mock data for the valence up quark PDF were generated from the GK model (red line)~\cite{Goloskokov:2006hr} by subjecting it to the inverse Fourier transform and inverse matching. The $x$ dependence was then reconstructed with the BG approach and matching (purple band), showing strong tension with the original model. Meanwhile, the ANN reconstruction (orange band) shows full agreement with the original model, see text for more details.}
\label{fig:comparison_BG}
\end{figure}

The obtained result is shown as the purple band in Fig.~\ref{fig:comparison_BG}, with the original GK model as the red line. This test reveals that the BG reconstruction hardly matches the input distribution, leading to large overestimation at intermediate $x$ and dramatic suppression at $x\gtrsim0.4$, hitting zero around $x=0.6$ and continuing as a nonphysical oscillation until $x=1$, where the reconstructed PDF actually hits negative values.
At the same time, the uncertainty of the reconstruction is in no way reflected by the error.
This mirrors the fact that the BG criterion, although model-independent, is a simple choice of a distribution that minimizes the variance of the solution with respect to the statistical variation of underlying data.
Such a criterion formally solves the inverse problem in the sense that only one distribution satisfies the criterion out of infinitely many distributions that are consistent with the discrete and truncated data.
One can suspect that these two features of the data make it impossible to arrive back at the initial GK model.
However, before discussing the details of our ANN framework, we also display the result of the same mock data test with ANNs used for the reconstruction of the $x$-dependence (orange band) in Fig.~\ref{fig:comparison_BG} and we conclude full agreement between the original curve and the reconstructed one.
Hence, it is not the quality of the lattice data that prevents a successful reconstruction with the BG approach -- advanced reconstruction methods like ANNs are able to perform well with the currently available quality of lattice data.
We note that the presence of the inverse problem is translated to uncertainties of the reconstruction, which is manifested by the larger width of the orange band with respect to the BG curve in purple, which only reflects statistical uncertainties imposed on the samples of mock data.

The next reason for ANNs being essential in our framework is the $t$ dependence of GPDs.
We will demonstrate that extending the reconstruction of the $x$ dependence to incorporate also the square of four-momentum transfer is relatively straightforward.
Meanwhile, without the flexibility of parametrization with ANNs, one is forced to reconstruct the $t$ dependence in a model-dependent way, e.g.\ by using multipole fits at fixed $x$.
This is, clearly, against the spirit of a first-principle lattice calculation.

\begin{figure}[!t]
\centering
\includegraphics[width=0.6\columnwidth]{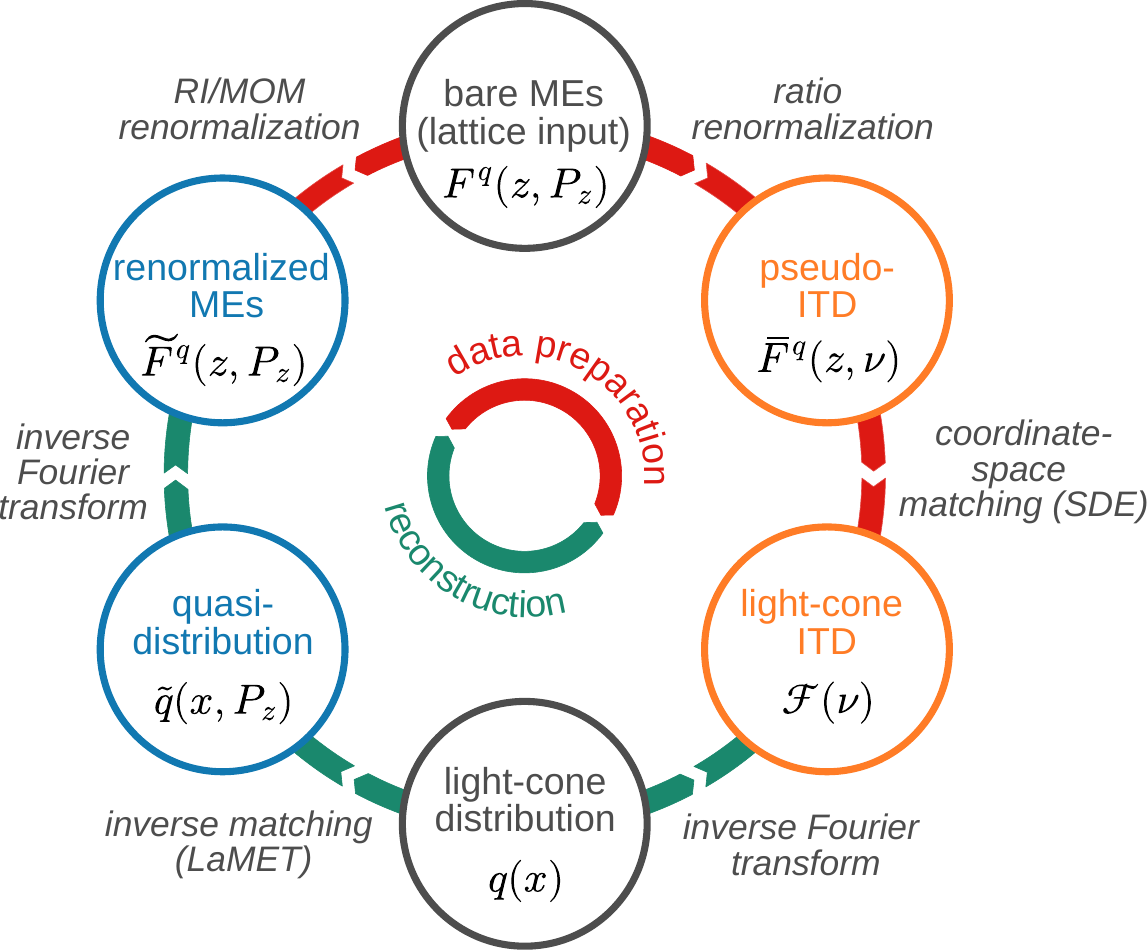}
\caption{Sketch of our unified framework combining the quasi- (LaMET) and pseudo-distribution (SDE) methods to reconstruct the $x$ dependence of light-cone \mbox{distributions}.}
\label{fig:method}
\end{figure}

Finally, it is also essential that ANNs offer a natural framework for a meaningful unification of the quasi- and pseudo-distribution approaches.
We explain this aspect by sketching our procedure in Fig.~\ref{fig:method}, for the PDF case.
The object that is sought after is the $x$-dependent light-cone PDF $q(x)$, represented by the bottom circle in the chart.
It is a neural-network parametrization trained in a single global fit to lattice inputs from both LaMET (quasi) and SDE (pseudo).
The pure lattice input are bare MEs $F^q(z,P_z)$ represented by the top circle.
Before they enter the ANN procedure, they are subjected to renormalization, RI/MOM for quasi and double ratio for pseudo, resulting in renormalized MEs $\widetilde{F}^q(z,P_z)$ and the pseudo-ITD $\bar{F}^q(z,\nu)$, respectively.
We perform the coordinate-space matching (SDE) on the latter, leading to the light-cone ITD $\mathcal{F}(\nu)$.
These stages of handling bare MEs are independent of the neural network and are evaluated unambiguously, which we emphasize with the red arrows labeled as ``data preparation''.
The neural network implements the map $x\!\mapsto\! q(x)$, i.e., it looks for a parametrization of the light-cone distribution (optimized with a genetic algorithm) that simultaneously minimizes residuals to the LaMET input (renormalized MEs) and the SDE input (light-cone ITD at short distance), ensuring that both factorization frameworks consistently constrain the same $q(x)$.
We note that this requires not only that the procedure connects momentum space and coordinate space via a Fourier transform, but also implements the inverse LaMET matching.
This is emphasized by the green arrows labeled as ``reconstruction''.
The mathematical details of our formulation are given in the next section.
Its essence is that LaMET constrains the $x$ dependence in its region of validity, between some $x_{\rm min}$ and $x_{\rm max}$, while SDE complements the $x$ dependence reconstruction in the remaining regions, also working in its domain of applicability of short distances, $z<z_{\rm max}$.
We remark that although Fig.~\ref{fig:method} illustrates the PDF case, the procedure for GPDs is analogous, obviously with lattice inputs depending additionally on the invariant four-momentum transfer $t$, and the $(x, t)$ dependence being reconstructed by the ANN.

\section{Reconstruction of parton distribution functions}
\label{sec:pdf}

\subsection{Functional form of light-cone PDFs}
\label{sec:pdf:form}

\begin{figure}
\centering
\includegraphics[scale=0.5]{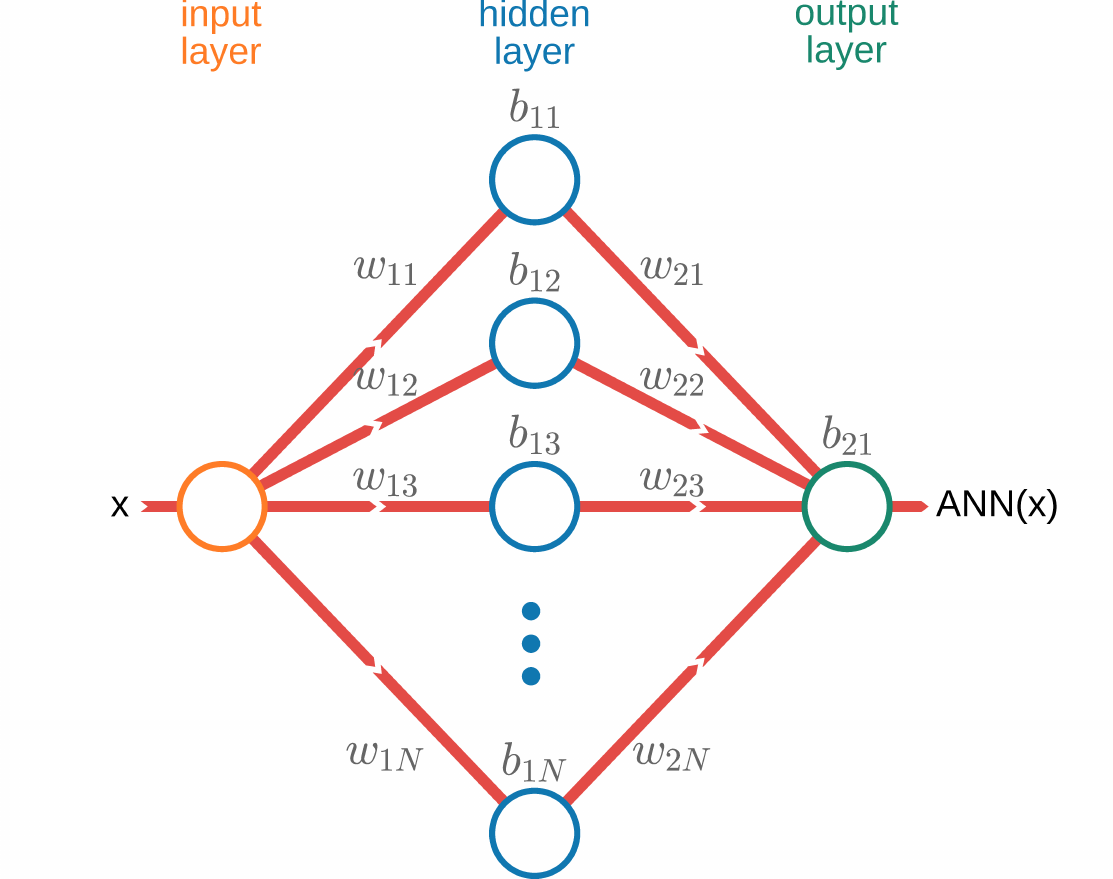}
\caption{Schematic structure of a generic neural network, $\mathrm{ANN}(x)$, used in this analysis, featuring $N$ neurons in the hidden layer, with weights $w_{jk}$ and biases $b_{jk}$ explicitly marked.}
\label{fig:ann}
\end{figure}

The primary light-cone PDFs that we want to extract are separately constructed for the valence and sea components.
For a given quark flavor, we have
\begin{align}
q^{(0)}(x) = x^{\delta^{(0)}}(1-x)^{\rho^{(0)}}\, \mathrm{ANN}^{(0)}(x) \,.
\label{eq:q0}
\end{align}
Here, \smash{$\mathrm{ANN}^{(0)}(x)$} represents a single artificial neural network. In this analysis, we utilize a simple feedforward network with only one hidden layer and \smash{$N^{(0)}$} neurons. This network is visualized in Fig.~\ref{fig:ann}, and the way it processes input information can be expressed by the following equation,
\begin{flalign}
\mathrm{ANN}^{(0)}(x) = \quad b_{21}^{(0)} + \sum_{i=1}^{N^{(0)}}\Big[w_{2i}^{(0)}\ln\left(1+\exp(b_{1i}^{(0)}+w_{1i}^{(0)}x)\right)\Big] \,.
\label{eq:ann}
\end{flalign}
Here, we implicitly used the weighted sum aggregation and the softplus activation function. The weights and biases, \smash{$w_{jk}^{(0)}$} and \smash{$b_{jk}^{(0)}$}, respectively, where $j$ and $k$ denote the layer and in-layer indices, are free parameters of the network. They are intended to be constrained, along with the powers \smash{$\delta^{(0)}$} and \smash{$\rho^{(0)}$}, by lattice data.

The power-behavior prefactor in Eq.~\eqref{eq:q0} is customary in analyses constraining PDFs, see, e.g., NNPDF parameterizations~\cite{NNPDF:2014otw}, but may still be regarded by some as a model bias. We note, however, that this bias can be ``softened'', for example by applying the power-behavior prefactor to each element of the sum in Eq.~\eqref{eq:ann}. We have explored this possibility and do not observe any significant effect. We also note that the choice of the artificial neural network architecture presented in Eq.~\eqref{eq:ann} is arbitrary. Different architectures should not lead to significant changes, assuming that flexibility in reproducing the stated problem is preserved.

We conclude the description of \smash{$q^{(0)}(x)$} by noting that for valence quarks the distributions are explicitly normalized as follows:
\begin{align}
   \int_{0}^{1}dx\,u_{\mathrm{val}}^{(0)}(x) = 2
   \quad
   \mathrm{and}
   \quad
   \int_{0}^{1}dx\,d_{\mathrm{val}}^{(0)}(x) = 1 \,.
\end{align}
In practice, the normalization is achieved in the reconstruction procedure by rescaling \smash{$w_{2j}^{(0)}$} and \smash{$b_{21}^{(0)}$} parameters by a common factor.

\begin{figure}
\centering
\includegraphics[width=0.49\columnwidth]{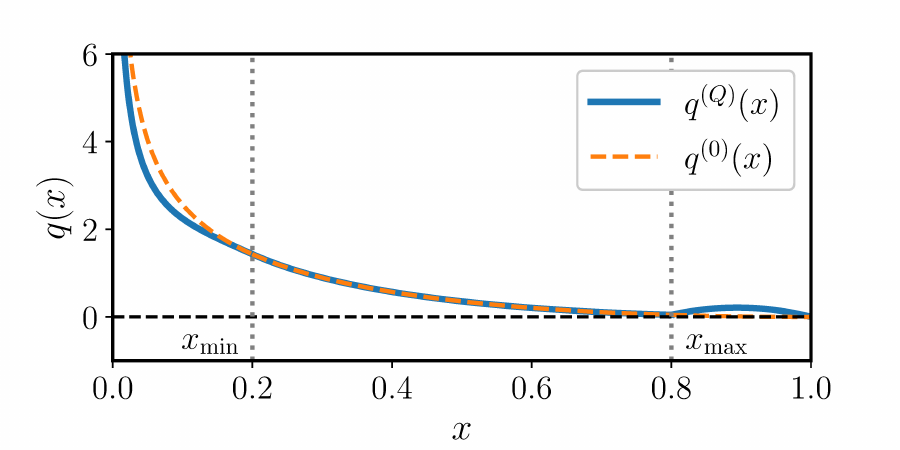}
\includegraphics[width=0.49\columnwidth]{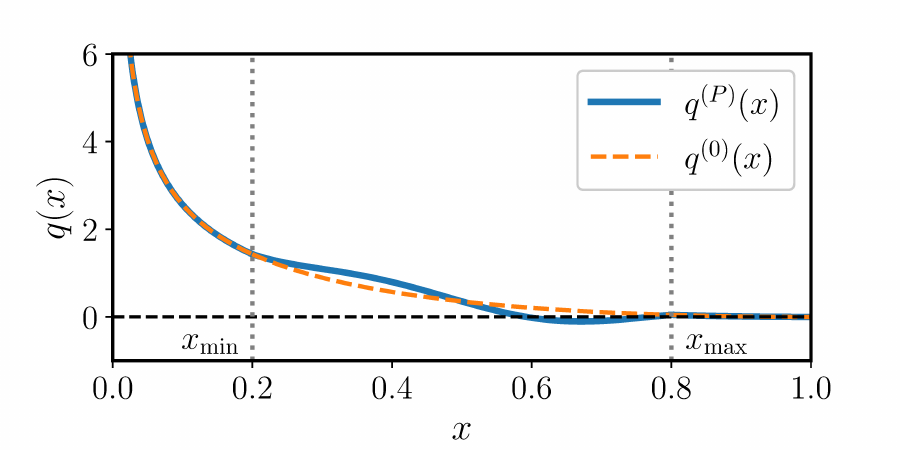}
\caption{Left panel: illustrative example of $q^{(Q)}(x)$, used for the reconstruction of $q^{(0)}(x)$ from quantities evaluated within the LaMET framework. Right panel: same, but for $q^{(P)}(x)$, used in the context of the SDE framework. The contribution of $q^{(0)}(x)$ is identical in both cases. By construction, $q^{(Q)}(x)$ coincides with $q^{(0)}(x)$ for $x\in[x_{\rm{min}},x_{\rm{max}}]$, and $q^{(P)}(x)$ coincides with $q^{(0)}(x)$ for $x\leq x_{\rm{min}}$ and $x\geq x_{\rm{max}}$.  The boundaries $x_{\rm{min}} = 0.2$ and $x_{\rm{max}} = 0.8$ are chosen for demonstration purposes.}
\label{fig:standard_q1234}
\end{figure}

In addition to \smash{$q^{(0)}(x)$}, we introduce three auxiliary functions, \smash{$q^{(1,2,3)}(x)$}, which describe deviations of the former from light-cone distributions directly extracted from lattice data. As described in Sec.~\ref{sec:pdfs_and_gpds_from_lattice}, these deviations arise due to power corrections of various origins and differ between the LaMET and SDE methods. We take
\begin{align}
&q^{(Q)}(x) = 
\begin{cases}
q^{(0)}(x) + q^{(1)}(x)\,,  &0 < x < x_{\rm{min}}\,, \\
q^{(0)}(x) + q^{(2)}(x)\,,  &x_{\rm{max}} < x < 1\,, \\
q^{(0)}(x)\,,                 &\mathrm{otherwise}\,,
\end{cases}
\label{eq:qQ}
\\
&q^{(P)}(x) = 
\begin{cases}
q^{(0)}(x) + q^{(3)}(x)\,,  &x_{\rm{min}} < x < x_{\rm{max}}\,, \\
q^{(0)}(x)\,,                 &\mathrm{otherwise}\,,
\end{cases}
\label{eq:qP}
\end{align}
where $x_{\rm{min}}$ and $x_{\rm{max}}$ are tentative limits beyond which the lattice data are believed to be dominated by power corrections when applying momentum-space factorization (LaMET), see Eq.~\eqref{eq:xquasi}. {\color{red}According to previous studies (see, e.g., Refs.~\cite{Gao:2022uhg,Ding:2024saz}), typical values of $x_{\rm{min}}=0.2$ and $x_{\rm{max}}=0.8$ are adopted in this work at the chosen $P_z$ (see Appendix \ref{app:lattice}). Specifically, with $P_z=1.67$ GeV, these boundaries imply that power correction terms are of $\mathcal{O}(0.8)$, which could not be ignored.} The key idea of our method is that LaMET (quasi) data are fitted only with \smash{$q^{(Q)}(x)$}, while SDE (pseudo) data are fitted simultaneously, but only with \smash{$q^{(P)}(x)$}. The primary light-cone PDF, \smash{$q^{(0)}(x)$}, the same in \smash{$q^{(Q)}(x)$} and \smash{$q^{(P)}(x)$}, is thus constrained by both in complementary domains of $x$. The idea is illustrated in Fig.~\ref{fig:standard_q1234}, where we schematically show how the combined PDFs, \smash{$q^{(Q)}(x)$} and \smash{$q^{(P)}(x)$}, are constructed from \smash{$q^{(0,1,2,3)}(x)$}.

To ensure continuity at $x_{\rm{min}}$ and $x_{\rm{max}}$ and vanishing of light-cone PDFs at $x=1$, the auxiliary functions satisfy the following boundary conditions:
\begin{gather}
q^{(1)}(x_{\rm{min}}) = 0 \,, \nonumber\\
q^{(2)}(x_{\rm{max}}) = q^{(2)}(1) = 0 \,, \nonumber \\
q^{(3)}(x_{\rm{min}}) = q^{(3)}(x_{\rm{max}}) = 0\,. 
\label{eq:pdf:q123_boundary}
\end{gather}
We implement these conditions by selecting the following functional forms analogous to Eq.~\eqref{eq:q0}:
\begin{align}
q^{(1)}(x) &= x^{\delta^{(1)}}(x_{\rm{min}}-x)^{\rho^{(1)}}\, \mathrm{ANN}^{(1)}(x) \,, \nonumber \\
q^{(2)}(x) &= (x-x_{\rm{max}})^{\delta^{(2)}}(1-x)^{\rho^{(2)}}\, \mathrm{ANN}^{(2)}(x) \,, \nonumber \\
q^{(3)}(x) &= (x-x_{\rm{min}})^{\delta^{(3)}}(x_{\rm{max}}-x)^{\rho^{(3)}}\, \mathrm{ANN}^{(3)}(x) \,. 
\label{eq:pdf:q123_form}
\end{align}
The architecture of the neural networks  \smash{$\mathrm{ANN}^{(1,2,3)}(x)$} is the same as for  \smash{$\mathrm{ANN}^{(0)}(x)$}. That is, each of these networks can be described by Eq.~\eqref{eq:ann}, with the weights and biases, along with the powers \smash{$\delta^{(1,2,3)}$} and \smash{$\rho^{(1,2,3)}$}, unique to each function given in Eq.~\eqref{eq:pdf:q123_form}.

\begin{figure*}[t!]
\centering
\includegraphics[width=0.49\columnwidth]{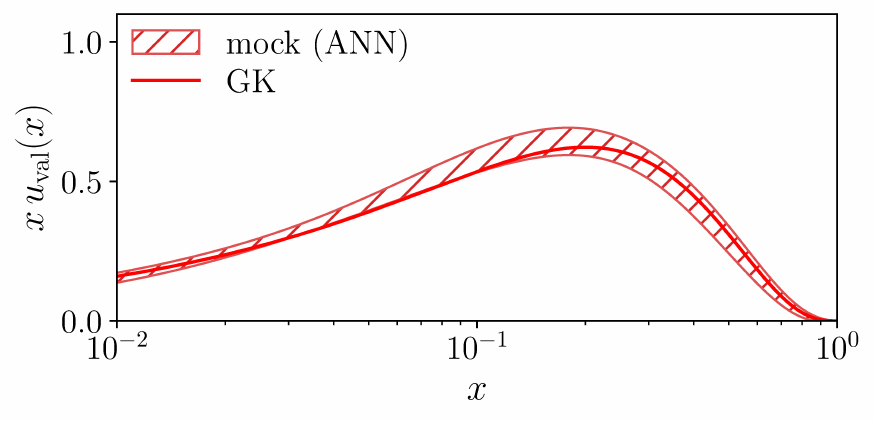}
\includegraphics[width=0.49\columnwidth]{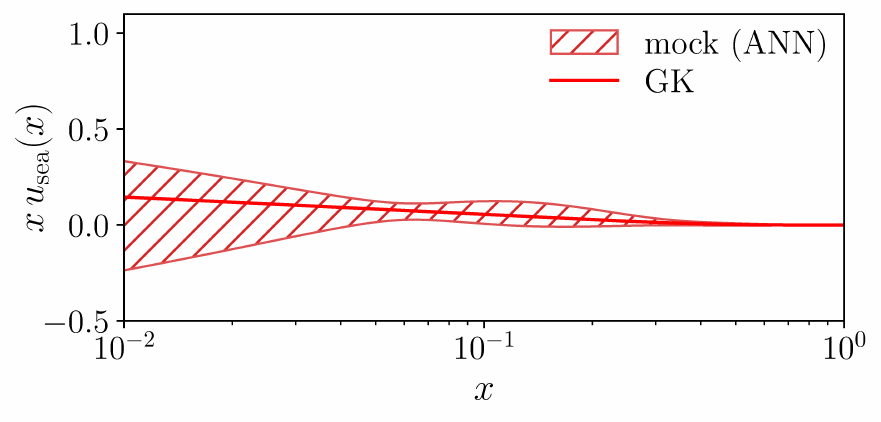}
\includegraphics[width=0.49\columnwidth]{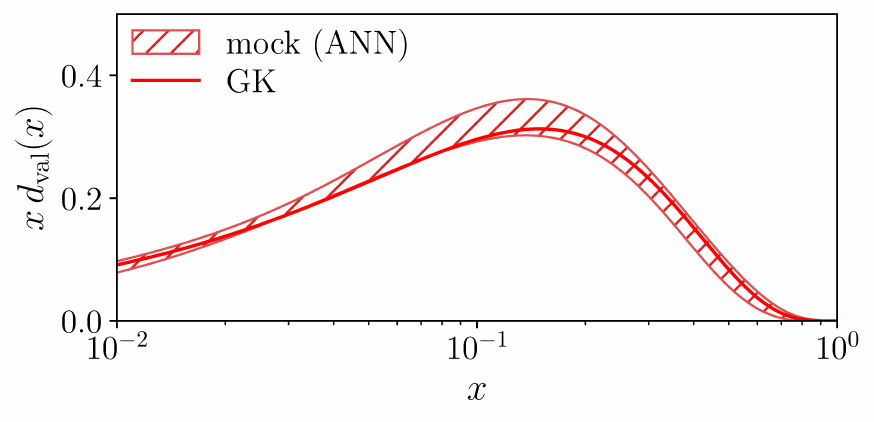}
\includegraphics[width=0.49\columnwidth]{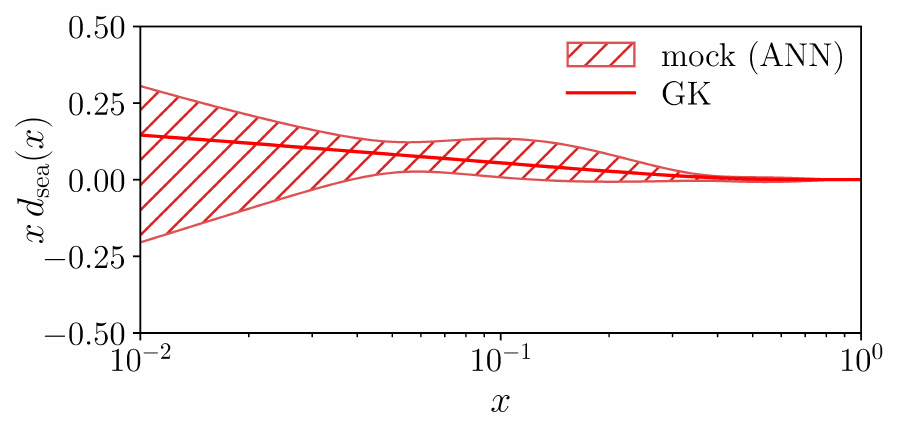}
\caption{Light-cone PDFs, $x q^{(0)}(x)$, reconstructed from mock data based on the GK model. The reference corresponding to the mock data is also shown (noted as ``GK''). Results for up (down) quarks are displayed in the upper (lower) row, with valence (sea) contributions shown in the left (right) columns.}
\label{fig:mock_result}
\end{figure*}

At this stage, before applying the method to actual lattice data for the reconstruction of the $x$ dependence and the unification of the LaMET/SDE approaches, we test the robustness of this ANN setup with a mock data test.
For the LaMET side, we use the same GK-based mock data as for the analogous test of the standard BG approach (see Fig.~\ref{fig:comparison_BG} and its description in the text), but supplement it with additional mock data corresponding to SDE.
In this case, we test all four combinations of $u/d$ flavors and their valence/sea components.
The reconstructed light-cone PDFs, \smash{$xq^{(0)}(x)$}, are shown as dashed bands in Fig.~\ref{fig:mock_result}. The corresponding GK model curves are shown in red. In all cases, the ANN reconstruction closely reproduces the GK PDFs, demonstrating the validity of our approach. 
We also emphasize a striking contrast with the BG reconstruction, which failed the test based on the same mock data.
Finally, we note that the ANN procedure of PDFs reconstruction translates the limitations of the data, i.e.\ their discreteness and truncation, to enhanced errors, reflecting the presence of the inverse problem.

\begin{figure*}
\centering
\includegraphics[width=0.49\columnwidth]{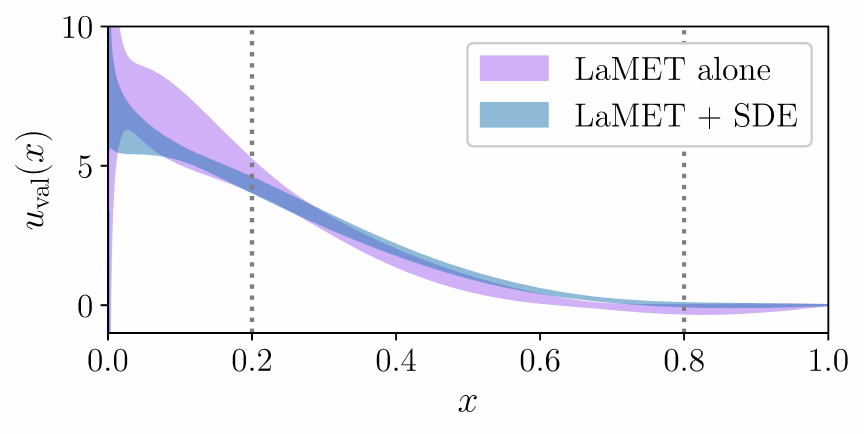}
\includegraphics[width=0.49\columnwidth]{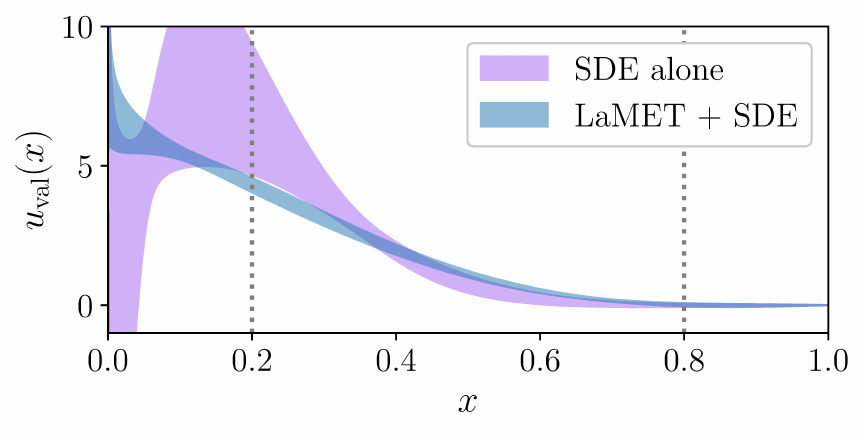}
\caption{Comparison of the default result (``LaMET+SDE'') from our unified framework for the up quark valence PDF with reconstruction scenarios relying solely on LaMET (left) and SDE (right) inputs.}
\label{fig:qp_com}
\end{figure*}

\subsection{Results}
\label{sec:pdf:results}
The key elements of our numerical framework are as follows: we use a genetic algorithm~\cite{10.5555/522098} to constrain the free parameters of \smash{$q^{(Q, P)}(x)$} from lattice data; the artificial neural networks consist of five neurons in their hidden layers, i.e., $N^{(0,1,2,3)} = 5$, which we found to be optimal for providing a flexible model while avoiding excessive freedom that could lead to overfitting; we explicitly assume a certain order of divergence of the primary light-cone PDFs as $x \to 0$, namely, \smash{$-1 < \delta^{(0)} < 0$} for valence quarks and \smash{$-2 < \delta^{(0)} < -1$} for sea quarks, which should be understood as a theory-driven constraint motivated by Regge theory, PDF evolution and empirical observations~\cite{Ewerz:2013kda, Collins:1989gx, NNPDF:2014otw}. The technical details of the reconstruction are summarized in Appendix~\ref{app:pdf:numerical_framework}.
{\color{red}It is important to emphasize here, nevertheless, that while exhaustive architectural sensitivity tests are crucial for the robust extraction of distributions like PDFs from lattice QCD data -- a task that will remain a formidable challenge for the foreseeable future -- such an extensive undertaking is beyond the scope of the current study. Importantly, such tests would not alter the fundamental methodology of combining the LaMET and SDE approaches, which remains the primary focus of this work.}

It is important to stress in the main text though, that while exhaustive architectural sensitivity
tests are crucial for a robust extraction of objects like PDFs from lattice QCD data, which is expected to remain a formidable challenge for the foreseeable future, such an extensive undertaking is beyond the scope of the current study. Importantly,
such tests would not alter the fundamental methodology of combining the LaMET and SDE approaches, which
remains the primary focus of this work.

\begin{figure*}
\centering
\includegraphics[width=0.49\columnwidth]{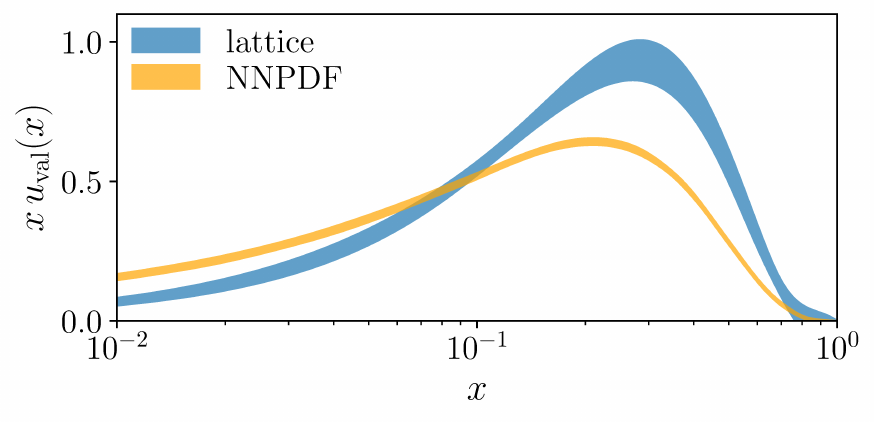}
\includegraphics[width=0.49\columnwidth]{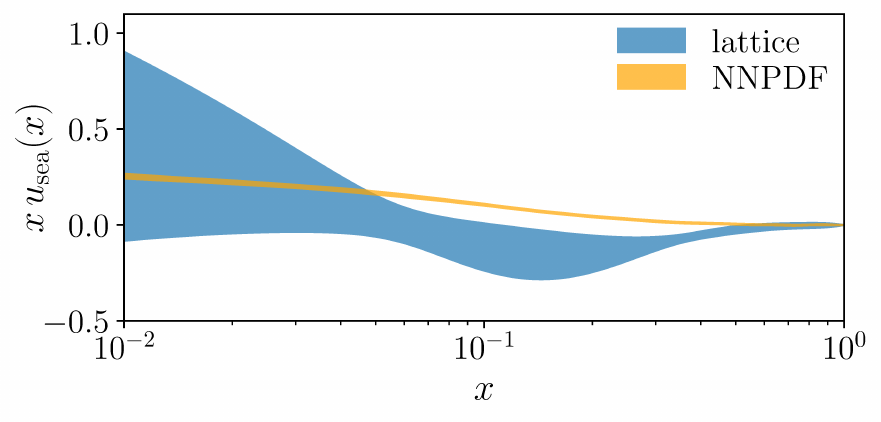}
\includegraphics[width=0.49\columnwidth]{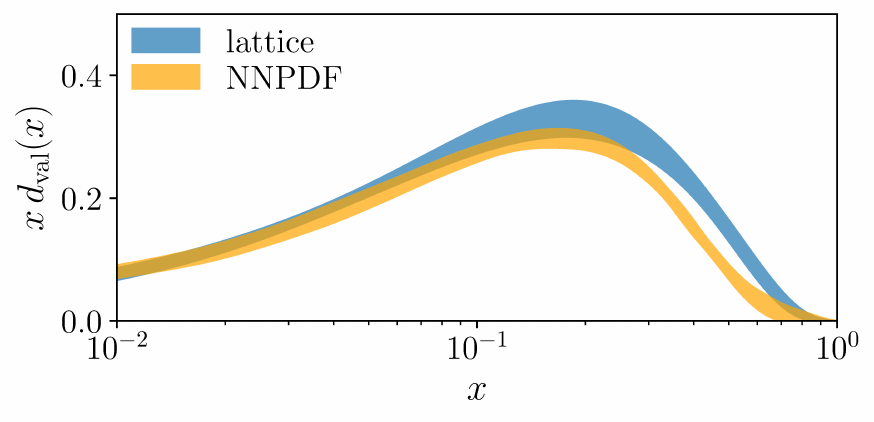}
\includegraphics[width=0.49\columnwidth]{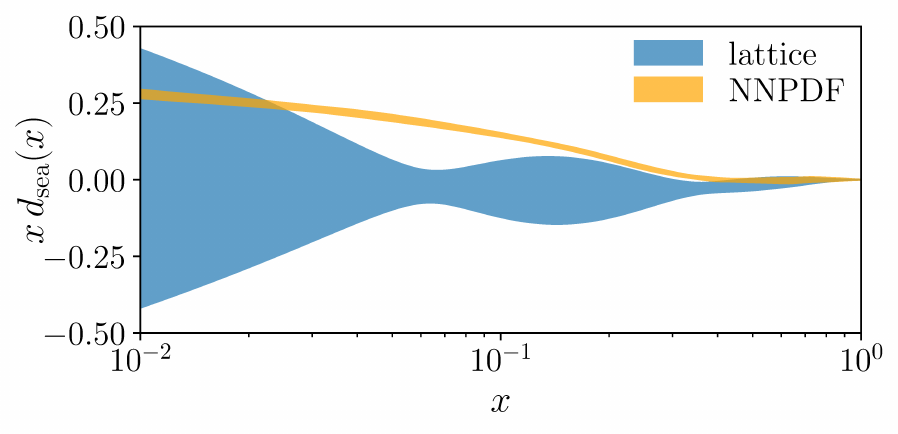}
\caption{Light-cone PDFs, $x q^{(0)}(x)$, reconstructed from our lattice data (blue bands). The NNPDF result~\cite{NNPDF:2014otw}, representing a phenomenological determination constrained by experimental data, is also shown (orange bands). Results for up (down) quarks are displayed in the upper (lower) row, with valence (sea) contributions shown in the left (right) columns.}
\label{fig:lat_result}
\end{figure*}

We apply the method to the lattice data described in Sec.~\ref{sec:pdfs_and_gpds_from_lattice}.
First, we investigate the effects of three different strategies of employing our ANN framework.
In two of them, LaMET or SDE are used alone, without any benefit from their possible complementarity.
The third scenario combines the quasi- and pseudo-distribution approaches in a unified ``LaMET+SDE'' fit.
The up quark valence PDF is shown in Fig.~\ref{fig:qp_com}. For LaMET alone (left panel), we observe a systematic departure from the unified result, which can be attributed to power corrections of $\mathcal{O}(1/x^2\Lambda_{\rm QCD}^2)$ and $\mathcal{O}(1/(1-x)^2\Lambda_{\rm QCD}^2)$ in the endpoint regions. A smaller but discernible mismatch also remains in the intermediate region $0.2\!=\!x_{\mathrm{min}}<x<x_{\mathrm{max}}\!=\!0.8$. For SDE alone (right panel), uncertainties inflate substantially: higher-twist contributions of $\mathcal{O}(z^{2}\Lambda_{\rm QCD}^2)$ necessitate restricting the analysis to short separations (i.e., $z\le z_{\max}\!=\!4a\approx0.37$ fm; meanwhile, LaMET uses data up to $z_{\rm cut}=9a\approx0.84$ fm), thereby reducing the information available to constrain the PDFs. Taken together, it indicates that neither using LaMET alone nor SDE alone suffices to determine the PDFs with high precision and stability.
In turn, their combination, ``LaMET+SDE'', displayed in both panels of Fig.~\ref{fig:qp_com}, yields a more reliable result, since the LaMET framework provides PDF-constraining information only in its domain of applicability, $x_{\mathrm{min}}<x<x_{\mathrm{max}}$, and the SDE method augments this information with short-distance data that can be viewed as global characteristics of the distribution, i.e., not attributed to any region of~$x$. In this way, we have shown that the unification of LaMET and SDE approaches provides stronger constraints for the reconstruction of light-cone PDFs. {\color{red}Crucially, this synergy is not merely a numerical reduction of errors but is grounded in theoretical consistency, leveraging the complementarity between the endpoint-sensitive LaMET corrections and the global constraints of SDE.}

In Fig.~\ref{fig:lat_result}, we show the LaMET+SDE results from our unified framework for the valence/sea components of up/down quarks, focusing only on the primary contribution \smash{$q^{(0)}(x)$} (the extracted by-products \smash{$q^{(1,2,3)}(x)$} are shown in Appendix~\ref{app:pdf_results}).
The blue bands represent the results from our lattice data and the orange bands correspond to the NNPDF parametrizations~\cite{NNPDF:2014otw} obtained through the LHAPDF library~\cite{Buckley:2014ana}. The ANN reconstruction performs well for the valence-quark distributions, whereas the sea-quark ones are subject to very large uncertainties. Obviously, the presented curves originating from lattice data should not to be treated fully quantitatively. They are still contaminated by unaddressed systematic uncertainties related to the lattice setup, such as discretization effects, a non-physical pion mass and effects of renormalization. 
{\color{red}These effects may be of similar magnitude for both valence and sea quarks, but the PDFs corresponding to the latter are strongly suppressed, leading to larger relative uncertainties for the sea sector. Moreover, the sea distributions originate from the difference of $q_{\rm val}+2q_{\rm sea}$ (determined by the imaginary part of MEs) and $q_{\rm val}$ (real part of MEs) and hence, are subject to somewhat larger statistical errors.
Thus, the overall uncertainties for sea quark PDFs are currently larger than their expected magnitudes, which is reflected in the right panels of Fig.~\ref{fig:lat_result}.
It is important to emphasize that, crucially, the remaining systematics is fully controllable and its account can be incorporated into our ANN analysis, which we plan to address in the future.
Then, we expect that quantitative statements about the sea sector may also be obtained, although the suppression of sea PDFs means that the relative uncertainties will always be more favorable for the valence sector.}

Detailed information regarding fitting methods, positivity constraints, and systematic analyses is provided in Appendix~\ref{app:pdf_results}. Specifically, we investigated the impact of enforcing positivity constraints on PDFs, which, in phenomenological analyses, strictly hold only at leading order~\cite{Altarelli:1998gn}. We find a negligible effect of this constraint on valence quarks, but a significant impact on the sea contribution. 
Furthermore, systematic studies involving treatments of real and imaginary components of lattice data, as well as sensitivity analyses of auxiliary parameters confirm that our method remains stable under variations in these parameters.
In particular, the results are robust against changing the $x_{\mathrm{min}}$, $x_{\mathrm{max}}$, and $z_{\mathrm{max}}$ parameters, indicating control over power corrections in the LaMET and SDE parts of our framework, and against changes in the $z_{\rm cut}$ parameter, implying small sensitivity to the large-$z$ truncation of the LaMET part.

\subsection{Conclusions}
\label{sec:pdf:conclusions}

In this section, we introduced an ANN-assisted parametrization of the light-cone PDF, $q^{(0)}(x)$, with a power-law prefactor enforcing endpoint behavior and exact valence-number normalization. In addition, three auxiliary components, $q^{(1,2,3)}(x)$, were used to localize power-correction effects and guarantee continuity. Benchmarks on mock data show faithful reproduction of the reference PDFs, validating the approach. Applied to present lattice inputs, the method yields well-constrained valence distributions but sizable uncertainties for sea quarks.
Finally, direct comparisons demonstrate that LaMET-only or SDE-only fits cannot simultaneously control endpoint biases and statistical noise, while the combined strategy -- using $q^{(Q)}(x)$ for LaMET and $q^{(P)}(x)$ for SDE -- achieves a more robust reconstruction.

\section{Reconstruction of generalized parton distributions at $\xi=0$}
\label{sec:gpd}

The reconstruction of GPDs at $\xi = 0$ from lattice-QCD data follows the same procedure as for PDFs, with the only formal difference being the introduction of dependence on $t$. The main challenge here is to impose theory-driven constraints that can be deduced from the interpretation of nucleon tomography: (i) the dependence on $t$ must yield an integrable function for the Fourier transform in Eqs.~\eqref{eq:nt:H} and~\eqref{eq:nt:E}; (ii) the dependence on $t$ must vanish as $x \to 1$, so that the Fourier transform reduces to a delta function, correctly localizing the active parton; and (iii) the distance between the active parton and the spectator system must remain finite. These constraints are discussed in detail in Appendix~\ref{app:tomography}.

\subsection{Functional form of light-cone GPDs at $\xi=0$}
\label{sec:gpds:form}

Because of the aforementioned theory-driven constraints, introducing dependence on $t$ must be done with care. At the level of constructing ANNs, one cannot simply combine $x$ and $t$ in a weighted sum and then apply an arbitrary activation function. Instead, the constraints must be incorporated directly into the network architecture. This difficulty has been addressed in the context of GPDs in Ref.~\cite{Dutrieux:2021wll}, though focusing on the interplay between $x$ and $\xi$ rather than between $x$ and~$t$.

We construct GPDs at $\xi=0$ as follows,
\begin{align}
q^{(0)}(x,t) = x^{\delta^{(0)}}(1-x)^{\rho^{(0)}}\, \mathrm{ANN}^{(0)}(x,t) \,,
\label{eq:t:q0}
\end{align}
where
\begin{flalign}
\mathrm{ANN}^{(0)}(x, t) = b_{21}^{(0)} + \sum_{i=1}^{N^{(0)}} w_{2i}^{(0)} \ln\left(1+\exp(b_{1i}^{(0)}+w_{1i}^{(0)}x)\right) \exp\left(s_{i}^{(0)}(1-x)^2t\right)
\label{eq:t:ann}
\end{flalign}
Here, each neuron in the original network used for PDFs (see Eq.~\eqref{eq:ann}) is now dependent on both $x$ and $t$, with the latter introduced through the slopes $\smash{s_{i}^{(0)}}$. This modeling approach, based on a series of exponential functions, is quite general owing to its relation to the Laplace transform and is, in particular, well suited for mimicking dipole-like behavior (see Ref.~\cite{Aschenauer:2025cdq} for further discussion).

The factor $(1-x)^2$ is introduced in Eq.~\eqref{eq:t:ann} to satisfy the theory-driven constraints outlined at the beginning of this section. Namely, it ensures that the dependence on $t$ vanishes as $x \to 1$, so that the Fourier transform of \smash{$q^{(0)}(x,t)$} with respect to $t$ reduces to a delta function in that limit, indicating that the active parton is located at the origin of the reference frame. Moreover, the choice of the second power of $(1-x)$ keeps the distance to the spectator system finite as $x \to 1$, whereas a smaller (larger) power would cause the nucleon to collapse (expand indefinitely).

In the following, we present results for the GPDs $H^{q}(x,0,t)$ and $E^{q}(x,0,t)$. For the latter and for valence quarks, we do not enforce the normalization of their forward limits specified in Eq.~\eqref{eq:forward_limit_E}, 
\begin{align}
\int_{0}^{1}dx\,e_{u_{\mathrm{val}}}^{(0)}(x) = \kappa_{u}
\quad
\mathrm{and}
\quad
\int_{0}^{1}dx\,e_{d_{\mathrm{val}}}^{(0)}(x) = \kappa_{d} \,,
\label{eq:forward_limit_E_norm}
\end{align}
where $\kappa_q$ are the contributions of partons to the anomalous magnetic moment of the proton. These constraints are omitted because $\kappa_q$ are not inputs to lattice-QCD calculations. Later, we comment on how the lattice data used reproduce the measured values of $\kappa_q$.

As in the case of PDFs, we use auxiliary functions $q^{(1,2,3)}(x,t)$ combined with $q^{(0)}(x,t)$ to obtain $q^{(Q, P)}(x,t)$, which we then fit to LaMET and SDE quantities, respectively (see the prescription given by Eqs.~\eqref{eq:qQ} and~\eqref{eq:qP}). The construction of $q^{(1,2,3)}(x,t)$ is again similar to $q^{(0)}(x,t)$, namely, the functional forms only differ by power prefactors, just like in Eq.~\eqref{eq:pdf:q123_form}, which is needed to satisfy the boundary conditions~\eqref{eq:pdf:q123_boundary}.

\subsection{Results}
\label{sec:gpds:results}

\begin{figure*}
\centering
\includegraphics[width=1\columnwidth]{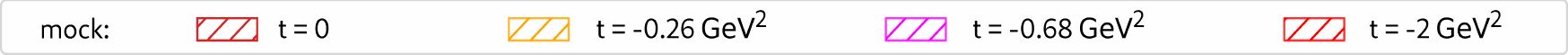} \\
\includegraphics[width=0.49\columnwidth]{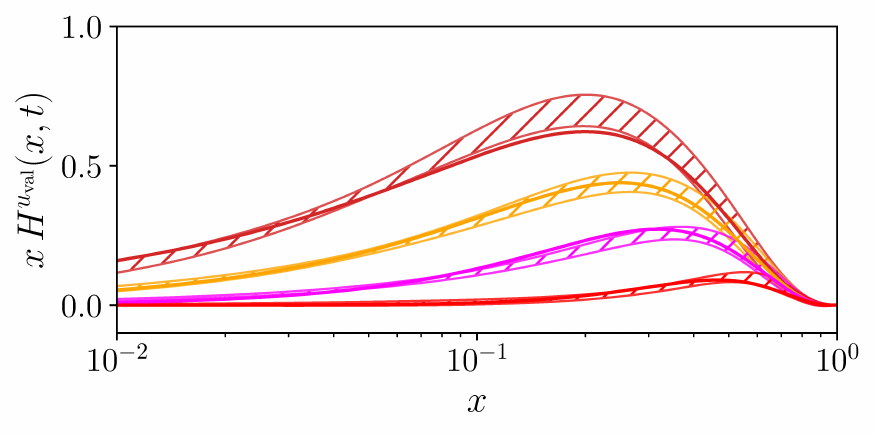}
\includegraphics[width=0.49\columnwidth]{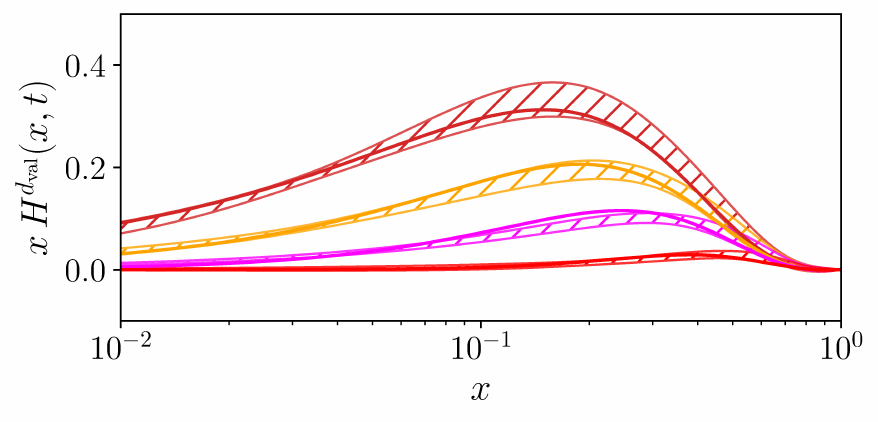}
\includegraphics[width=0.49\columnwidth]{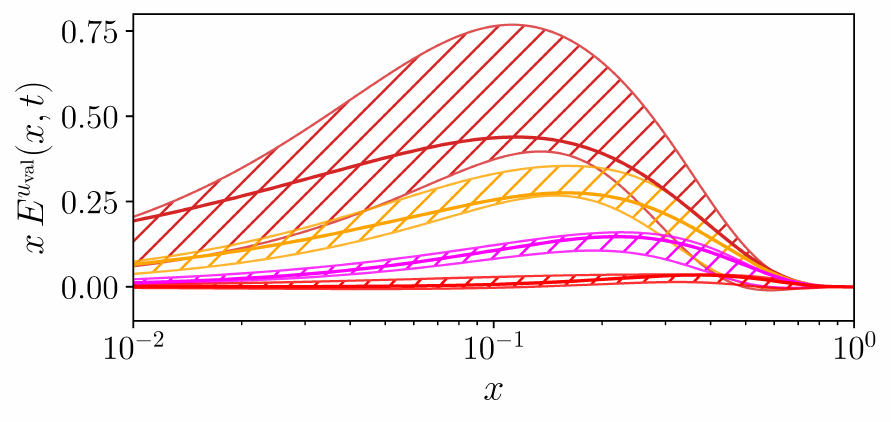}
\includegraphics[width=0.49\columnwidth]{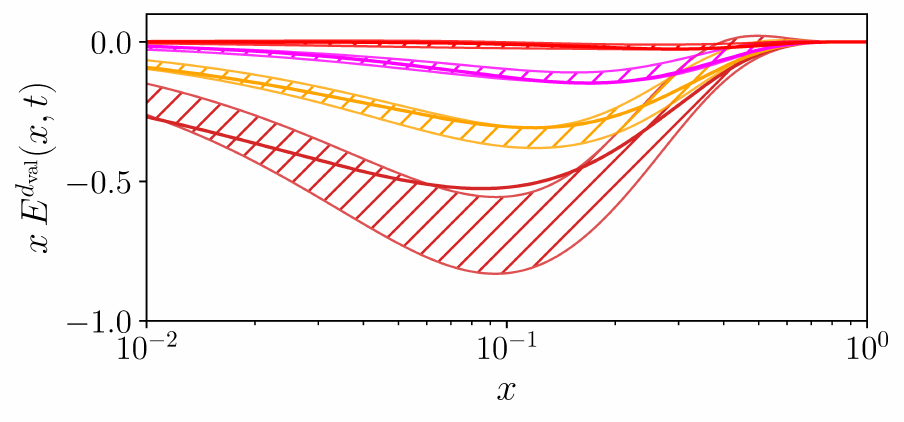}
\includegraphics[width=1\columnwidth]{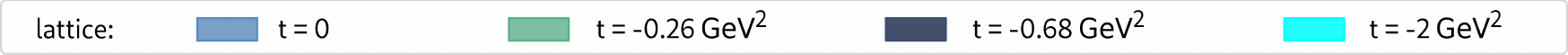} \\
\includegraphics[width=0.49\columnwidth]{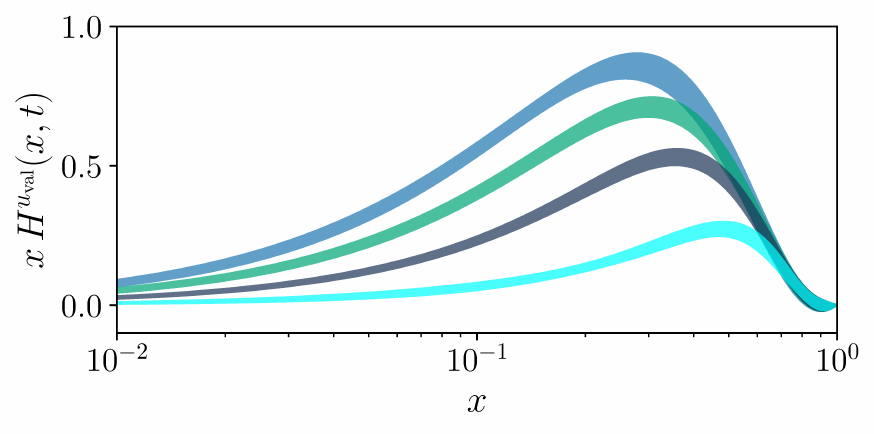}
\includegraphics[width=0.49\columnwidth]{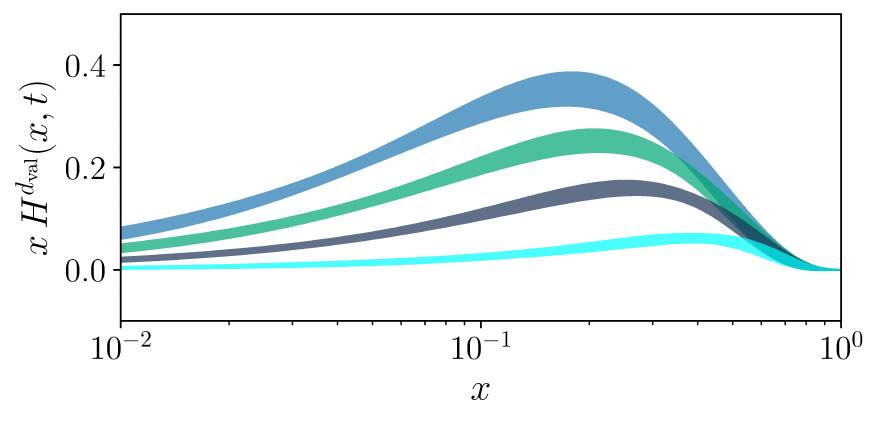}
\includegraphics[width=0.49\columnwidth]{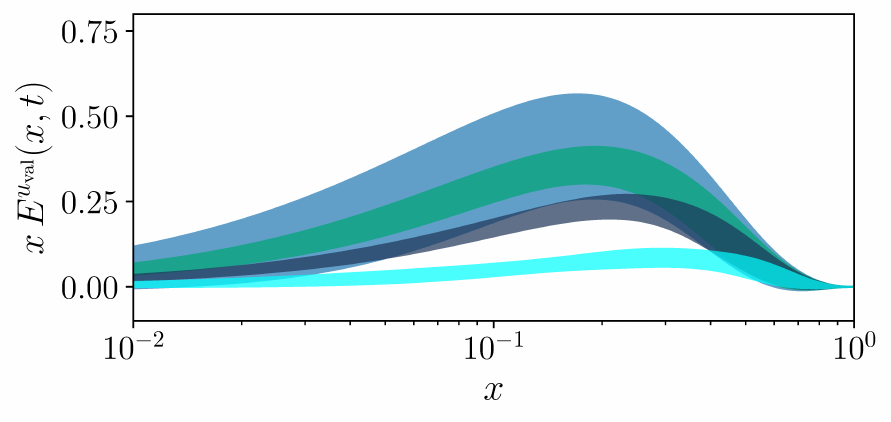}
\includegraphics[width=0.49\columnwidth]{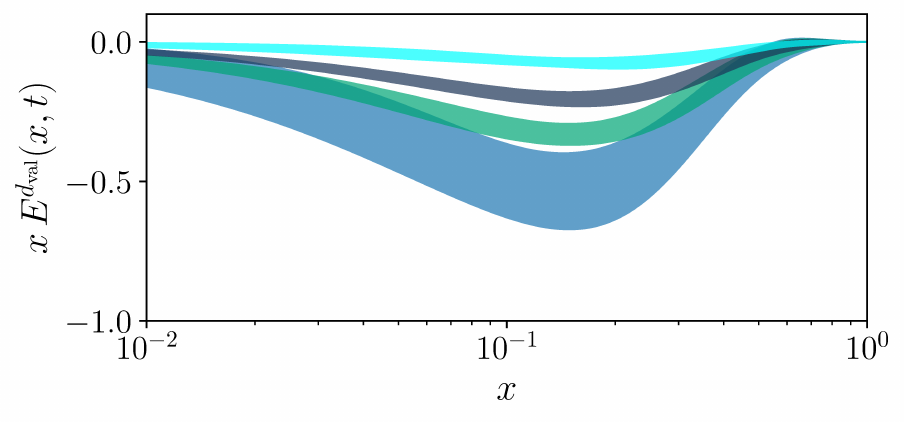}
\caption{GPDs $H^{q}(x,0,t)$ and $E^{q}(x,0,t)$ for valence up (left) and down (right) quarks as functions of $x$ at several fixed values of $t$, reconstructed from mock data (two upper rows) and lattice data (two lower rows). Solid lines denote the reference for the mock data.}
\label{fig:gpd:pdf}
\end{figure*}

Because of the difficulties to reconstruct the sea contribution from current lattice data (see the previous section), we focus only on the valence components in the demonstration of our method. In Fig.~\ref{fig:gpd:pdf}, we show GPDs $H^{q}(x,0,t)$ and $E^{q}(x,0,t)$ for up and down quarks as functions of $x$ at several fixed values of $t$, reconstructed from mock data and lattice data. 
The former are once again generated from the GK model. As one can see, we are able to correctly reproduce the reference for the mock data, once more indicating the validity of our reconstruction method.  

\begin{figure*}
\centering
\includegraphics[width=0.49\columnwidth]{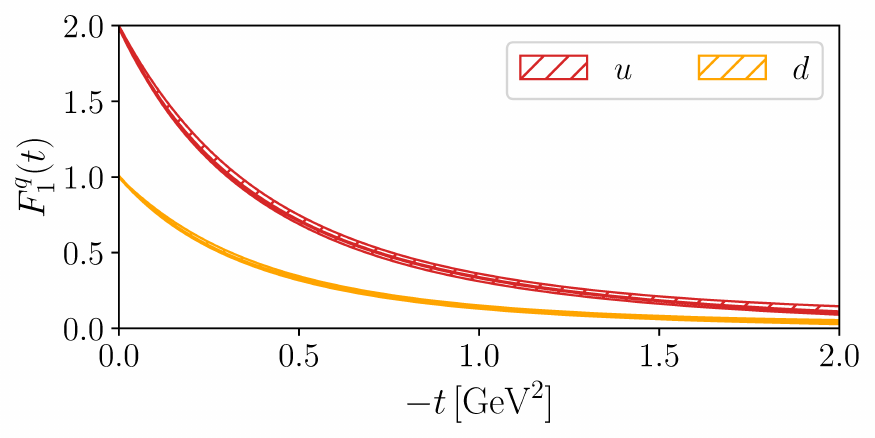}
\includegraphics[width=0.49\columnwidth]{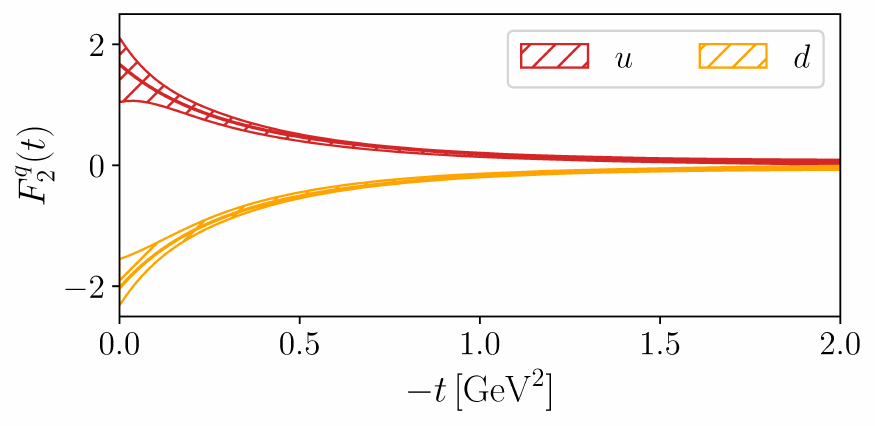}
\includegraphics[width=0.49\columnwidth]{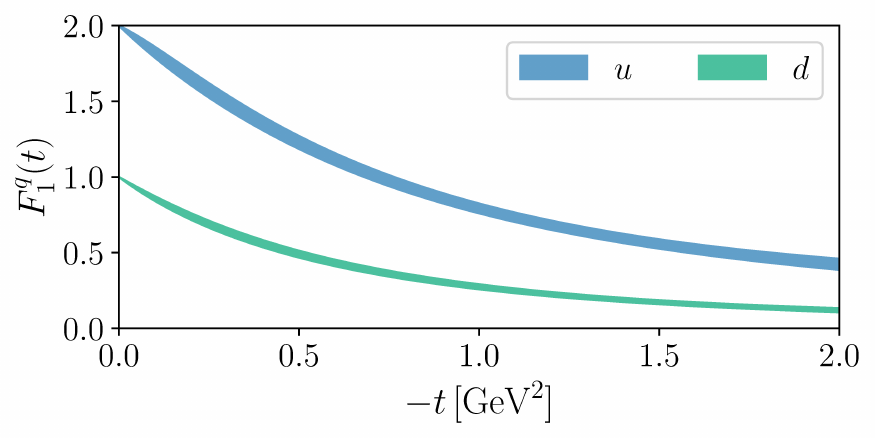}
\includegraphics[width=0.49\columnwidth]{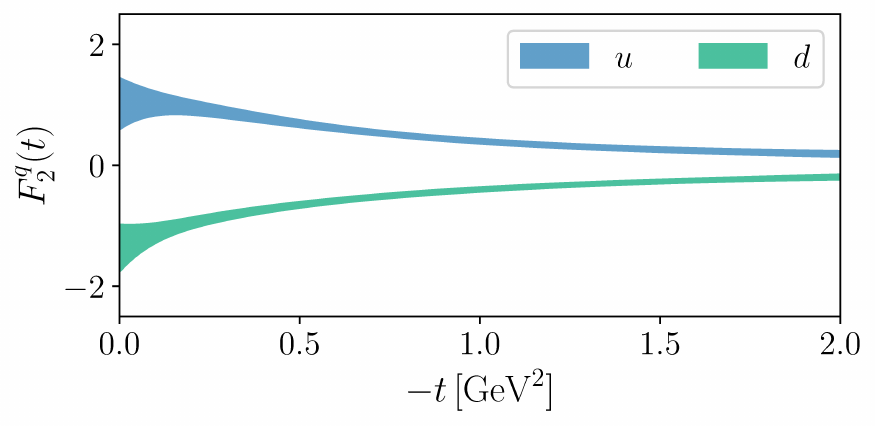}
\caption{Partonic contributions to Dirac (left) and Pauli (right) elastic form factors, reconstructed from mock data (upper row) and lattice data (lower row). The reference for the mock data is shown as solid lines.}
\label{fig:gpd:eff}
\end{figure*}

Among the presented values of $t$ in Fig.~\ref{fig:gpd:pdf}, one finds the curves for $t=0$, which correspond to the forward limits of the considered GPDs, see Eqs.~\eqref{eq:forward_limit_H} and~\eqref{eq:forward_limit_E}. 
An important consistency check for the GPD $H^{q}(x,0,t)$ is to compare it with the unpolarized PDF calculated explicitly.
We present this comparison in Appendix \ref{app:gpd_results}, concluding full agreement in the whole range of $x$.
This is highly non-trivial, since the lattice data for GPDs and PDFs come from an independent calculation with possibly different systematic effects.
The forward limit of the GPD $E^{q}(x,0,t)$ is of particular interest, as it is a vital but elusive component in modeling GPDs, carrying information about the total angular momentum, cf. Eq.~\eqref{eq:basics_ji}. In general, the uncertainties for GPDs $E^q(x,0,t)$ are larger than those for $H^q(x,0,t)$, owing to the fact that the normalization~\eqref{eq:forward_limit_E_norm} is not enforced.

The anomalous magnetic moment of the proton obtained from lattice data is $\kappa_p = \nicefrac{2}{3}\kappa_u - \nicefrac{1}{3}\kappa_d = 1.14 \pm 0.26$, with the partonic contributions $\kappa_u = 1.03 \pm 0.43$ and $\kappa_d = -1.37 \pm 0.39$. This should be compared with the experimental value $\kappa_p^{\mathrm{exp.}} = 1.792847350(9)$~\cite{BASE:2014drs}. The values of total angular momenta evaluated from Ji's sum rule for lattice data are $J^{u_{\mathrm{val}}}=0.281 \pm 0.028$ and $J^{d_{\mathrm{val}}}=-0.0137 \pm 0.023$. Other estimates one may find in the literature include, for instance, $J^{u_{\mathrm{val}}} = 0.195 \pm 0.010$ and $J^{d_{\mathrm{val}}} = 0.0173 \pm 0.0046$~\cite{Cichy:2024afd}; and $J^{u} = 0.229 \pm 0.002^{+0.008}_{-0.012}$, $J^{u_{\mathrm{sea}}} = 0.015 \pm 0.003^{+0.001}_{-0.000}$, $J^{d} = -0.007 \pm 0.003^{+0.020}_{-0.005}$, $J^{d_{\mathrm{sea}}} = 0.022 \pm 0.005^{+0.001}_{-0.000}$~\cite{Bacchetta:2011gx} (for more estimates, see Ref.~\cite{Hashamipour:2021kes}). It is important to note that all the aforementioned estimates of total angular momentum are subject to unknown model uncertainties. At present, the values obtained in our analysis are also still affected by systematic uncertainties (e.g., discretization effects and effects of a nonphysical pion mass). Thus, some of the nucleon characteristics are not reproduced quantitatively. However, this is expected to improve over time, with dedicated calculations to quantify all systematic effects.

In Fig.~\ref{fig:gpd:eff}, we show the partonic contributions to the Dirac and Pauli elastic form factors (see Eqs.~\eqref{eq:theory:eff_H} and~\eqref{eq:theory:eff_E}), for both mock and lattice data. Once again, we are able to reproduce the reference for the mock data. From the figures for the Pauli form factor, $F_2(t)$, we observe a growth of uncertainties as $t \to 0$, which is a consequence of not enforcing the normalization~\eqref{eq:forward_limit_E_norm} for GPDs $E^q(x,0,t)$. The increase of uncertainties in this limit reflects the unbiased nature of our reconstruction, which naturally leads to larger uncertainties in regions not constrained by lattice data, where we rely on the extrapolation.

\begin{figure*}
\centering
\includegraphics[width=1\columnwidth]{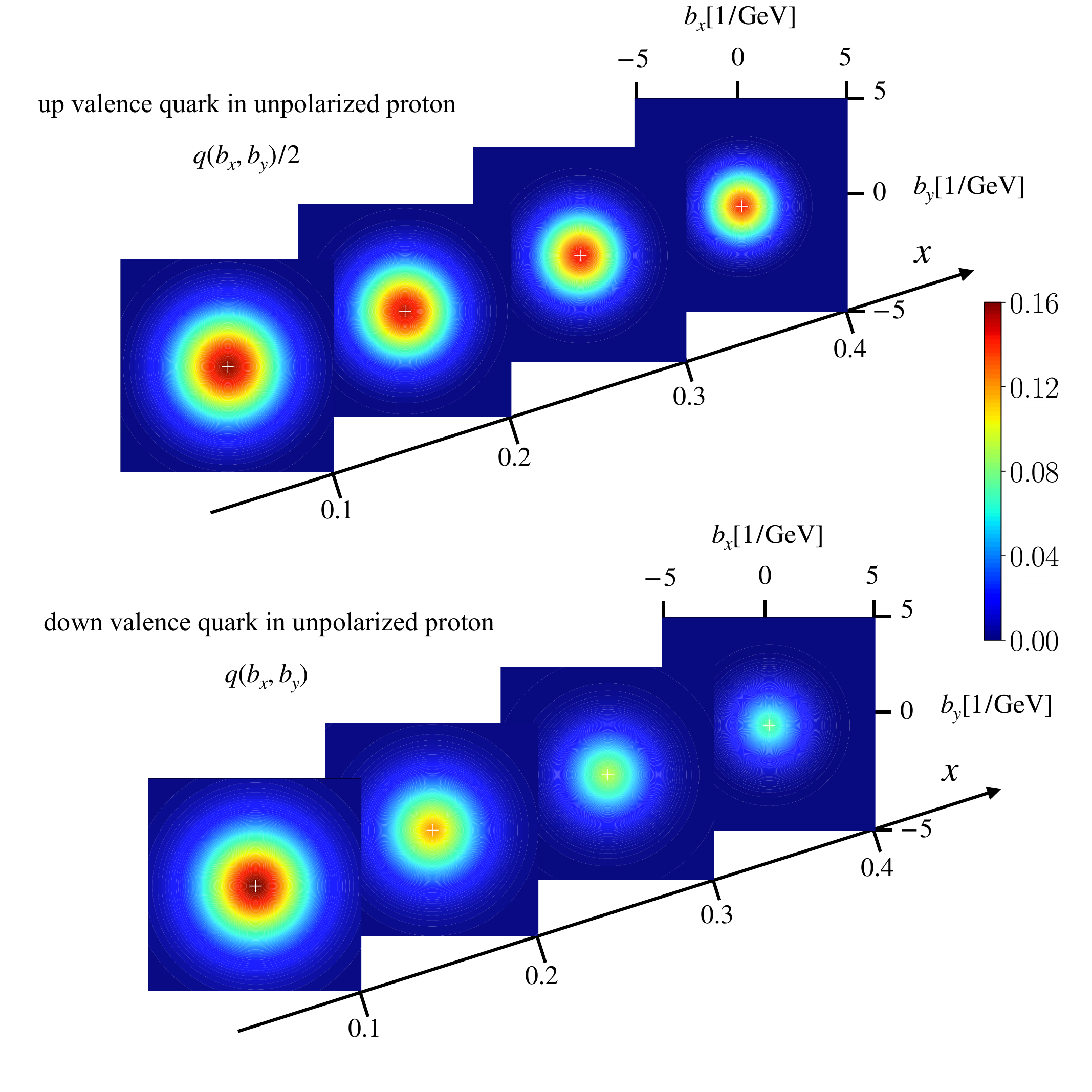}
\includegraphics[width=0.24\columnwidth]{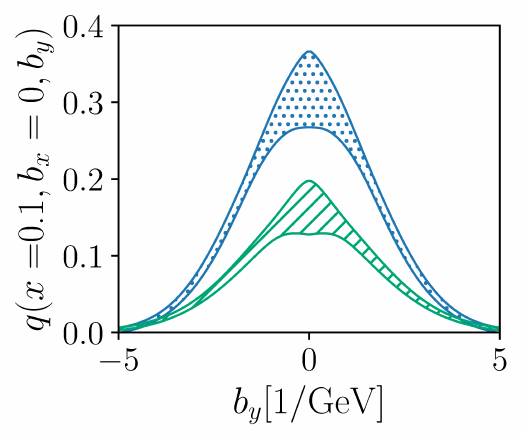}
\includegraphics[width=0.24\columnwidth]{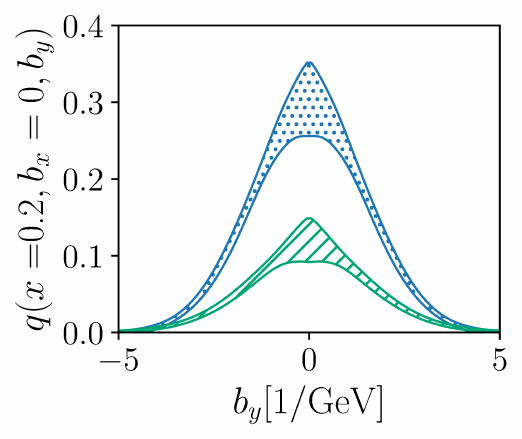}
\includegraphics[width=0.24\columnwidth]{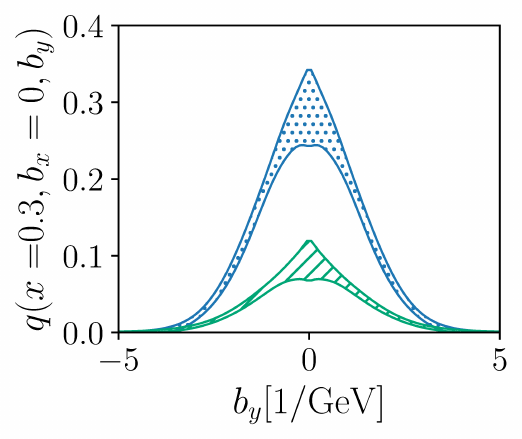}
\includegraphics[width=0.24\columnwidth]{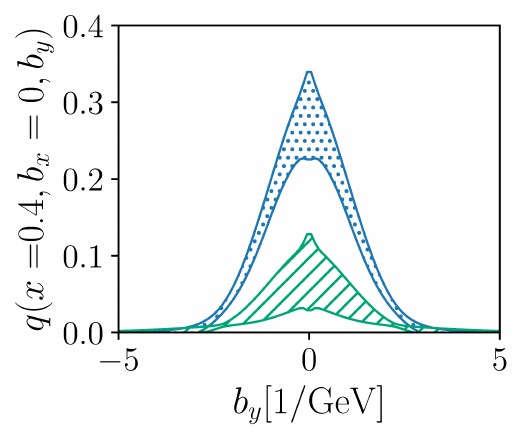}
\caption{Nucleon tomography images for the unpolarized proton and the up and down valence quarks. Results are presented for $x = 0.1, 0.2, 0.3$ and $0.4$ as 2D profiles, with the coordinate system origin indicated by small white crosses. The bottom panels display the corresponding sections along the Y-axis at fixed $b_x = 0$ for up (blue dotted band) and down (green dashed band) valence quarks. For clarity, the up quark distributions in the 2D profiles are rescaled by a factor of $\nicefrac{1}{2}$.}
\label{fig:gpd:nt_1}
\end{figure*}

\begin{figure*}
\centering
\includegraphics[width=1\columnwidth]{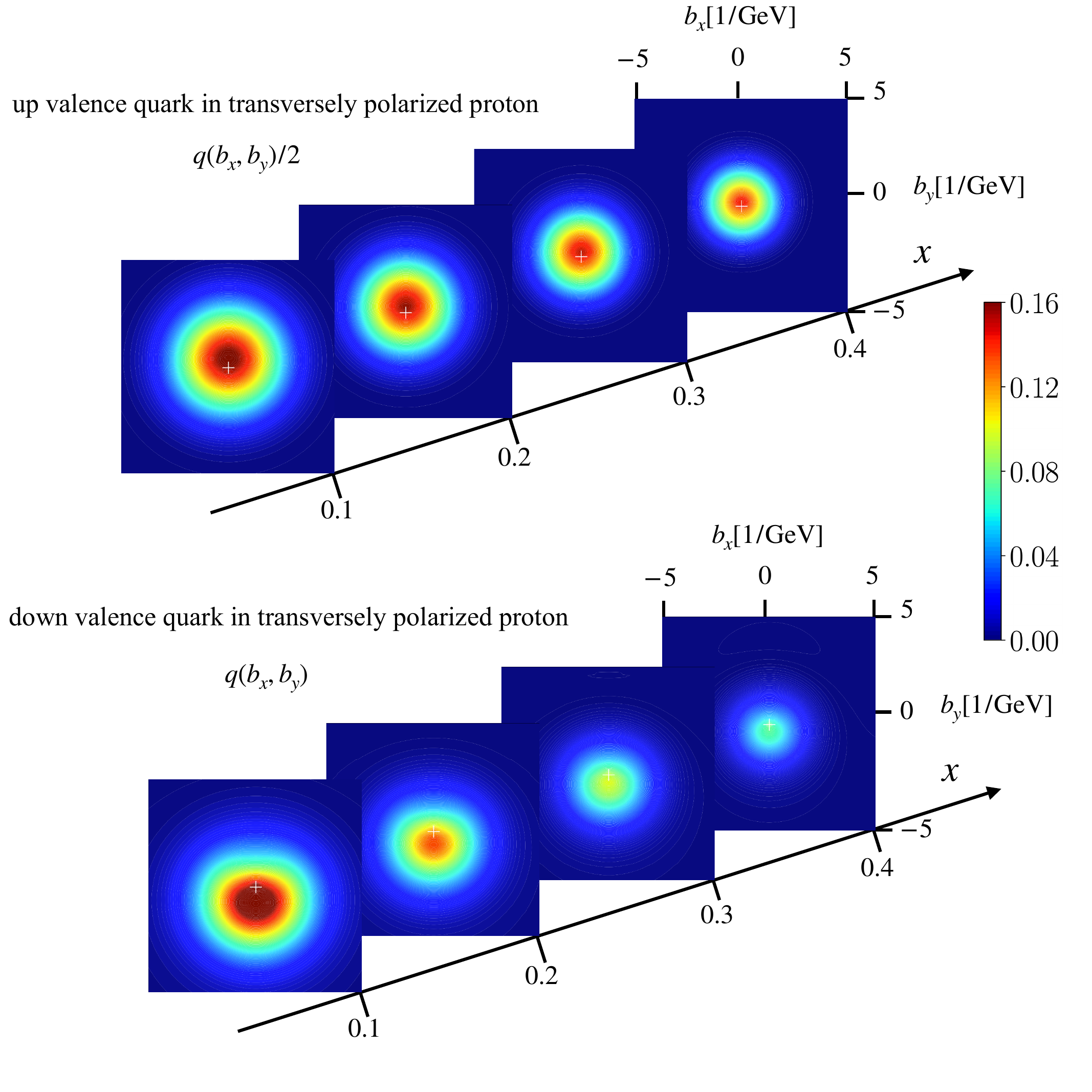}
\includegraphics[width=0.24\columnwidth]{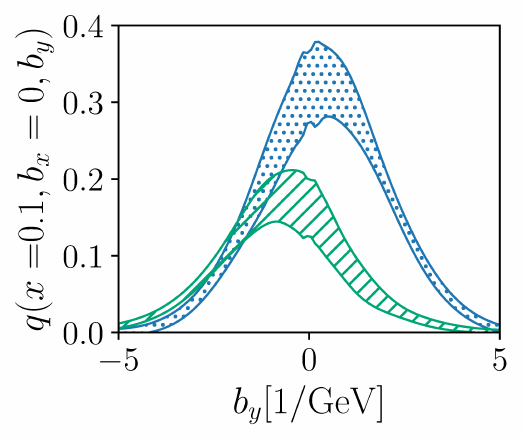}
\includegraphics[width=0.24\columnwidth]{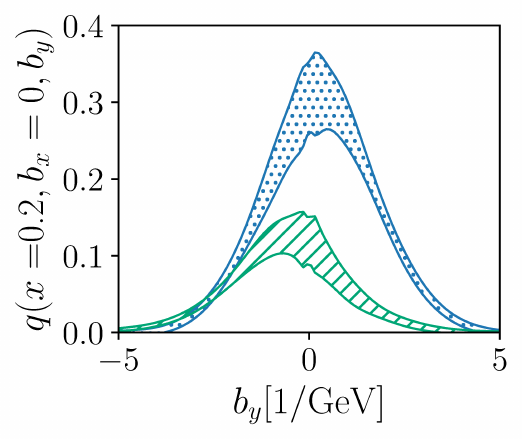}
\includegraphics[width=0.24\columnwidth]{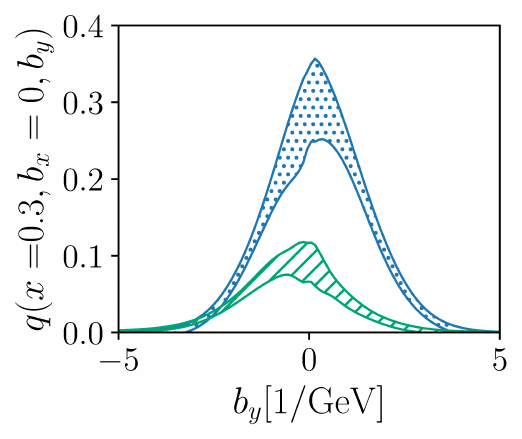}
\includegraphics[width=0.24\columnwidth]{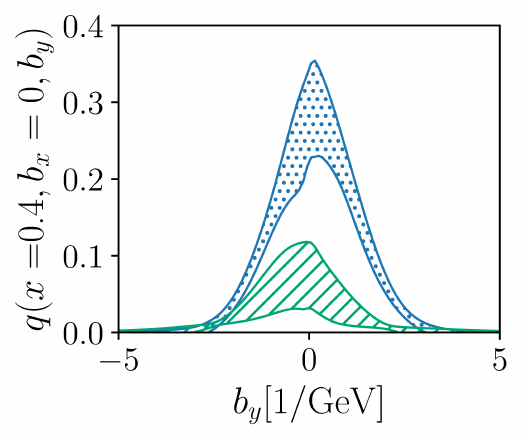}
\caption{Nucleon tomography images for the transversely polarized proton. See Fig.~\ref{fig:gpd:nt_1} for more details.}
\label{fig:gpd:nt_2}
\end{figure*}

Finally, the resulting tomographic pictures from the lattice analysis are presented in Figs.~\ref{fig:gpd:nt_1} and \ref{fig:gpd:nt_2}. We show both one-dimensional profiles for the unpolarized and transversely polarized proton, which allow us to indicate the associated uncertainties, as well as two-dimensional profiles. The deformation of parton densities in the case of a transversely polarized proton is clearly visible. Taking into account the inability of the lattice data used to reproduce PDFs and EFFs, we note that these plots should be regarded only as a demonstration of what will be achievable with more precise lattice QCD calculations, preferably in analyses augmented by measurements such as those anticipated at the EIC~\cite{Aschenauer:2025cdq}. 

Further discussion of the GPD results is provided in Appendix~\ref{app:gpd_results}. In particular, we investigate the quality of the reconstruction of lattice data, present a comparison between the forward limit of GPD $H^q(x,0,t)$ and results directly obtained for PDFs, and provide additional results on nucleon tomography.

\subsection{Conclusions}
\label{sec:gpd:conclusions}

In this section, we extended the ANN-based reconstruction to $\xi = 0$ GPDs by embedding the $t$-dependence neuron-wise through an exponential decay function. Unifying LaMET and SDE approaches in a similar way as employed for PDFs, we reconstructed $H^q(x,0,t)$ and $E^q(x,0,t)$ for the valence quarks. Benchmarks on mock data are accurately reproduced, once again validating the adapted approach. The absence of an enforced forward-limit normalization for GPD $E^q$ leads to higher uncertainties in this case. Using current lattice inputs, we successfully constrain the GPDs, enabling the calculation of derived quantities, such as nucleon tomography and total angular momentum. {\color{red}It is noticed that, these results, including nucleon tomography, are primarily intended as a demonstration of the proposed unified framework. Since this work is fundamentally an exploratory study focusing on methodology, these results are framed as a proof-of-principle rather than definitive physics predictions.}

\clearpage

\section{Summary}
\label{sec:summary}

This work presents a unified neural-network framework that extracts light-cone PDFs and zero-skewness GPDs directly from lattice-QCD matrix elements. It fits two complementary inputs at the same time: quasi-distributions (in the LaMET approach) and pseudo-distributions (in the SDE approach). The framework reconstructs the target light-cone distributions and then applies a Fourier transform and the LaMET matching. A genetic algorithm trains the parameters and enforces agreement between the two approaches on the same object, the PDF $q(x)$ or the GPD $q(x,t)$.

To handle the inverse problem from discrete and truncated Euclidean data, we use flexible feed-forward network with power-law factors that reflect behavior at small- and large-$x$. Auxiliary components represent power corrections at selected $x$ regions. Tests on mock data show that the neural-network reconstruction, unlike the commonly used Backus-Gilbert method, reproduces the known input distribution (the GK model). When applied to lattice data, the joint LaMET+SDE fit outperforms either input alone: LaMET constrains the mid-$x$ region, SDE adds short-distance control, and together they yield stable valence PDFs with quantified uncertainties. Sea-quark features remain limited by current lattice systematics. Robustness studies with respect to analysis choices (e.g., $x_{\min}, x_{\max}$) demonstrate that the unified LaMET+SDE reconstruction can consistently accommodate future, higher-precision inputs. 

We extend the framework to GPDs at zero skewness by introducing the $t$ dependence at the neuron level through an exponential term. The neural-network reconstruction recovers $H^q(x,0,t)$ and $E^q(x,0,t)$ from the GK model and, with present lattice data, yields derived observables, elastic form factors, nucleon tomographic picture, as well as estimates of the quark total angular momentum via Ji's sum rule. Overall, the neural-network reconstruction informed by both LaMET and SDE offers a practical route from lattice-QCD correlators to nucleon imaging and can incorporate future experimental inputs.

{\color{red}There are several directions for future research within our framework.
An important thread is to quantify all sources of systematic uncertainties.
These include lattice-specific effects related to the finite lattice spacing and volume, possible contamination by excited states or effects of the non-physical pion mass.
It is also important to obtain data corresponding to larger nucleon boosts, leading to a better control over higher-twist effects.
Moreover, effects stemming from renormalization and matching need to be assessed, by adopting improved renormalization and matching schemes with resummations to account for possible perturbative contaminations.
Our present study has explored a possible ANN framework, but improvements and extensions thereof are, obviously, possible.
We aim to test modifications of the framework, such as different neural-network setups and optimization algorithms.
The foreseen extensions include nonzero skewness, additional channels (e.g., polarized ones or at higher twist) and incorporation of experimental constraints and data (e.g., from DVCS).
Consequently, building on the advantages of the framework that are already clear from the present study, it is highly promising to employ the neural-network reconstruction for a comprehensive study of GPDs, including the $\xi$ dependence.} Overall, it can provide crucial information for a better understanding of the nucleon's structure in synergy with planned experiments, such as the Electron-Ion Collider.

\acknowledgments
M.-H.~C. and K.~C. are supported by the National Science Centre (Poland) grant OPUS No.\ 2021/43/B/ST2/00497.
M.~C. acknowledges financial support by the U.S. Department of Energy, Office of Nuclear Physics,  under Grant No.\ DE-SC0025218.
P.~S. and J.~W.
were supported by the grant No.~2024/53/B/ST2/00968
of the National Science Centre, Poland.

We gratefully acknowledge Polish high-performance computing infrastructure PLGrid (HPC Center: ACK Cyfronet AGH) for providing computer facilities and support within computational grant No.\ PLG/2025/018071.
This research was carried out with the support of the Interdisciplinary Centre for Mathematical and Computational Modelling at the University of Warsaw (ICM UW).
Computations were also partially performed at the Poznan Supercomputing and Networking Center (Eagle/Altair supercomputer), using the computers of Centre of Informatics Tricity Academic Supercomputer Network (Tryton and Tryton Plus supercomputers) and facilities of the USQCD Collaboration, funded by the Office of Science of the U.S. Department of Energy. 
This research used resources of the National Energy Research Scientific Computing Center, a DOE Office of Science User Facility supported by the Office of Science of the U.S. Department of Energy under Contract No. DE-AC02-05CH11231 using NERSC awards NP-ERCAP0022961 and NP-ERCAP0027642.
This research used resources of the Oak Ridge Leadership Computing Facility, which is a DOE Office of Science User Facility supported under Contract DE-AC05-00OR22725.
The gauge configurations have been generated by the Extended Twisted Mass Collaboration on the KNL (A2) Partition of Marconi at CINECA, through the Prace project Pra13\_3304 ``SIMPHYS".
Inversions were performed using the DD-$\alpha$AMG solver~\cite{Frommer:2013fsa} with twisted mass support~\cite{Alexandrou:2016izb}. 

\section*{Appendix}
\appendix
\section{Interpretation of nucleon tomography}
\label{app:tomography}

As stated in Sec.~\ref{sec:theory}, nucleon tomography allows for a determination of the position of the active parton in the plane perpendicular to the nucleon's motion. The position is given by the impact parameter, ${\bf b_\perp} = (b_x, b_y)$, which is defined in the coordinate system whose origin is set by the center of momentum of all partons~\cite{Burkardt:2002hr}, such that 
\begin{equation}
\underbrace{\vphantom{\sum}x\, {\bf b_\perp}}_{\substack{\mathrm{active} \\ \mathrm{parton}}}
+ \underbrace{\sum x'\, {\bf b'_\perp}}_{\substack{\mathrm{spectator} \\ \mathrm{partons}}}
= {\bf 0} \,,
\label{eq:nt:momentum_center}
\end{equation}
where $x'$ and ${\bf b'_\perp}$ are the momentum fractions and transverse positions of the spectator partons, respectively. 

If one of the partons carries most of the nucleon's momentum, the origin of the coordinate system under discussion shifts toward that parton, such that in the limiting case one has 
\begin{equation}
q(x=1, {\bf b_\perp}) = \delta({\bf b_\perp})\,.
\label{eq:nt:xTo1_q}
\end{equation}
Therefore, we will  observe the nucleon tomography profile to converge to a point as $x \to 1$. This should not be interpreted as a ``shrinkage'' of the nucleon’s size. Instead, it reflects a kinematic effect originating from the choice of the coordinate system.

\begin{figure*}[!ht]
\centering
\includegraphics[width=0.8\textwidth]{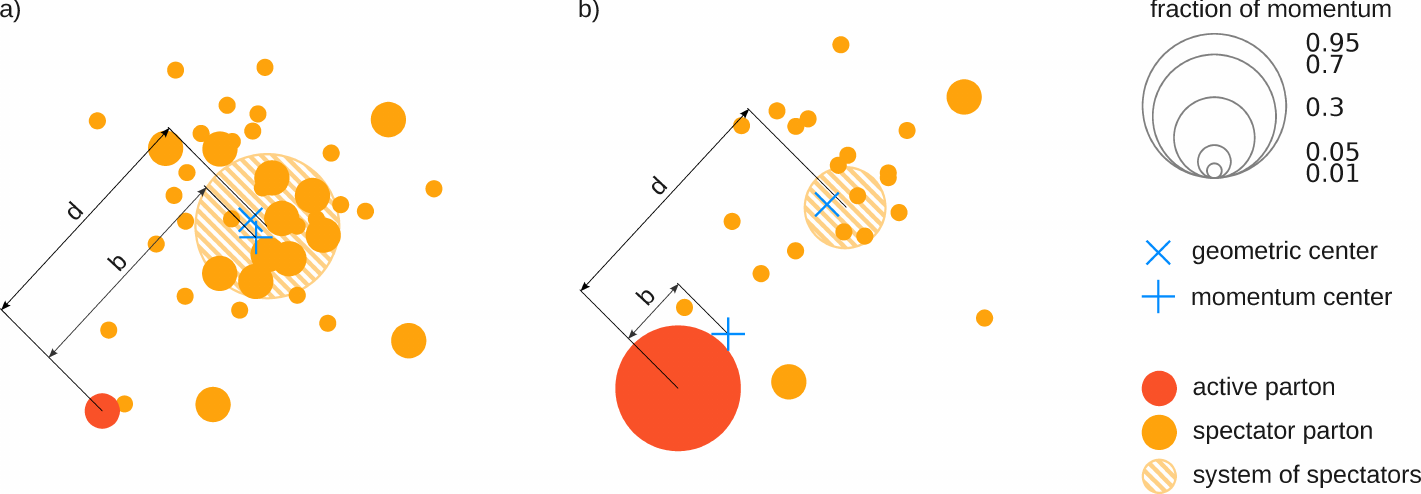}
\caption{Cartoon illustrating the consequences of using the coordinate system in which nucleon tomography is defined. (a) Momenta of all partons are small: the geometric and momentum centers are close to each other. (b) Momentum of one of the partons is large: the momentum center is shifted toward this parton. More details are provided in the text.}
\label{fig:nt_scheme}
\end{figure*}

Let us now define a collective system of spectators. It is characterized by the momentum fraction $1 - x$ and is located at the position $-x\,{\bf b_\perp}/(1 - x)$.  The distance between the active parton and the collective system of spectators is then
\begin{equation}
    |{\bf d_\perp}| = \frac{\bf |b_\perp|}{1 - x} \,.
    \label{eq:nt:d}
\end{equation}
Because of the coordinate system used, studying nucleon tomography as a function of $|{\bf d_\perp}|$, rather than $|{\bf b_\perp}|$, is more meaningful when $x \to 1$. We illustrate this with the help of Fig.~\ref{fig:nt_scheme}, which shows a schematic cartoon of two scenarios for the spatial and momentum distributions of partons. In Fig.~\ref{fig:nt_scheme}a), the parton momenta are small, as are the individual contributions to Eq.~\eqref{eq:nt:momentum_center}, placing the center of momentum close to the geometric center. The later is defined such that
\begin{equation}
\underbrace{\vphantom{\sum}{\bf b_\perp}}_{\substack{\mathrm{active} \\ \mathrm{parton}}}
+ \underbrace{\sum {\bf b'_\perp}}_{\substack{\mathrm{spectator} \\ \mathrm{partons}}}
= {\bf 0} \,.
\label{eq:nt:geometric_center}
\end{equation}
In this case, the interpretation of ${\bf b_\perp}$ as the position of the active parton is straightforward. In Fig.~\ref{fig:nt_scheme}b), one of the partons carries a large fraction of the nucleon’s momentum (here, $x = 0.7$), and as a result, the center of momentum is shifted toward this parton. The center of the spectator system, however, remains close to the geometric center. In such situation, the information carried by $|{\bf d_\perp}|$ tends to be more physically meaningful.

The consequences of using the coordinate system in which nucleon tomography is defined must be taken into account when modeling GPDs. Namely, to respect Eq.~\eqref{eq:nt:xTo1_q}, the dependence of the GPDs $H^q(x, 0, t)$ on the variable $t$ must ``vanish'' when $x \to 1$,
\begin{equation}
\lim_{x \to 1} \frac{d}{dt} H(x, 0, t) = 0 \,.
\label{eq:nt:xTo1_H}
\end{equation}
To guarantee that $|{\bf d_\perp}|$ in Eq.~\eqref{eq:nt:d} remains finite in the limit $x \to 1$, we impose a condition on the average relative distance squared between the active parton and the spectator system~\cite{Diehl:2004cx},
\begin{equation}
\langle d^2 \rangle_x^q = \frac{\langle b^2 \rangle_x^q}{(1 - x)^2} \,.
\label{eq:nt:ave_d}
\end{equation}
Here,
\begin{equation}
\langle b^2 \rangle_x^q = \frac{\int d^2 {\bf b_\perp} \, {\bf b}_\perp^2 q(x, {\bf b_\perp})}{\int d^2 {\bf b_\perp} \, q(x, {\bf b_\perp})} 
\label{eq:nt:ave_b}
\end{equation}
describes the average width of the parton distribution at a given~$x$.

Let us also comment on the interpretation of the distortion of parton densities in a transversely polarized nucleon. As indicated in Sec.~\ref{sec:theory}, when the nucleon is polarized along the $X$-axis, the distortion is visible in the perpendicular direction. It can be quantified as follows~\cite{Burkardt:2002hr}:
\begin{align}
\Delta_{X}^q = \int_{-1}^1 dx\int d^2{\bf b_\perp}\,
b_y x\, q_X(x,{\bf b_\perp}) = \frac{1}{2M} \int_{-1}^1 dx\, x E_q(x,0,0)\,.
\end{align}
This quantity vanishes when summed over all partons (including gluons),
\begin{align}
\Delta_{X}^g + \sum_q \Delta_{X}^q = 0 \,.
\end{align}
Such vanishing can be associated with the behavior of anomalous gravitomagnetic moments in composite systems such as the nucleon~\cite{Brodsky:2000ii}, but is also evident when combining Ji’s sum rule~\eqref{eq:basics_ji} with the momentum sum rule expressed by
\begin{align}
   1 & = \int_{0}^1 dx H^{g}(x,0,0) + \sum_q \int_{-1}^1 dx\,x H^{q}(x,0,0)  \nonumber \\
     & = \int_{0}^1 dx xg(x) + \sum_q \int_{0}^1 dx x \left(q_{\rm val}(x) + 2q_{\rm sea}(x)\right) \,.
\end{align}

\section{More details on the lattice framework}
\label{app:lattice}

In this appendix, we include some details of our lattice setup. For a more thorough account of this setup, we refer to \cite{Alexandrou:2019lfo} (for PDFs) and \cite{Alexandrou:2020zbe,Bhattacharya:2022aob} (for GPDs).

We extract proton matrix elements from Euclidean correlators, which are defined from two- and three-point functions. For a proton carrying momentum $\mathbf{p}$, the two-point function is
\begin{equation}
C_{\text{2pt}}(\Gamma_0;\mathbf{p};t)=\sum_{\mathbf{x}}\,
\langle J(t,\mathbf{x})\,J^\dagger(0,\mathbf{0})\rangle\,e^{-i\mathbf{p}\cdot\mathbf{x}}\,,
\end{equation}
where $J$ is the proton interpolating operator and $\Gamma_0$ is the unpolarized parity projector.
The three-point function for a nonlocal bilinear operator ${\cal O}_\mu$ with a straight Wilson line of length $z$ and the $\gamma_\mu$ Dirac matrix ($\mu=0$ for PDFs and $\mu=0,1,2$ for GPDs), and the parity projector $\Gamma_\kappa$ ($\kappa=0$ -- unpolarized, $\kappa\neq0$ -- polarized in the $\kappa$-direction) is
\begin{equation}
C_{\text{3pt},\mu}(\Gamma_\kappa; \mathbf{p}',\mathbf{p};z, t_s,\tau) = \sum_{\mathbf{x}_s,\mathbf{x}} \langle 0 | J(t_s,\mathbf{x}_s){\cal O}_\mu(\tau,\mathbf{x})J^\dagger(0,\mathbf{0}) |  0 \rangle  e^{-i\mathbf{p}'\cdot\mathbf{x}_s}\,e^{i(\mathbf{p'-p})\cdot\mathbf{x}}\,.
\end{equation}
We note that the source position is chosen to be $(0,\mathbf{0})$, and thus, the insertion and sink times are denoted by $\tau$ and $t_s$. The momentum transfer between the initial and final states, $\mathbf{p}'-\mathbf{p}$, is such that it is assigned to the initial state of the proton; this is indicated as an asymmetric frame~\cite{Bhattacharya:2022aob}, which is computationally efficient, as the momentum of the final state is fixed. The calculation of $C_{\text{3pt}}$ is done with the fixed-sink sequential method \cite{Martinelli:1988rr}. Using the two- and three-point correlation functions, one forms the optimized ratio
\begin{align}
R_\mu(\Gamma_\kappa;\mathbf{p}',\mathbf{p};z,t_s,\tau)&=
\frac{C_{\text{3pt},\mu}(\Gamma_\kappa;\mathbf{p}',\mathbf{p};z,t_s,\tau)}{C_{\text{2pt}}(\Gamma_0;\mathbf{p}';t_s)} \times \nonumber\\
&\sqrt{\frac{
C_{\text{2pt}}(\Gamma_0;\mathbf{p};t_s{-}\tau)\,C_{\text{2pt}}(\Gamma_0;\mathbf{p}';\tau)\,C_{\text{2pt}}(\Gamma_0;\mathbf{p}';t_s)
}{
C_{\text{2pt}}(\Gamma_0;\mathbf{p}';t_s{-}\tau)\,C_{\text{2pt}}(\Gamma_0;\mathbf{p};\tau)\,C_{\text{2pt}}(\Gamma_0;\mathbf{p};t_s)
}}\,,
\end{align}
which cancels overlap factors and time dependence and allows a plateau extraction of the ground state in $\tau$ away from contact regions. Kinematically equivalent momenta with identical $p^2$ or $(p')^2$ are averaged to improve statistics. 
The correlators are calculated on an ensemble of $N_f=2{+}1{+}1$ twisted-mass fermions \cite{Frezzotti:2000nk,Frezzotti:2003ni} with a clover term \cite{Sheikholeslami:1985ij}, Iwasaki gauge action \cite{Iwasaki:1983iya}, $a\simeq0.093$\,fm, $32^3\times64$, $m_\pi\simeq260$\,MeV \cite{Alexandrou:2018egz}. The time separation between the initial and final state is chosen to be $t_s=10a\approx0.93$ fm, which provides suppression of excited states effects below statistical precision \cite{Alexandrou:2019lfo}.

To improve the signal quality, we apply three techniques: momentum smearing~\cite{Bali:2016lva} on quark fields, APE smearing~\cite{APE:1987ehd} on the gauge field and stout smearing~\cite{Morningstar:2003gk} on the Wilson line. The momentum smearing modifies the standard Gaussian smearing~\cite{Alexandrou:1992ti} and is applied to the source and sink proton states to enhance overlap with boosted ground states. We use 50 iterations of the Gaussian smearing with $\alpha_G=4$ and the momentum smearing parameter $\zeta=-0.6$ was optimized to achieve the best error reduction. We also use 50 iterations of APE smearing with $\alpha_{\rm APE}=0.5$. The stout smearing is applied to the gauge links of the nonlocal operator, which results in suppression of the statistical noise. In particular, we use five steps of stout smearing with parameter $\rho=0.15$. While stout modifies both the bare matrix elements and renormalization factors, the renormalized matrix elements remain independent of the number of stout steps. For more details on these techniques and studies of the impact of their parameters, we refer to the original papers and studies in the context of quasi-distributions, e.g., \cite{Alexandrou:2019lfo}.

The bare operators under study contain logarithmic divergences, as well as divergences due to the Wilson line \cite{Ishikawa:2017faj,Ji:2017oey}. Therefore, a renormalization scheme is necessary to eliminate them. Depending on the methodology used (LaMET or SDE), the renormalization differs. In particular, for LaMET we use the RI$'$ scheme adapted to non-local operators~\cite{Alexandrou:2017huk,Alexandrou:2019lfo}. This factor is determined nonperturbatively using the momentum-source method~\cite{Gockeler:1998ye,Alexandrou:2015sea} by matching the amputated lattice vertex to its tree value at a reference momentum $p^2=\bar\mu_0^2$, with the quark-field renormalization fixed from the propagator. Results are obtained at several pion masses and chirally extrapolated to remove residual mass dependence. We then convert results to the $\overline{\text{MS}}$ scheme and evolve to a scale $\mu$ of 2~GeV, controlling discretization via a linear $(ap)^2\!\to\!0$ extrapolation within a safe momentum window. More information on how to optimize the values of $(ap)^2$ can be found in Refs.~\cite{Alexandrou:2015sea,Alexandrou:2017huk,Alexandrou:2019lfo}. The final light-cone quantities are, thus, quoted in the standard $\overline{\text{MS}}$ scheme at 2~GeV after application of the matching kernel defined within LaMET, see Appendix \ref{app:matching}.

In the SDE approach, the proton matrix elements are analyzed as functions of the Ioffe time $\nu=z\cdot P$. The UV divergences and some higher-twist effects are suppressed by the double ratio \cite{Orginos:2017kos}, which for the PDF reads
\begin{equation}
\bar{F}(z,\nu)=\frac{F(z,\nu)/F(0,\nu)}{F(z,0)/F(0,0)}\,,
\end{equation}
where the flavor index is suppressed.
In the case of GPDs, a similar double ratio is constructed, with the PDF ME $F(z,\nu)$ replaced by the GPD one $F_H(z,\nu)$ or $F_E(z,\nu)$ (the other ingredients of the ratio remain as PDF MEs).
The procedure is applicable for multiplicatively renormalizable operators such as $\gamma_0$ considered in this work. After constructing $\bar{F}$ over multiple $(z,\nu)$ combinations, coordinate-space matching can be performed, see Appendix \ref{app:matching}, resulting in final distributions in the $\overline{\text{MS}}$ scheme at 2~GeV. 

As mentioned previously, the calculation is performed in the asymmetric kinematic frame \cite{Bhattacharya:2022aob}, where the final momentum is fixed to $\vec{p}'=(0,0,P_z)$, while the initial momentum carries the momentum transfer. The details of the kinematic setup are shown in Table~\ref{tab:stat}.

\begin{table}[t!]
\caption{\small Details of our lattice setup: invariant four-momentum transfer $-t$, nucleon boost $P_z$ and the total number of measurements. The LaMET part of our framework uses only the largest boost, $P_z=1.67$ GeV, while the SDE part employs all boosts.}
\begin{center}
\renewcommand{\arraystretch}{1.0}
\begin{tabular}{
P{0.2\columnwidth}
P{0.2\columnwidth}
P{0.2\columnwidth}
}
\toprule
$-t$\,[GeV$^2$] & $P_z$\,[GeV] & $N_{\rm meas}$ \\
\midrule
0  & $\pm$0.83   &1600\\
0  & $\pm$1.25   &8608\\
0  & $\pm$1.67   &32384\\
\midrule[0px]
0.17  & $\pm$0.83  &6400\\
0.17  & $\pm$1.25  &17216\\
0.17  & $\pm$1.67  &129536\\
\midrule[0px]
0.34 & $\pm$0.83  &6400 \\
0.34 & $\pm$1.25 &12480 \\
0.34 & $\pm$1.67 &129536\\
\midrule[0px]
0.65  & $\pm$0.83 &6400\\
0.65  & $\pm$1.25 &17216\\
0.65  & $\pm$1.67 &129536\\
\midrule[0px]
0.81 & $\pm$0.83 &12800 \\
0.81 & $\pm$1.25 &24960 \\
0.81 & $\pm$1.67 &259072 \\
\midrule[0px]
1.24 & $\pm$0.83 &6400 \\
1.24 & $\pm$1.25 &12480 \\
1.24 & $\pm$1.67 &129536\\
\midrule[0px]
1.38  & $\pm$0.83 &6400\\
1.38  & $\pm$1.25 &17216\\
1.38  & $\pm$1.67 &129536\\
\midrule[0px]
1.52 & $\pm$0.83 &12800 \\
1.52 & $\pm$1.25 &24960 \\
1.52 & $\pm$1.67 &259072 \\
\midrule[0px]
2.29  & $\pm$0.83 &6400\\
2.29  & $\pm$1.25 &17216\\
2.29  & $\pm$1.67 &129536\\
\bottomrule
\end{tabular}
\label{tab:stat}
\end{center}
\end{table}

\section{Matching between Euclidean and Minkowski distributions}
\label{app:matching}

For completeness, we detail our implementation of the matching equations that relate light-cone distributions to lattice inputs. We present these details while deliberately avoiding the explicit use of the plus-prescription for clarity. The matching equations given below are written for unpolarized PDFs and $\xi=0$ GPDs. In the latter case, the variable $t$ enters only as a parameter of the input and output distributions.

In our procedure, the light-cone distribution is parametrized by the neural network.
Hence, for the LaMET approach, we need the inverse matching that translates light-cone distributions in the $\overline{\text{MS}}$ scheme to quasi-PDFs/GPDs corresponding to a finite momentum $P_z$ and renormalized in the RI$'$ scheme. We use the procedure derived in~\cite{LatticeParton:2018gjr},
\begin{align}
\tilde{q}(x,P_z,p_z^R,\mu_R)=\int_{-1}^{1}\frac{dy}{|y|}C\left(\frac{x}{y},r,\frac{yP_z}{\mu},\frac{yP_z}{p_z^R}\right)q(y,\mu)+\mathcal{O}\left(\frac{M^2}{P_z^2},\frac{\Lambda^2_{\rm{QCD}}}{x^2P_z^2},\frac{\Lambda^2_{\rm{QCD}}}{(1-x)^2P_z^2}\right) \,.
\label{eq:match}
\end{align}
Here, $q(y,\mu)$ denotes the light-cone distribution that we aim to reconstruct from LaMET quantities, with the renormalization scale $\mu$ indicated explicitly, while $\tilde{q}(x,P_z,p_z^R,\mu_R)$ represents the inverse-matched quasi-distribution. The scales $p_z^R$ and $\mu_R$ denote the parton momentum and the renormalization scale specific to the RI$'$ scheme, respectively. Both quantities serve as parameters of that scheme. The function $C$ is the perturbatively computed matching coefficient, where $r=\mu_R^2/(p_z^R)^2$. The power corrections $\mathcal{O}(M^2/P_z^2,\Lambda^2_{\rm{QCD}}/x^2P_z^2)$, $\mathcal{O}(\Lambda^2_{\rm{QCD}}/(1-x)^2P_z^2)$ arise from the nucleon mass and higher-twist contributions, and are suppressed by the large hadron momentum $P_z$. {\color{red}In addition, the matching for GPDs at zero skewness involves another source of power corrections arising from the momentum transfer, denoted as $\mathcal{O}(|t|/P_z^2)$. In our kinematic setup, the largest value $|t|=2.3\;\rm{GeV}^2$ yields corrections comparable in magnitude to those at $x_{\rm min}$ and $x_{\rm max}$. However, for the simplicity of ANN framework, these $t$-dependent power corrections are not explicitly considered in the current study.}

For brevity, in the subsequent expressions we omit the dependence on $P_z$, $p_z^R$, $\mu_R$ and $\mu$ in the distributions both before and after the matching. We adopt the one-loop perturbative matching coefficient, which consists of a bare term $\tilde{q}^{(11)}(x)$ and an additional RI$'$ counterterm $\tilde{q}^{(12)}(x)$. Thus, Eq.~\eqref{eq:match} becomes:
\begin{align}
\tilde{q}(x)=q(x)+\frac{\alpha_sC_F}{2\pi}\left[\tilde{q}^{(11)}(x)+\tilde{q}^{(12)}(x)\right].
\end{align}
We use the one-loop $\overline{\rm MS}$ value of $\alpha_s$ at $\mu=2$ GeV. The original forms of $\tilde{q}^{(11)}(x)$ and $\tilde{q}^{(12)}(x)$ are provided elsewhere~\cite{LatticeParton:2018gjr}. Here, we present these functions only after performing the substitution of variables $\zeta = x/y$,
\begin{align}
\tilde{q}^{(11)}(x)&=\left[\int_{-\infty}^{-|x|}+\int_{|x|}^{\infty}\right]d\zeta\left[f_{1,\rm{ps}}\left(\zeta,\frac{xP_z}{\zeta\mu}\right)\frac{q\left(\displaystyle\frac{x}{\zeta}\right)}{|\zeta|}\right.\left.-f_{1,\rm{ps}}\left(\zeta,\frac{xP_z}{\mu}\right)q(x)\right] \,,\label{eq:matching:q11}\\
\tilde{q}^{(12)}(x)&=\left[\int_{-\infty}^{-|x|}+\int_{|x|}^{\infty}\right]d\zeta\frac{|x|P^z}{\zeta^2p_z^R}f_{2,\rm{ps}}\left(1+\frac{xP_z}{\zeta p_z^R}(\zeta-1),r\right)q\left(\frac{x}{\zeta}\right)-\nonumber\\
&\left[\int_{-\infty}^{-|x|}+\int_{|x|}^{\infty}\right]d\zeta\frac{|x|P^z}{p_z^R}f_{2,\rm{ps}}\left(1+\frac{xP_z}{p_z^R}(\zeta-1),r\right)q(x) \,.\label{eq:matching:q12}
\end{align}
The explicit expressions for $f_{1,\mathrm{ps}}$ and $f_{2,\mathrm{ps}}$ are:
\begin{align}
f_{1,\mathrm{ps}}\left(x,\lambda\right)=\frac{\alpha_s C_F}{2\pi}
\begin{cases}
\displaystyle\frac{1+x^2}{1-x}\ln\frac{x}{x-1}+1\,, & x>1\,, \\
\displaystyle\frac{1+x^2}{1-x}\ln\left(4x(1-x)\lambda^2\right)-\frac{x(1+x)}{1-x}\,, & 0<x<1\,, \\
\displaystyle -\frac{1+x^2}{1-x}\ln\frac{x}{x-1}-1\,, & \mathrm{otherwise}\,,
\end{cases}
\end{align}
\begin{align}
&f_{2,\mathrm{ps}}(x,r)=\frac{\alpha_s C_F}{2\pi}\times\nonumber\\
&\begin{cases}
\displaystyle\frac{3-3r-2x}{2(r-1)(x-1)}+\frac{4rx-8x^2+8x^3}{(r-4x+4x^2)^2}+\frac{2-2r-rx+2x^2}{(r-1)^{3/2}(x-1)}\tan^{-1}\frac{\sqrt{r-1}}{2x-1}\,, & x>1\,, \\
\displaystyle\frac{3-3r-2x+4x^2}{2(r-1)(1-x)}+\frac{-2+2r+rx-2x^2}{(r-1)^{3/2}(1-x)}\tan^{-1}\sqrt{r-1}\,, & 0<x<1\,, \\
\displaystyle-\frac{3-3r-2x}{2(r-1)(x-1)}-\frac{4rx-8x^2+8x^3}{(r-4x+4x^2)^2}-\frac{2-2r-rx+2x^2}{(r-1)^{3/2}(x-1)}\tan^{-1}\frac{\sqrt{r-1}}{2x-1}\,, & \mathrm{otherwise}\,.
\end{cases}
\end{align}

Since the integrals in Eqs.~\eqref{eq:matching:q11} and~\eqref{eq:matching:q12} extend over infinite ranges, we introduce a cutoff parameter $\zeta_c$ to numerically approximate the integration limits. An additional difficulty arises from the divergences at $\zeta = 1$, which are regulated by introducing a small parameter $\delta$ and separating the integral into two regions, namely $\int_a^b \to \int_a^{1-\delta}+\int_{1+\delta}^b$. Thus, the integration domains appearing in $\tilde{q}^{(11)}(x)$ and $\tilde{q}^{(12)}(x)$ are modified accordingly:
\begin{align}
\int_{-\infty}^{-|x|}+\int_{|x|}^{\infty}\to
\begin{cases}
\displaystyle\int_{-\zeta_c}^{-|x|}+\int_{1+\delta}^{\zeta_c}+\int_{|x|}^{1-\delta}\,, & |x|<1, \\
\displaystyle\int_{-\zeta_c}^{-|x|}+\int_{|x|}^{\zeta_c}\,, & \mathrm{otherwise} \,.
\end{cases}
\end{align}
We have verified numerically that our results exhibit negligible dependence on the regulators $\zeta_c$ and $\delta$, as long as they are sufficiently large and small, respectively.

Within the SDE approach, we perform the matching in Ioffe-time space for the pseudo–MEs prior to the ANN reconstruction. Using the one-loop matching coefficient~\cite{Radyushkin:2018cvn}, the matched ITD reads
\begin{align}
\mathcal{F}(\nu) = \bar{F}(z,\nu) - \frac{\alpha_s C_F}{2\pi}\int_0^1 \! du\, C_\nu(u,z^2\mu^2)\,\big[ \bar{F}(z,u\nu)-\bar{F}(z,\nu) \big]+\mathcal{O}\left(z^2\Lambda^2_{\rm{QCD}}\right),
\end{align}
where $\bar{F}(z,\nu)$ denotes the (unmatched) pseudo-ITD and $C_\nu(u,z^2\mu^2)$ is the one-loop kernel,
\begin{align}
C_\nu(u,z^2\mu^2) = \frac{1+u^2}{u-1}\,\ln\!\left(\frac{z^2\mu^2 e^{2\gamma_E+1}}{4}\right) + \frac{4\ln(1-u)}{u-1} - 2(u-1).
\end{align}
Same as in the LaMET-based matching, we take the one-loop $\overline{\rm MS}$ value of $\alpha_s$. In principle, the matched ITDs $\mathcal{F}(\nu)$ should be independent of $z$. In practice, however, performing the matching at different values of $z$ may introduce residual effects, which is caused by power corrections. Accordingly, the matched ITD $\mathcal{F}(\nu)$ serves as the input to our ANN reconstruction.

\section{Numerical framework}
\label{app:pdf:numerical_framework}

As described in Sec.~\ref{sec:pdf}, in the reconstruction we constrain the free parameters of \smash{$q^{(0,1,2,3)}(x)$} using lattice data. These are the powers \smash{$\delta^{(0,1,2,3)}$} and \smash{$\rho^{(0,1,2,3)}$}, the biases \smash{$b_{21}^{(0,1,2,3)}$} and \smash{$b_{1i}^{(0,1,2,3)}$}, and the weights \smash{$w_{2i}^{(0,1,2,3)}$} and \smash{$w_{1i}^{(0,1,2,3)}$}. In our demonstration of the method, we use five neurons in the hidden layers of \smash{$\mathrm{ANN}^{(0,1,2,3)}(x)$}, such that in Eq.~\eqref{eq:ann} we have \smash{$N^{(0,1,2,3)} = 5$}. For a given quark flavor, this results in a total of $4 \times (3 + 5 \times 3) = 72$ free parameters for the valence and sea contributions, each.

Such a large number of free parameters is a common feature of machine learning techniques, which distinguishes them from analyses utilizing traditional parametric Ans{\"a}tze. We note that this large number of free parameters cannot be constrained in a standard fit based on, e.g., the Minuit library~\cite{JAMES1975343}. Instead, a dedicated algorithm must be used. 

{\color{red}
In the case of artificial neural networks, various implementations of the backpropagation algorithm are typically employed, relying on either the analytical or numerical evaluation of the loss function gradient. In our numerical framework, however, backpropagation-based methods are not easily applied directly, as they would require, for instance, knowledge of how quasi-distributions vary with respect to the free parameters of the network -- something that is difficult to evaluate due to the complexity of RI/MOM matching (cf. Appendix~\ref{app:matching}). 
}

We therefore use an implementation of the genetic algorithm~\cite{10.5555/522098}. In the context of machine learning, it is a popular alternative to backpropagation. The genetic algorithm is particularly useful when standard minimization techniques fail, and it has the distinct advantage of being able to avoid local minima. Moreover, since it does not rely on minimizing the loss function along a path in the hyperspace of the network’s parameters, it is well-suited for manually enforcing certain constraints, such as positivity (see, e.g., Ref.~\cite{Dutrieux:2021wll}). For brevity, we do not describe the genetic algorithm in detail here, but instead refer the reader to Ref.~\cite{Moutarde:2019tqa} for more information{\color{red}, including specific details on its implementation}.

When using the genetic algorithm, we must specify allowed ranges for the constrained parameters. For the parameters of the neural networks, we use the following: {\color{red}$-10 < b_{21}^{(0,1,2,3)}$, $b_{1j}^{(0,1,2,3)},\, w_{2j}^{(0,1,2,3)},\, w_{1j}^{(0,1,2,3)} < 10$}. The choice of these ranges is not physically motivated and is made arbitrarily. They are wide enough to render the results insensitive to the choice, yet not so wide as to impair the reconstruction performance.
We note that if one wishes to enforce strict positivity of the PDFs, then with the softplus activation function we use, it is sufficient to require \smash{$b_{21}^{(0,1,2,3)},\, w_{2j}^{(0,1,2,3)} > 0$}.

The selection of ranges for the powers requires more careful consideration. For the primary light-cone PDFs, we use \smash{$-1 < \delta^{(0)} < 0$} for valence quarks and \smash{$-2 < \delta^{(0)} < -1$} for sea quarks. That is, we explicitly require the primary light-cone PDFs to diverge as $x \to 0$, with the power-law behavior motivated by both theoretical considerations from QCD and phenomenological observations. For the powers controlling the behavior as $x \to 1$, we use \smash{$1 < \rho^{(0)} < 10$} for both valence and sea quarks. For the auxiliary functions, we use \smash{$-2 < \delta^{(1)} < 10$}, \smash{$1 < \delta^{(2,3)} < 10$}, and \smash{$1 < \rho^{(1,2,3)} < 10$}. This means that all auxiliary functions vanish smoothly at their endpoints, except for $q^{(1)}(x)$, which may (but is not required to) diverge as $x \to 0$.

We constrain the free parameters separately for up and down quarks, as well as for the real and imaginary parts of the LaMET and SDE quantities. Specifically, we first use only the real part, which allows us to fix the free parameters associated with the valence contribution. In the second stage, we use the imaginary part to constrain the sea contribution. We found this two-stage method to be more stable and less susceptible to systematic impurities, which appear to affect the real and imaginary parts differently. In Appendix~\ref{app:pdf_results}, we discuss this issue in more detail and also present a comparison with a version of the reconstruction in which the real and imaginary parts are used simultaneously.

The reconstruction is performed separately for each replica obtained in the lattice-QCD analysis. The collected results are then used to evaluate the mean value and uncertainty of a specific quantity at a given point. The use of the replication method allows us to correctly propagate uncertainties from the original analysis, account for correlations between data points, and avoid possible double counting caused by evaluating LaMET and SDE quantities from the same bare matrix elements. Since, for a given replica, our goal is to exactly reproduce the data points related to that replica without using any information from other replicas, our goodness function is
\begin{equation}
    \chi^2 = \sum_i \left(v_i - v_i^{\mathrm{ref}}\right)^2 \,,
\end{equation}
where $v_i$ and $v_i^{\mathrm{ref}}$ represent a data point evaluated in the reconstruction procedure and the corresponding reference, respectively.

The numerical framework used for the reconstruction of GPDs at $\xi=0$ remains essentially the same as for PDFs. The only difference is the introduction of new parameters, namely, the slopes $\smash{s_{ij}^{(0,1,2,3)}}$. In the demonstration of our method, we again use five neurons in the hidden layers of \smash{$\mathrm{ANN}^{(0,1,2,3)}(x, t)$}. Additionally, we allow the slopes to vary between $0 < \smash{s_{ij}^{(0,1,2,3)}} < 10\,\mathrm{GeV}^{-2}$.

Note that with this selection of slope values, strict positivity of parton densities in impact-parameter space can be maintained for an unpolarized nucleon just by requiring positive PDFs. This condition is straightforward to implement, as described above. For a transversely polarized nucleon, the situation is more complex, as one must account for the interplay between the GPDs $H^q(x,0,t)$ and $E^q(x,0,t)$ (cf.~Eq.~\eqref{eq:nt:E}). Ensuring positivity in this case using only analytical considerations is challenging. Nevertheless, it can still be enforced numerically, as described in Ref.~\cite{Dutrieux:2021wll}. In our modeling approach, this task is particularly simple, as positivity can be, for instance, checked on a neuron-by-neuron basis. The problem then reduces to a set of relatively simple inequalities, since the relevant Fourier transforms and derivatives can be performed analytically.

\section{Further discussions of PDF results}
\label{app:pdf_results}

\begin{figure*}[p!]
\centering
\includegraphics[width=0.49\columnwidth]{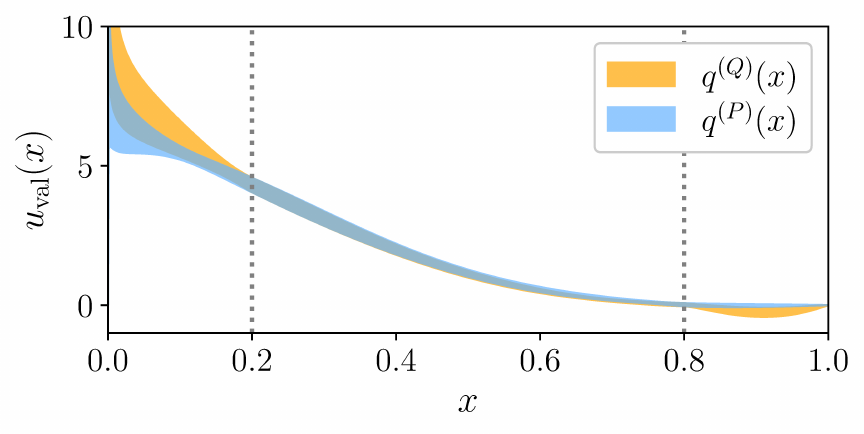}
\includegraphics[width=0.49\columnwidth]{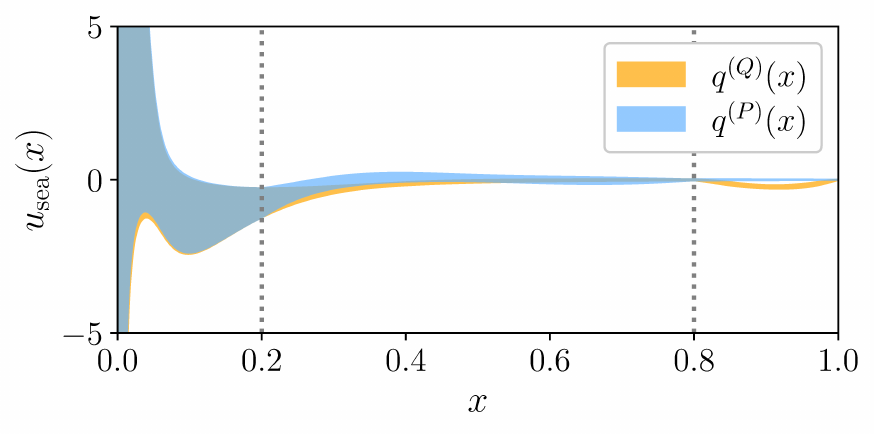}
\includegraphics[width=0.49\columnwidth]{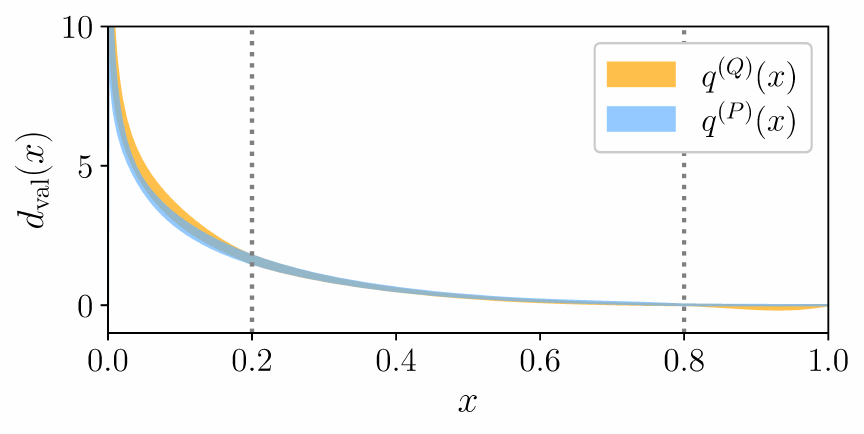}
\includegraphics[width=0.49\columnwidth]{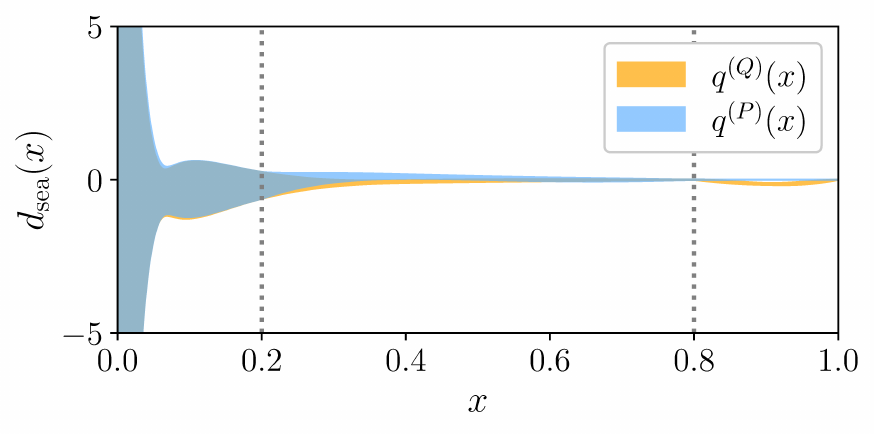}
\caption{Distributions $q^{(Q)}(x)$ and $q^{(P)}(x)$, fitted to LaMET and SDE quantities, respectively. The primary PDF $q^{(0)}(x)$ equals $q^{(Q)}(x)$ for $x_{\rm{min}} \leq x \leq x_{\rm{max}}$, and $q^{(P)}(x)$ for $0 < x < x_{\rm{min}}$ and $x_{\rm{max}} < x < 1$, cf.~Eqs.~\eqref{eq:qQ}~and~\eqref{eq:qP}.
}
\label{fig:q01234}
\end{figure*}
\begin{figure*}[p!]
\centering
\includegraphics[width=1\columnwidth]{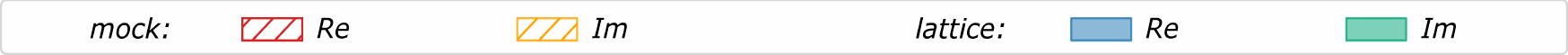} \\ 
\includegraphics[width=0.49\columnwidth]{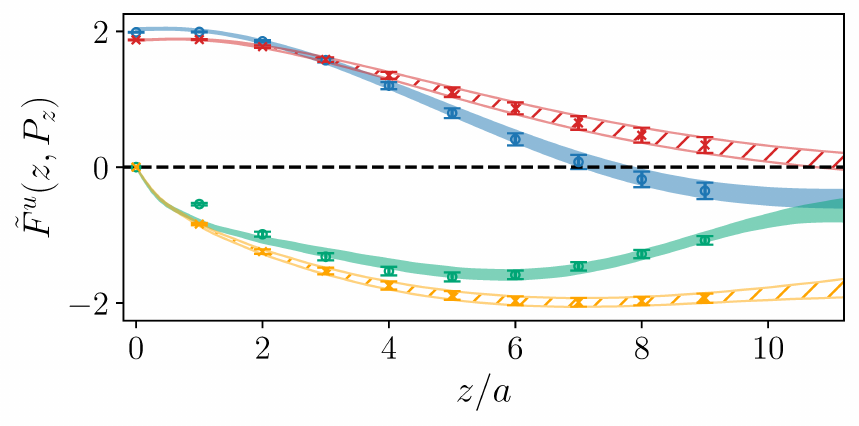}
\includegraphics[width=0.49\columnwidth]{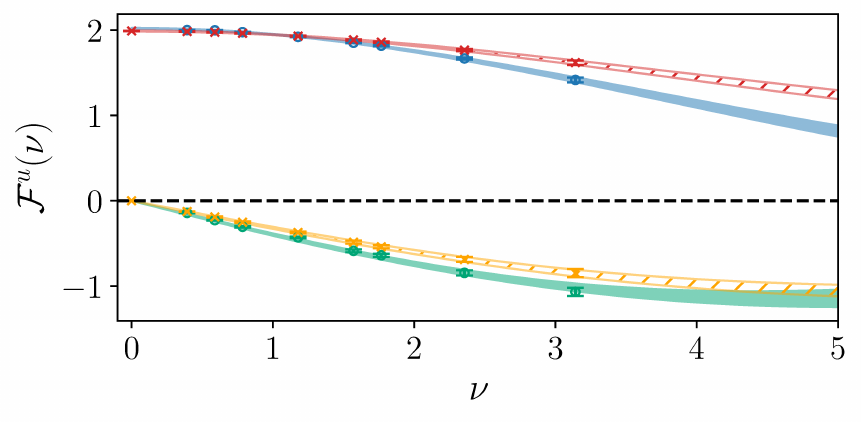}
\includegraphics[width=0.49\columnwidth]{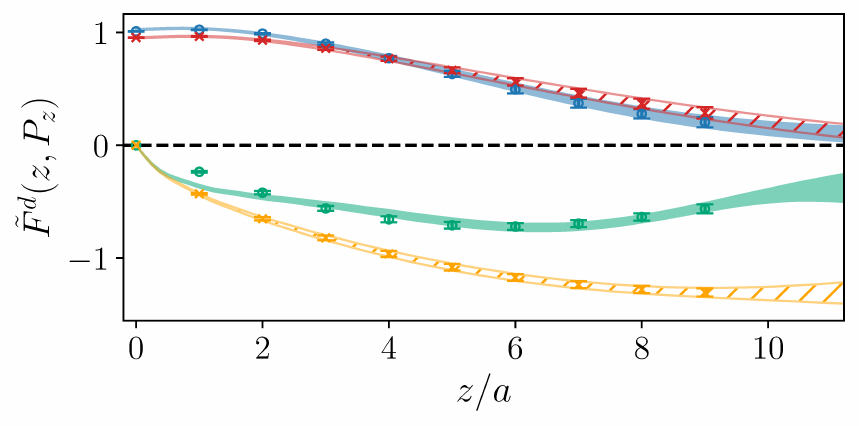}
\includegraphics[width=0.49\columnwidth]{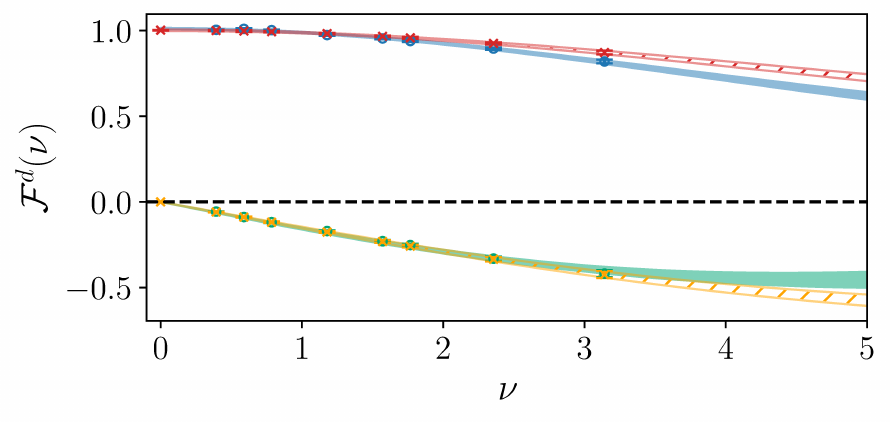}
\caption{Real and imaginary parts of $\widetilde F^q(z, P_z)$ evaluated within the LaMET framework (left) and $\mathcal{F}^q(\nu)$ evaluated within the SDE framework (right) for up (upper row) and down (lower row) quarks, shown for mock and lattice data. The bands represent the fitted quantities derived from $q^{(Q, P)}(x)$. The data points with uncertainties denote the input of mock or lattice data.}
\label{fig:fit}
\end{figure*}
\begin{figure*}
\centering
\includegraphics[width=0.49\columnwidth]{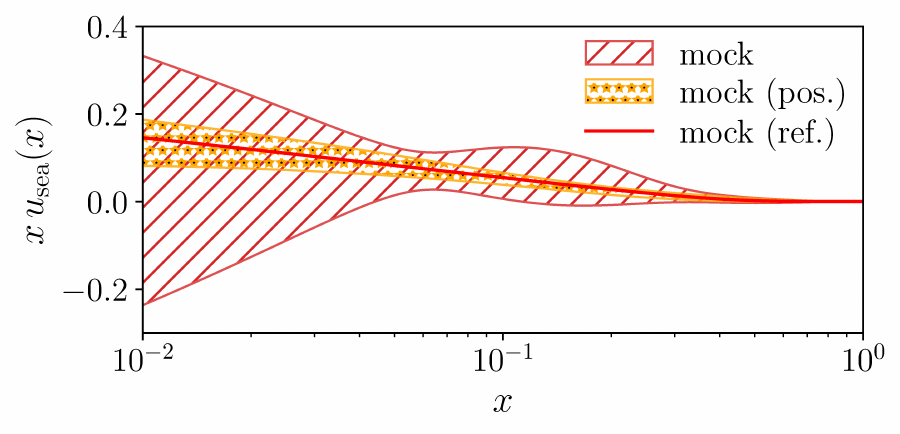}
\includegraphics[width=0.49\columnwidth]{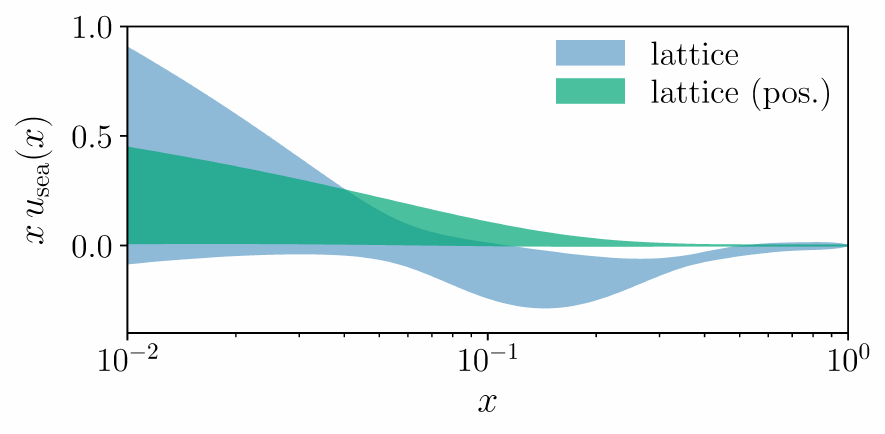}\\
\includegraphics[width=0.49\columnwidth]{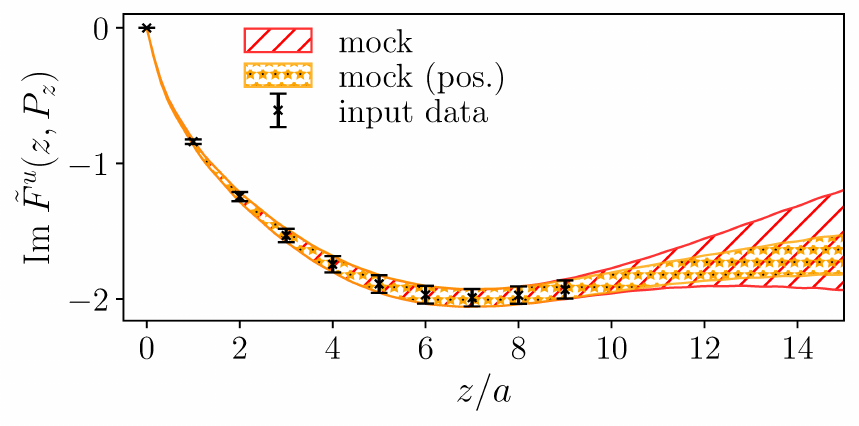}
\includegraphics[width=0.49\columnwidth]{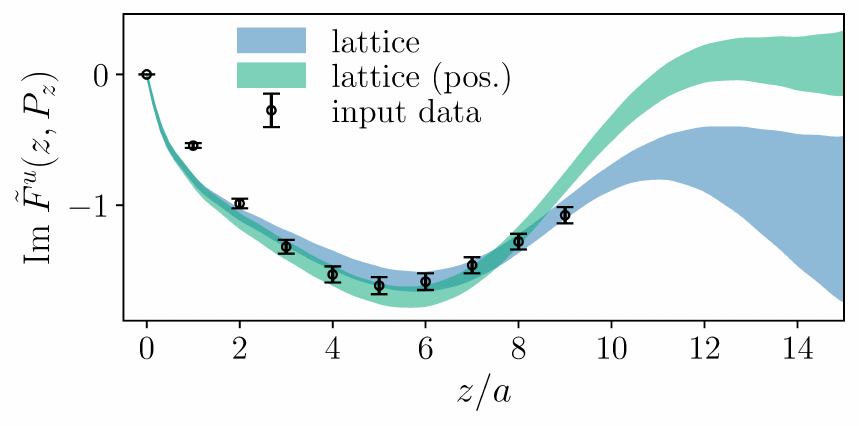}\\
\includegraphics[width=0.49\columnwidth]{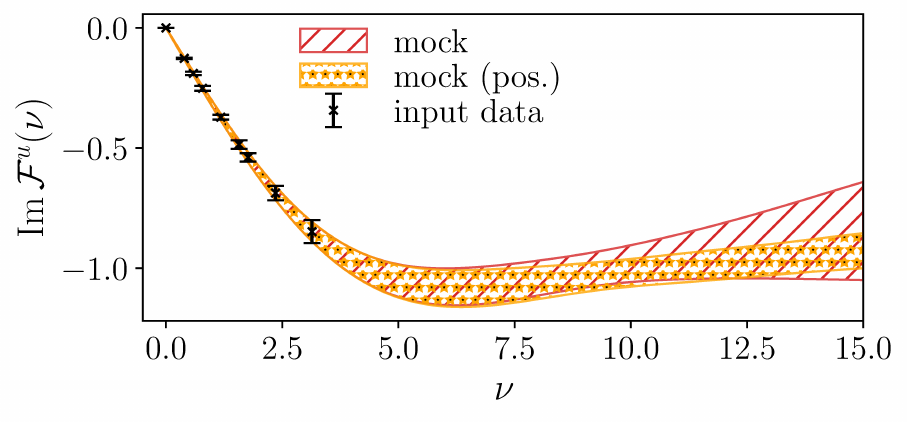}
\includegraphics[width=0.49\columnwidth]{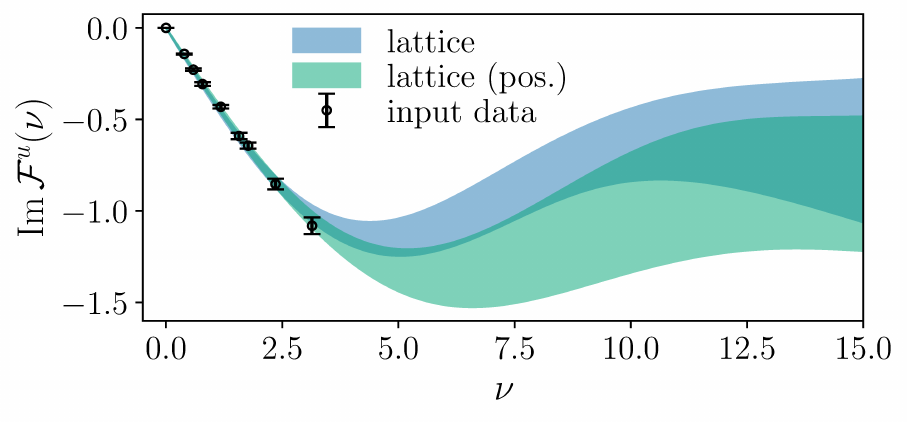}
\caption{The impact of enforcing strict positivity in the reconstruction of PDFs for up sea quarks from mock (left column) and lattice (right column) data. Results with positivity enforced are indicated by the ``pos.'' index, while the reference for mock data is denoted by ``ref.''. The first row shows the effect at the level of the reconstructed distributions, whereas the remaining two rows show the impact of enforcing positivity at the level of the fitted quantities, emphasizing changes in the unconstrained domains.
}
\label{fig:sea}
\end{figure*}
\begin{figure*}
\centering
\includegraphics[width=0.49\columnwidth]{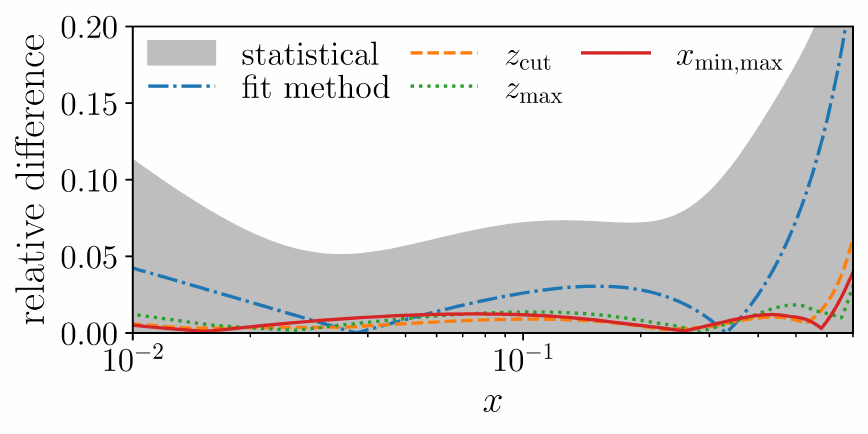}
\includegraphics[width=0.49\columnwidth]{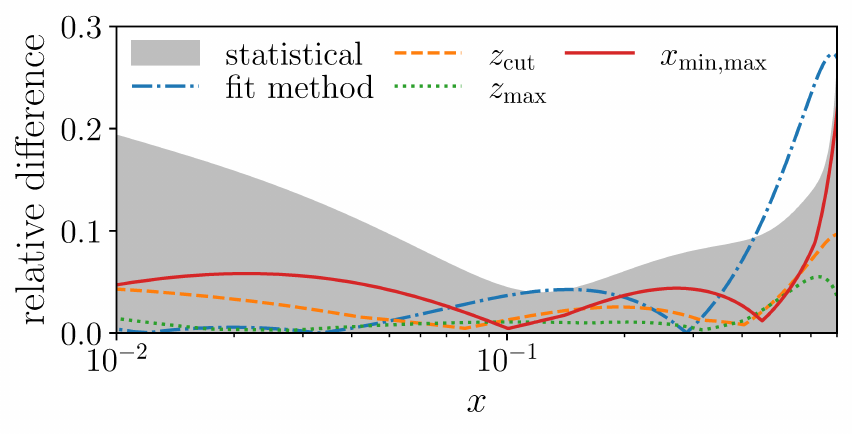}
\includegraphics[width=0.49\columnwidth]{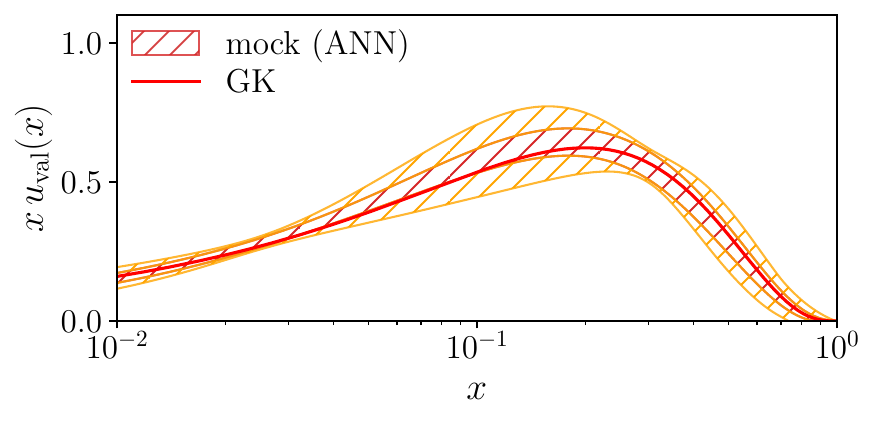}
\includegraphics[width=0.49\columnwidth]{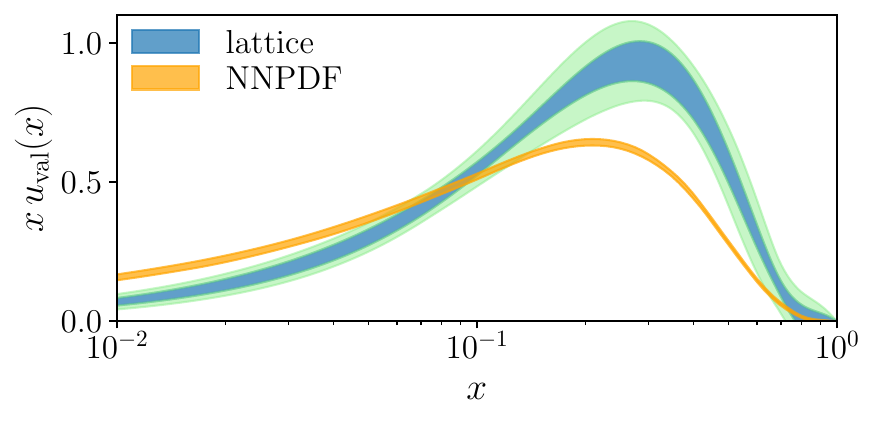}
\caption{Systematic studies related to the reconstruction of PDFs for up valence quark from mock (left panels) and lattice (right panels) results. In the upper panels, the gray bands represent the relative statistical uncertainties, while the curves denote relative differences (with respect to the default extraction) obtained in several systematic checks (see text for details). {\color{red}The lower panels show results including systematic uncertainties, where red (bottom-left) and blue (bottom-right) bands denote the statistical uncertainties, which are combined in quadrature with the systematic contributions shown as orange (bottom-left) and light green (bottom-right).}}
\label{fig:systematics}
\end{figure*}

This appendix presents further investigations of PDF results. 

Firstly, we complement the detailed information for the PDF reconstruction with Figs.~\ref{fig:q01234} and~\ref{fig:fit}. The former illustrates the functions directly used in the fits, namely, \smash{$q^{(Q, P)}(x)$}, and how they differ from \smash{$q^{(0)}(x)$} (cf. Eqs.~\eqref{eq:qQ} and~\eqref{eq:qP}). These functions are shown only for the lattice data, since in the case of mock data, the contributions of \smash{$q^{(1,2,3)}(x)$} are expected to be small (which we have confirmed). This is because the same equations are essentially used both to generate the mock data and to analyze them. Figure~\ref{fig:fit}, on the other hand, allows for a visual confirmation of the excellent fit quality. The reconstruction only struggles to reproduce the points at $z/a=1$ in the imaginary parts of the quantities evaluated on the lattice when using the LaMET framework. We confirm that this issue can be attributed to systematic effects present in the sea component of the lattice data used. This was verified using our approach, which enables a detailed study of such effects by first reconstructing the real part and subsequently subtracting the valence contribution.

Second, according to Figs.~\ref{fig:mock_result} and~\ref{fig:lat_result} in the main text, one can observe very large uncertainties for the sea quarks, caused by the limited ranges of $z$ and $\nu$ available. One may ask what happens if we require strict positivity of the sea quarks, which we implement as described in Appendix~\ref{app:pdf:numerical_framework}. The effect of using such a constraint is substantial, as we demonstrate with the help of Fig.~\ref{fig:sea} (for brevity, only for up quarks). For the mock data, we observe good agreement with the reference GK model. We also see that, after requiring strict positivity, the uncertainties shrink in the domains unconstrained by data, namely, at $z/a>9$ and $\nu > 3$. For the lattice-QCD data, these features are not as pronounced. This is because, as mentioned earlier, the sea contribution contains significant systematics. In addition, it is generally compatible with zero, to the extent that, after enforcing strict positivity, the fit for some replicas of the lattice-QCD data yields significantly worse values of $\chi^2$. Requiring strict positivity for the valence quarks has no practical significance. This is another practical consequence of enforcing the normalization, but it also shows that extracting valence PDFs from the real parts of LaMET and SDE quantities is generally easier than constraining the sea components from the imaginary parts.

Finally, we have performed a comprehensive systematic study addressing several sources of uncertainties. Specifically, the systematic effects we have investigated include:
\begin{itemize}
    \item \textit{Fitting methods}: While in the main analysis we separately fit the real and imaginary parts of the data, we have also performed a joint fit to quantify potential systematic differences.
    \item \textit{$z_{\rm cut}$ for the quasi matrix element}: $z_{\rm cut}$ represents the largest $z$ value in used data of $\widetilde{F}^q(z,P_z)$. The analysis displayed in the main text employs $z_{\rm cut}=9a$, whereas alternative choices: $z_{\rm cut}=8a$ and $10a$ were studied. The systematic uncertainty is estimated as the maximum deviation of two alternative choices from the central result.
    \item \textit{$z_{\rm max}$ for matched ITDs}: $z_{\rm max}$ denotes the maximum value of $z$ in used data of $\bar{F}^q(z,\nu)$. The results with $z_{\rm max}=4a$ are represented in the main text and here, we vary it to $3a$ and $5a$. The maximum deviation of these two choices from the central choice is taken as the systematic uncertainty.
    \item \textit{Sensitivity to $x_{\rm min}$ and $x_{\rm max}$}: $x_{\rm min}$ and $x_{\rm max}$ constrain the region of validity of LaMET. Our baseline values are set as $\{x_{\rm min}, x_{\rm max}\} = \{0.2, 0.8\}$. Additional sets, specifically $\{0.15, 0.85\}$ and $\{0.25, 0.75\}$, were considered to assess sensitivity. The resulting systematic uncertainty is taken as the maximum deviation among these two choices from the central result.
\end{itemize}

The results of this systematic analysis are summarized in Fig.~\ref{fig:systematics} for both mock and lattice data. In each case, for brevity, we present only the results obtained for up valence quarks. The quantity shown is the relative difference between the default and alternative extractions as a function of the momentum fraction $x$. The relative statistical uncertainty, also evaluated with respect to the default extraction, is indicated in {\color{red}the upper panels of} Fig.~\ref{fig:systematics} as a band. We find that all considered systematic uncertainties are generally smaller than the statistical uncertainty and therefore remain under control. The only exception arises in large $x$ region ($x\gtrsim0.7$) from the choice of fitting methods in the case of lattice data. This effect is expected due to the poor quality of lattice results for imaginary contributions, which, in a combined fit, affects the extraction of PDFs for valence quarks. {\color{red}In addition, we display the up valence quark results with statistical and systematic uncertainties combined in the same plot, as shown in the lower panels of Fig.~\ref{fig:systematics}.} However, the values of PDFs are strongly suppressed in this region and thus, the effect is of rather small practical significance. {\color{red} Since the primary objective of this work is to demonstrate the proposed methodology, our systematic analysis focuses on the parameter choices within the framework itself. A comprehensive study of systematic effects inherent to the lattice data, such as discretization and finite-volume effects, will be investigated in future work.}
\section{Further discussions of GPD results}
\label{app:gpd_results}

\begin{figure*}
\centering
\includegraphics[width=0.49\columnwidth]{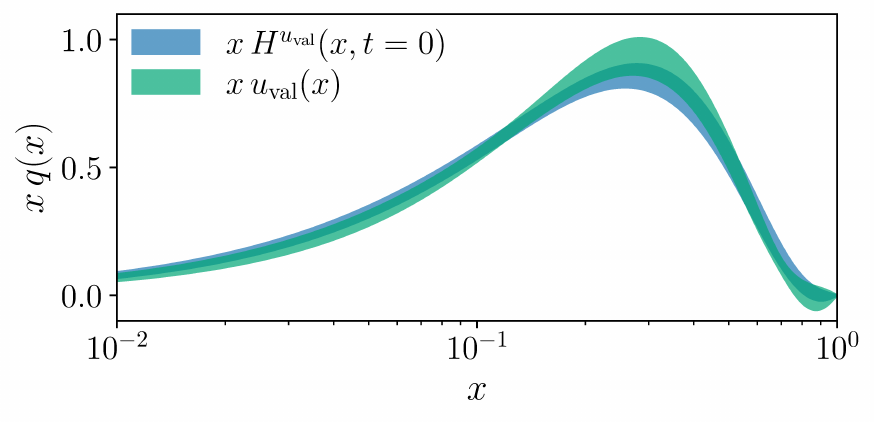}
\includegraphics[width=0.49\columnwidth]{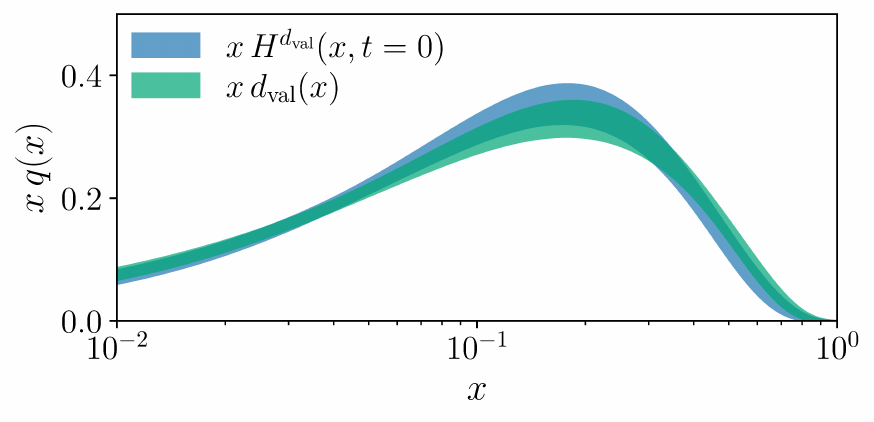}
\caption{Comparison of PDFs for up (left) and down (right) valence quarks obtained from the two lattice setups used in this study. Specifically, the green bands denote distributions obtained from lattice data explicitly calculated for PDFs, while the blue bands correspond to data calculated for GPDs at $\xi=0$. In the latter case, $t=0$ limits are shown.}
\label{fig:gpd:forward_limit}
\end{figure*}

This appendix presents additional studies of the GPD reconstruction. We begin by comparing the forward limits ($t \to 0$) of the reconstructed GPDs $H$ with the results directly obtained for the PDFs, as reported in Sec.~\ref{sec:pdf}. This comparison is essential, because the momentum smearing employed in the lattice calculations differs between the $t=0$ and $t \neq 0$ ensembles. Examining the $t \to 0$ limit of the $t \neq 0$ reconstruction therefore serves both to test the practical equivalence of these lattice setups and to assess the consistency of our reconstruction strategy across PDFs and $t$-dependent GPDs. Since current lattice inputs do not yet constrain sea-quark distributions reliably, we restrict this comparison to the valence distribution, as shown in Fig.~\ref{fig:gpd:forward_limit}. The observed agreement between $q_{\rm val}(x)$ and $H^q_{\rm val}(x, t=0)$ supports the compatibility of the lattice inputs and demonstrates the robustness of our method.

\begin{figure*}
\centering
\includegraphics[width=0.49\columnwidth]{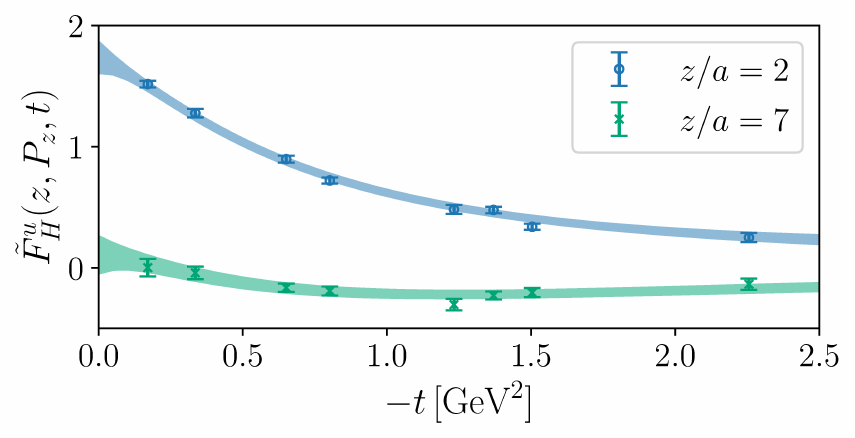}
\includegraphics[width=0.49\columnwidth]{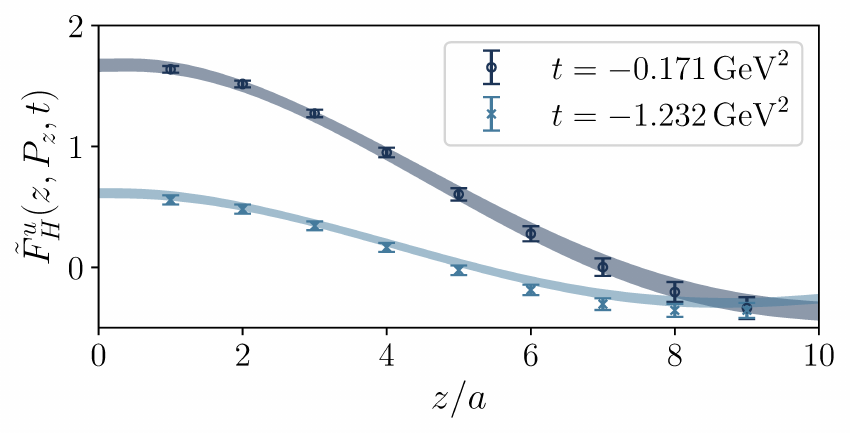}
\includegraphics[width=0.49\columnwidth]{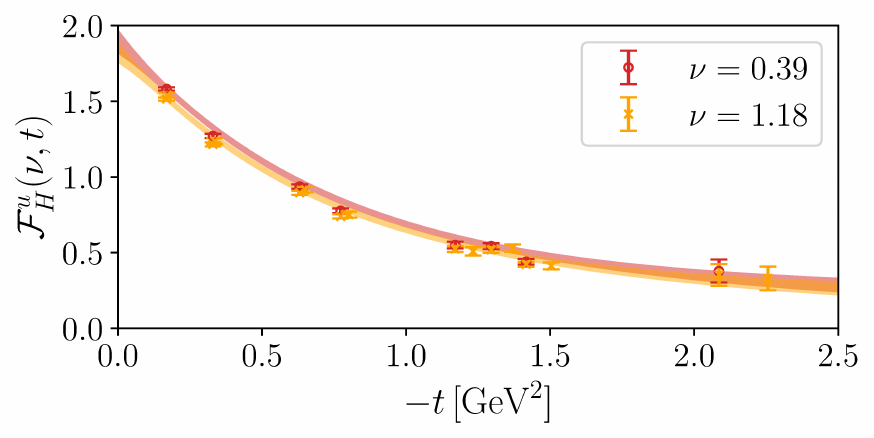}
\includegraphics[width=0.49\columnwidth]{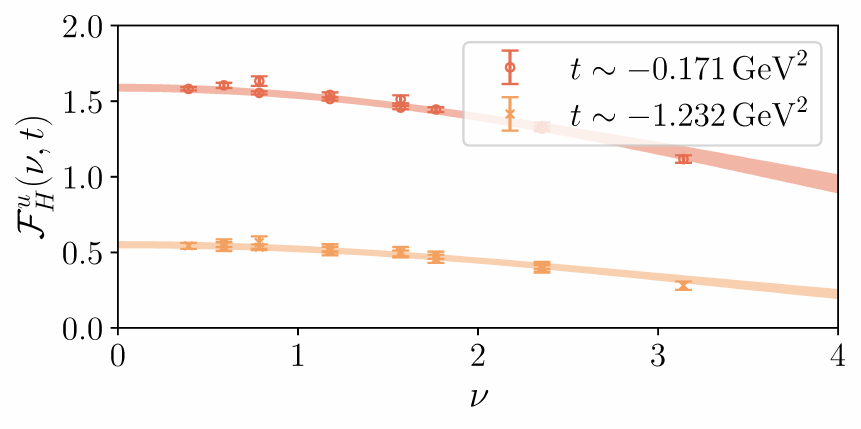}
\caption{Quality of the fits to LaMET (upper row) and SDE (lower row) quantities for mock and lattice data related to the GPD $H$ for up valence quarks. The quantities (points with error bars) are shown together with the fit results (bands) for several values of $z$ or $\nu$ and $t$, as functions of $t$ and of $z$ or $\nu$, respectively.}
\label{fig:gpd:fit}
\end{figure*}

Next, we quantify the quality of the GPD reconstruction from lattice data. Figure~\ref{fig:gpd:fit} compares the fitted results with the lattice inputs from two complementary viewpoints. The two left panels display the quasi matrix elements and the matched ITDs as functions of the $t$, whereas the two right panels show the corresponding dependence on $z$ and on the Ioffe time, respectively. Across all panels we find very good agreement, which validates the effectiveness of our unified approach utilizing neural-network reconstruction.

\begin{figure*}
\centering
\includegraphics[width=0.49\columnwidth]{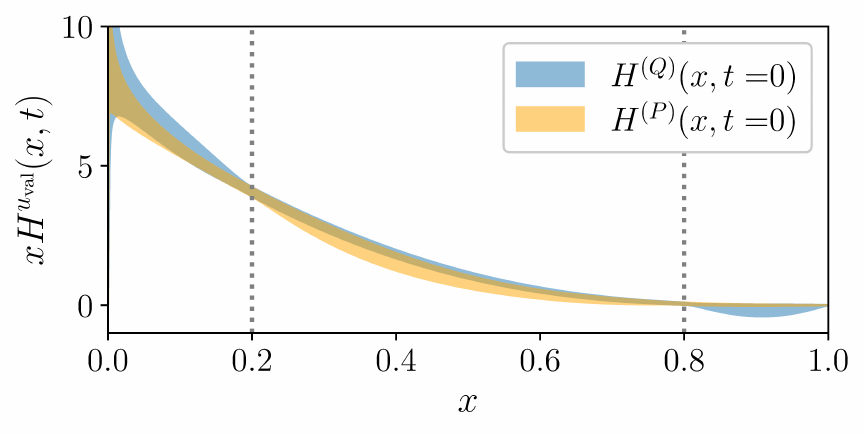}
\includegraphics[width=0.49\columnwidth]{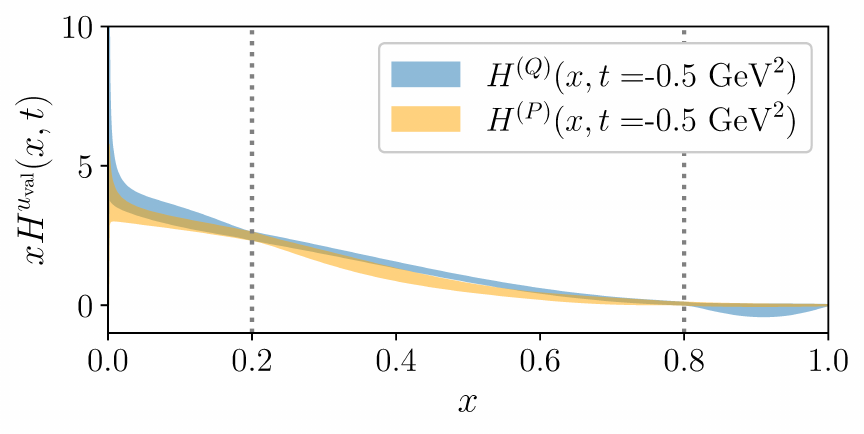}
\includegraphics[width=0.49\columnwidth]{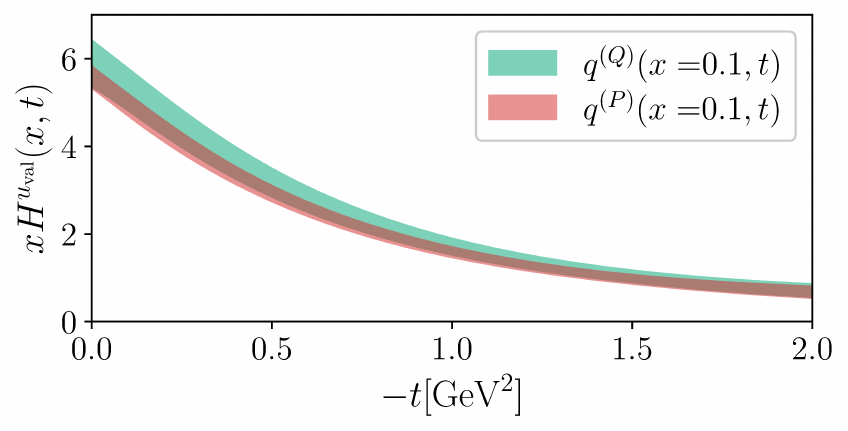}
\includegraphics[width=0.49\columnwidth]{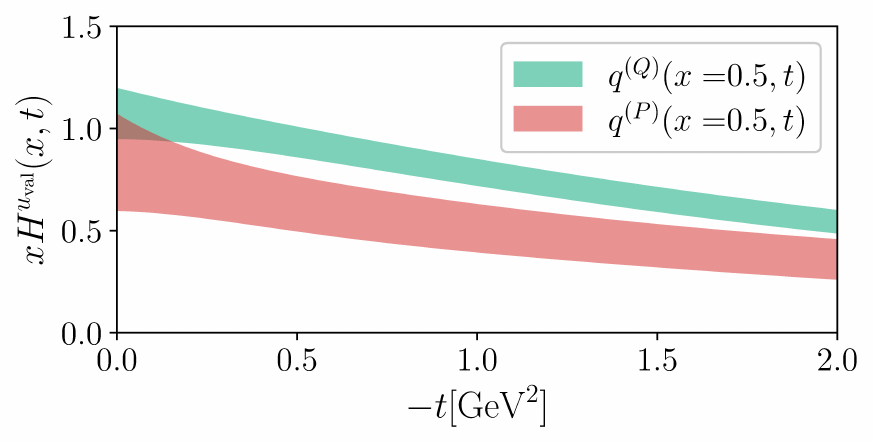}
\caption{Distributions $H^{u_{\mathrm{val}},(Q)}(x,0,t)$ and $H^{u_{\mathrm{val}},(P)}(x,0,t)$, fitted to LaMET and SDE quantities, respectively. The distributions are evaluated for two fixed values of $t$ (upper row) and $x$ (lower row), and are shown as functions of $x$ and $t$, respectively.}
\label{fig:gpd:HuQP}
\end{figure*}

We complement the analysis based on lattice inputs by presenting $q^{(Q)}(x,t)$ and $q^{(P)}(x,t)$ for the GPD $H$ of the valence up quark in Fig.~\ref{fig:gpd:HuQP}. The upper row displays the results at fixed values of $t$. As in the PDF case, $q^{(Q)}(x,t)$ primarily constrains $q^{(0)}(x,t)$ in the central region, while $q^{(P)}(x,t)$ governs the outer regions. Their complementarity demonstrates that combining LaMET and SDE information yields a controlled and reliable result. The lower row of Fig.~\ref{fig:gpd:HuQP} shows the $t$ dependence at selected fixed values of $x$, chosen to represent different kinematic regions. The pronounced differences between $q^{(Q)}(x,t)$ and $q^{(P)}(x,t)$ at fixed $x$ further highlight their distinct roles and reinforce this conclusion.

\begin{figure*}
\centering
\includegraphics[width=0.49\columnwidth]{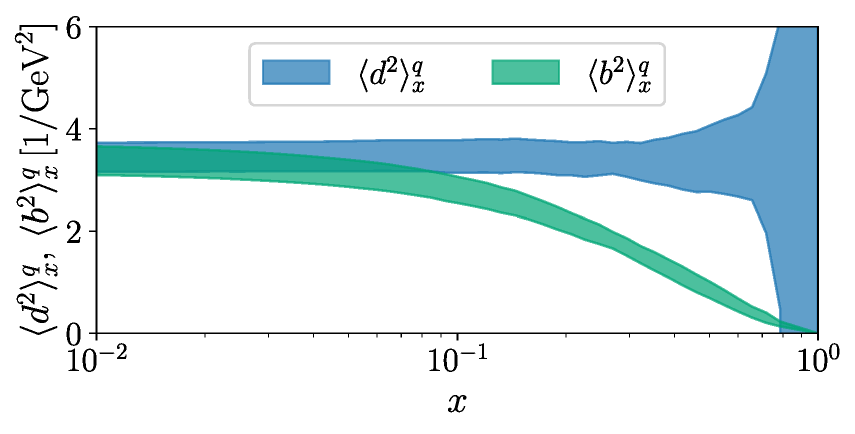}
\caption{The average squared distance between the active quark and the center of momentum, $\langle b^2 \rangle_x^q$, and between the active quark and the spectator system, $\langle d^2 \rangle_x^q$, as obtained from lattice data for up valence quarks.}
\label{fig:gpd:r2d2}
\end{figure*}

Finally, in Fig.~\ref{fig:gpd:r2d2} we present the estimates of the average squared distance between the active quark and the center of momentum, $\langle b^2 \rangle_x^q$, and between the active quark and the spectator system, $\langle d^2 \rangle_x^q$, obtained from lattice data for up valence quarks (see Eqs.~\eqref{eq:nt:ave_b} and~\eqref{eq:nt:ave_d}, respectively). This figure demonstrates that, with the careful choice of our Ansatz for the $t$-dependence, we ensure that in the limit $x \to 1$ the active parton is located at the origin of the reference frame, which is a strong theory-driven constraint (see Appendix~\ref{app:tomography}). Moreover, we guarantee that in this limit the nucleon size remains finite, i.e., the spectator system neither collapses onto nor separates to the infinity from the active parton. This is unfortunately not very well pronounced due to large uncertainty.

\bibliographystyle{JHEP}
\bibliography{bibliography}
\end{document}